\documentclass[11pt,a4paper]{article}
\usepackage{jheppub}
\usepackage{bvrshorthand}
\usepackage{tikz}
\usepackage{bbm}
\usepackage{siunitx}
\usepackage{adjustbox}
\usepackage{bbold}
\usepackage{marvosym}
\usepackage{multirow}
\usetikzlibrary{shapes.misc}
\usetikzlibrary{decorations.markings}
\usetikzlibrary{decorations.pathmorphing, patterns,shapes}
\usetikzlibrary{calc,shapes.geometric}
\usepackage[most]{tcolorbox}
\definecolor{lightgray}{gray}{0.9}
\usepackage[compat=1.1.0]{tikz-feynman}

\tikzset{cross/.style={cross out, draw=black, minimum size=2*(#1-\pgflinewidth), inner sep=0pt, outer sep=0pt},
	cross/.default={1pt}}

\title{On the $(\text{Fib} \boxtimes \text{Fib}) \rtimes S_2$ fusion category}
\date{}

\author{Maddalena Ferragatta, Balt C. van Rees}
\affiliation{CPHT, CNRS, Ecole Polytechnique, Institut Polytechnique de Paris, Palaiseau, France}
\abstract{There might exist non-rational Virasoro CFTs in two dimensions with a $(\text{Fib} \boxtimes \text{Fib}) \rtimes S_2$ categorical symmetry. We calculate the necessary ingredients for a modular conformal bootstrap analysis of these theories. After reviewing the basics of fusion categories, we present the irreducible representations, the lasso maps that intertwine between different Hilbert spaces, and finally the 22-by-22 modular S matrix. We highlight the peculiarities introduced by the non-invertible nature of the symmetry. This paper is written in a pedagogical manner and can therefore serve as an accessible entry point into the literature.}

\begin{document}

\maketitle

\section{Introduction}
The main motivation for this work is to cast light on the possible existence of non-rational Virasoro CFTs in two dimensions. We define these as unitary CFTs with $c > 1$, a normalizable vacuum, a discrete spectrum, a modular invariant torus partition function, and, most importantly, no conserved currents outside those of the Virasoro identity module.

Unlike their rational cousins, which are solvable \cite{Belavin:1984vu,Verlinde:1988sn,Moore:1988qv,Vafa:1988ag}, non-rational CFTs in two dimensions will most likely be as difficult to solve as higher-dimensional CFTs. The non-rational CFTs with only Virasoro currents are however significantly more enigmatic because to date the evidence for even their existence is scant: almost all the known two-dimensional RG flows are widely believed to either end in a gapped phase, at a rational theory, or perhaps at a non-rational theory with a bigger symmetry as in some supersymmetric theories. Other well-understood constructions seem to lead to a continuous spectrum, as in Liouville theory.

In the search for non-rational Virasoro CFTs, some of the more promising constructions are the putative fixed points of RG flows around $N$ coupled unitary Virasoro minimal models. The discussion was recently brought to the forefront by the putative perturbative constructions of \cite{Antunes:2022vtb,Antunes:2024mfb,Antunes:2025erb}, but perturbative and numerical calculations for such setups had been done before \cite{Dotsenko:1995dy,Dotsenko:1998gyp}, albeit with a different motivation. The work \cite{Dotsenko:2001zb} highlights (for the first time as far as we could tell) the potential non-rationality of the central charge of a closely related construction.

In \cite{Chang:2018iay} it was stressed that some of these constructions preserve part of the non-invertible \emph{symmetry category} \cite{Bhardwaj:2017xup} of the UV fixed point. This leads to the compelling idea that there might be non-rational Virasoro CFTs with a big categorical symmetry, which merits further exploration.

In \cite{Antunes:2025huk} the authors searched for numerical evidence for the existence of theories whose categorical symmetry contains several copies of the Fibonacci (or Fib) fusion category. Their starting point was the so-called Golden anyon chain \cite{Feiguin:2006ydp}, which is renowned for its manifest Fib symmetry and can be used to reach either the tricritical Ising or the three-state Potts model without any fine-tuning. The authors of \cite{Antunes:2025huk} suggested taking $n$ such chains and couple them with the single relevant operator that preserves $\text{Fib}^{\boxtimes n} \rtimes S_n$ times a small global symmetry that depends on the model.\footnote{Here $\boxtimes$ denotes the Deligne product where lines in different copies are transparent to each other.} Indeed, this construction is a natural discretization of the continuum flows between CFTs that had been proposed in the literature discussed above \cite{Dotsenko:1998gyp,Chang:2018iay}. For $n=2$ and $3$ the numerical investigations of \cite{Antunes:2025huk} (see also \cite{Blakeney:2025ext}) led to an interesting phase portrait, with a promising possible fixed point at $c \approx 1.77$ at $n=2$, and another possible fixed point with $c \approx 2.10$ at $n=3$. Neither of these appear to correspond to known theories.

\paragraph{}

In this work we make a first step toward a numerical conformal bootstrap analysis of these two-dimensional CFTs, which are characterized by a ``big'' non-invertible symmetry category. The numerical conformal bootstrap methods that began with \cite{Rattazzi:2008pe} provide an excellent tool with which candidate non-rational Virasoro CFTs should be investigated. Our work envisages a modular conformal bootstrap analysis using the method proposed in \cite{Hellerman:2009bu} (see \cite{Fitzpatrick:2023lvh} and \cite{Chiang:2023qgo} for a refinement which partially takes into account the integrality of the coefficients). This approach has been used to impressive precision for large $c$ theories \cite{Afkhami-Jeddi:2019zci}, is related to sphere-packing \cite{Hartman:2019pcd, Afkhami-Jeddi:2020hde}, and can be extended to theories with anomalous symmetries \cite{Lin:2019kpn,Lin:2021udi,Lanzetta:2022lze} and to BCFTs \cite{Collier:2021ngi}. Most importantly for our purposes, it is relatively straightforward to take into account non-invertible symmetries: this was first done in \cite{Lin:2023uvm} and more recently also in \cite{Albert:2025umy}.\footnote{One could also try to analyze four-point functions in order to try to bootstrap non-rational Virasoro CFTs, following the preliminary analysis in \cite{Kousvos:2024dlz}. A comprehensive analysis will however be more involved in that case, because of the complicated nature of the Virasoro conformal blocks and the need to work out tensor products of representations of the non-invertible symmetry. It would also be interesting to investigate the annulus partition function, requiring the techniques of \cite{Choi:2024tri}, or combine setups as in \cite{Meineri:2025qwz}.}

Concretely the goal of this paper is to present the modular T and S matrices for the fusion category
\begin{equation}
	(\text{Fib} \boxtimes \text{Fib}) \rtimes S_2\,,
\end{equation}
and to use the opportunity to give a detailed introduction to unitary fusion categories, their tube algebra \cite{ocneanu1994chirality}, and the computation of their representations. We will in particular explain how to understand the direct and semidirect product construction, how lasso maps can intertwine representations of different dimensions and discuss briefly the connection to the Drinfeld center. Our example contains several new structures that are absent from more canonical introductory examples like Fib and Ising. We therefore believe that this paper can nicely complement the existing lecture notes \cite{Schafer-Nameki:2023jdn,Shao:2023gho} and pedagogical explanations \cite{Lin:2022dhv,Bartsch:2023wvv,Aasen:2020jwb} aimed at a physics audience, and will thereby further lower the bar to new entrants in the field. At the same time, our work should perhaps also be seen as an appeal to the community of experts for the development of software, akin to GAP for groups \cite{GAP4}, which will automate such computations in the future.

The actual numerical modular bootstrap results will be reported elsewhere \cite{toappear}.

\section{Review of Fib}
In this section we will review the basic properties of the Fibonacci fusion category. Informally speaking, a fusion category is defined by a collection of oriented topological line defects on the plane and a prescription for how they can fuse together. See, \emph{e.g.}, \cite{Chang:2018iay,etingof2015tensor} for a more formal definition.

\begin{figure}[ht]
	\centering
	\includegraphics{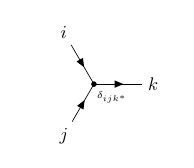}
	\caption{Junction $ i \times j \to k$.}
	\label{fig:junction}
\end{figure}

Our topological line defects, or ``simple objects'' will be denoted by $i, j, k,\ldots$. The line $i$ with its orientation reversed will be denoted with $i^*$. We denote with $\delta_{ijk}$ the number of trivalent vertices where three ingoing lines of type $i$, $j$, and $k$ can meet. The fusion rules are then\footnote{Unless stated otherwise, whenever a sum over defect lines appears, it is implicitly understood to run over the simple objects of the fusion category.}
\begin{equation}
	i \times j = \sum_{k} \delta_{ijk^*} k\,,
\end{equation}
which we capture graphically in figure \ref{fig:junction} (note the counterclockwise ordering of the indices). In our case $\delta_{ijk}$ will always be either 0 or 1. The identity line $1 = 1^*$ corresponds to no line at all, and therefore $\delta_{i i^* 1} = 1$ always. Finally to each network of lines there corresponds an orientation-reversed network and therefore $\delta_{ijk} = \delta_{j^*i^*k^*}$. 

The vacuum expectation value of a line of type $i$ is called the quantum dimension and denoted $d(i)$.\footnote{In this work we assume that there is no anomaly, so the expectation value of a closed loop on the plane is the same as that wrapping the cylinder in the vacuum state.}
 In correlation functions this means that any circular line of type $i$ that does not wrap around any cycles or operators can be replaced with $d(i)$ times the identity operator. This directly implies that the quantum dimensions must also obey the fusion rules:
\begin{equation}
	\label{eq: dimension}
d(i) d(j) = \sum_{k} \delta_{ijk^*} d(k)\, .
\end{equation}
The total dimension of the category is
\begin{equation}
\mathcal{D} = \sqrt{\sum_{k} d(k)^2} \, .
\end{equation}

The fusion rules stipulate what happens if we fuse, say, two loops without vertices wrapping the cylinder. The introduction of vertices allows for more complicated networks of lines. It is important to unambiguously define the vertex normalization. We will follow the usual conventions and define it through the so-called bubble move:
\begin{equation}
\adjincludegraphics[valign=c,scale=0.8]{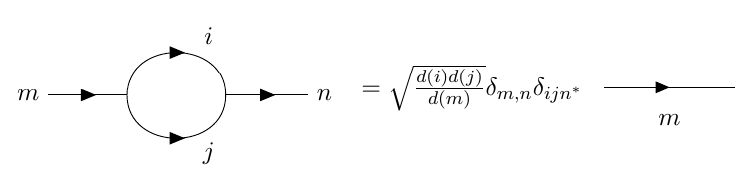}\, .
\end{equation}
We can then simplify some simple networks, for example:
\begin{equation}
\adjincludegraphics[valign=c,scale=0.8]{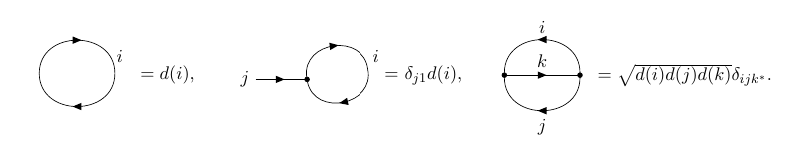}
\end{equation}

Associativity of the fusion operation is captured by the existence of so-called F symbols, defined graphically as follows (F move):
\begin{equation}
	\begin{aligned}
\adjincludegraphics[valign=cx]{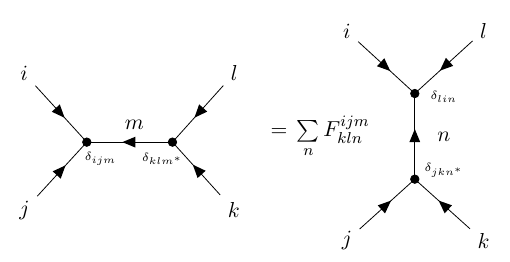}\, .
\end{aligned}
\end{equation}
The F symbols themselves must obey the famous pentagon identity, see for example \cite{Moore:1989vd}.

\subsection{Fib}
Let us exemplify our general discussion with the Fibonacci (Fib) category. It contains only one non-trivial line $\tau$ with a simple fusion rule:
\begin{align}
	\tau \times \tau &= 1 + \tau \,.
\end{align}
Since $\tau$ is self-dual, it has no orientation. The dimensions of the lines are $d(1) = 1$, $d(\tau) \equalscolon \xi$. The fusion rules impose $\xi^2= 1 + \xi$ and the solution with the golden ratio, $\xi = \frac{1 + \sqrt{5}}{2}$, defines the Fibonacci category.\footnote{The other solution is not consistent with unitarity, see subsection \ref{subsubsect:reflectionpositivity}.}

There are only two non-trivial F moves:
\begin{equation}
\label{FmovesFibgraphical}
\adjincludegraphics[valign=c,scale=0.6]{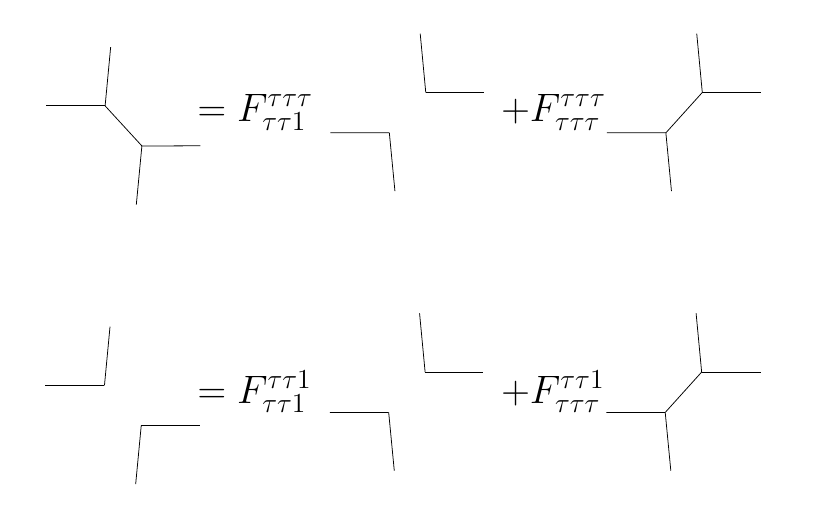}\,.
\end{equation}
It is easy to bootstrap the actual values of the four F symbols in these equations. For example, by contracting the four external legs pairwise we find constraints of the form:
\begin{equation}
\adjincludegraphics[valign=c,scale=0.7]{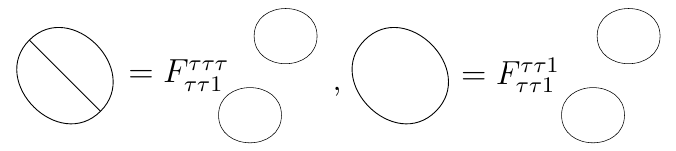}\, ,
\end{equation}
which are easily translated into numbers because of the graphical equalities given previously. Altogether we find the well-known result:
\begin{equation}
	\label{eq:FFib}
	\begin{pmatrix}
		F^{\tau \tau 1}_{\tau \tau 1} & F^{\tau \tau 1}_{\tau \tau \tau} \\
		F^{\tau \tau \tau}_{\tau \tau 1} & F^{\tau \tau \tau}_{\tau \tau \tau} \\
	\end{pmatrix}
	= 
		\begin{pmatrix}
		\frac{1}{\xi} & \frac{1}{\sqrt{\xi}} \\
		 \frac{1}{\sqrt{\xi}} & -\frac{1}{\xi}  \\
	\end{pmatrix} \, .
\end{equation}

\subsection{Hilbert spaces}
Let us continue our general discussion of fusion categories. In this section we will consider networks of lines on the cylinder. The basic building blocks for such networks are as follows:
\begin{center}
\includegraphics[scale=0.5]{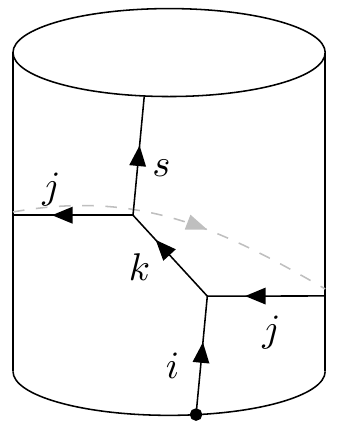}\,.
\end{center}
For a two-dimensional QFT on the cylinder this network corresponds to a so-called \emph{lasso map}
\begin{equation}
	\label{lasso map}
	\cL_{s \leftarrow i}^{jk} : \cH_i \to \cH_s 
\end{equation}
where $\cH_i$ is the (twisted) Hilbert space on the circle in the presence of the line $i$, and analogously for $\cH_s$.\footnote{The term `lasso map' is conventionally used for the action of the symmetry on the space of operators. Here we use it for the action on the (twisted) cylinder Hilbert space because we are concerned with CFTs for which the operator-state correspondence identifies the two.} Importantly, this map commutes with the translation symmetries and therefore preserves the energy and spin of the state. Our aim in this section will be to understand the twisted sector Hilbert spaces and the way they are constrained by the existence of the maps $\cL_{s \leftarrow i}^{jk}$.\footnote{If the QFT is also a CFT then the states in these Hilbert spaces are of course in one-to-one correspondence with local (non-topological) operators on which the topological lines can end.}

\subsubsection{The finite group case}
To see the usefulness of understanding the lasso maps, let us briefly recall what happens for a group-like symmetry. In that case there exists a line for every element $g$ of some finite group $G$, and the fusion rules just correspond to group multiplication.

First, consider the operator $\cL_{s \leftarrow i}^{jk}$ and choose $i = g$, $k = h$ with $g, h \in G$. Then necessarily $j = g^{-1}h$ and $s = h^{-1}g h$. The map
\begin{equation}
	\cL^{g^{-1}h \,\,  h}_{h^{-1}g h \leftarrow g}	
\end{equation}
then relates the $g$-twisted Hilbert space to the $h^{-1} g h$-twisted Hilbert spaces. Since the inverse map also exists, we reproduce the well-known result that the twisted Hilbert spaces of conjugate elements in $G$ are equivalent.

Second, consider maps from $\cH_g$ to itself by setting $s = i = g$ for some $g \in G$. If $j = h$ then the existence of the map requires $gh = h g$, meaning $h$ must lie in the centralizer $C_g$ of $g$. In what follows, we will denote the maps $\cL_{g \leftarrow g}^{h\, gh}$ by $\cL_g^h$.

The maps $\cL_g^h$ can then be combined into projectors onto the irreducible representations $\pi_g$ of $C_g$. Such projectors take the well-known form
\begin{align}
	\label{Projgroup}
	\hat P_{g}^{\pi_g} = \frac{\text{dim}(\pi_g)}{|C_g|} \sum \limits_{h \in C_g} \bar \chi_{\pi_g}(h) \cL_g^h\, ,
\end{align}
where $\chi_{\pi_g}\left(h\right)$ is the character of the representation of $h$ $\rho_{\pi_g}(h)$. Altogether we recover the familiar result that the twisted sector Hilbert spaces decompose into orthogonal subspaces corresponding to the irreducible representations of the centralizer group  \cite{dijkgraaf1989operator}.

\subsubsection{The tube algebra}
The first step in understanding the physics of the lasso maps is to understand the so-called \emph{tube algebra}. Its elementary relations read:
\begin{equation}
	\label{algebra_Lasso}
	\mathcal{L}_{i \leftarrow s}^{j k} \times \mathcal{L}_{s' \leftarrow i'}^{j' k'}   = \delta_{s s'} \sum_{m,n} \sqrt{\frac{d(j) d(j')}{d(m)}} F^{j k^* s}_{j' k'^* n^*} F^{i^* j^* k}_{j'^* n m^*} F^{ i' j' k'^*}_{j n^* m} \, \mathcal{L}_{i \leftarrow i'}^{m n}  \, .
\end{equation} 
This equation is found by concatenating two lasso maps on the cylinder and reducing the resulting network to a linear combination of single lasso maps, as follows:
\begin{equation}
	\begin{aligned}
	\adjincludegraphics[valign=cx,scale=0.7]{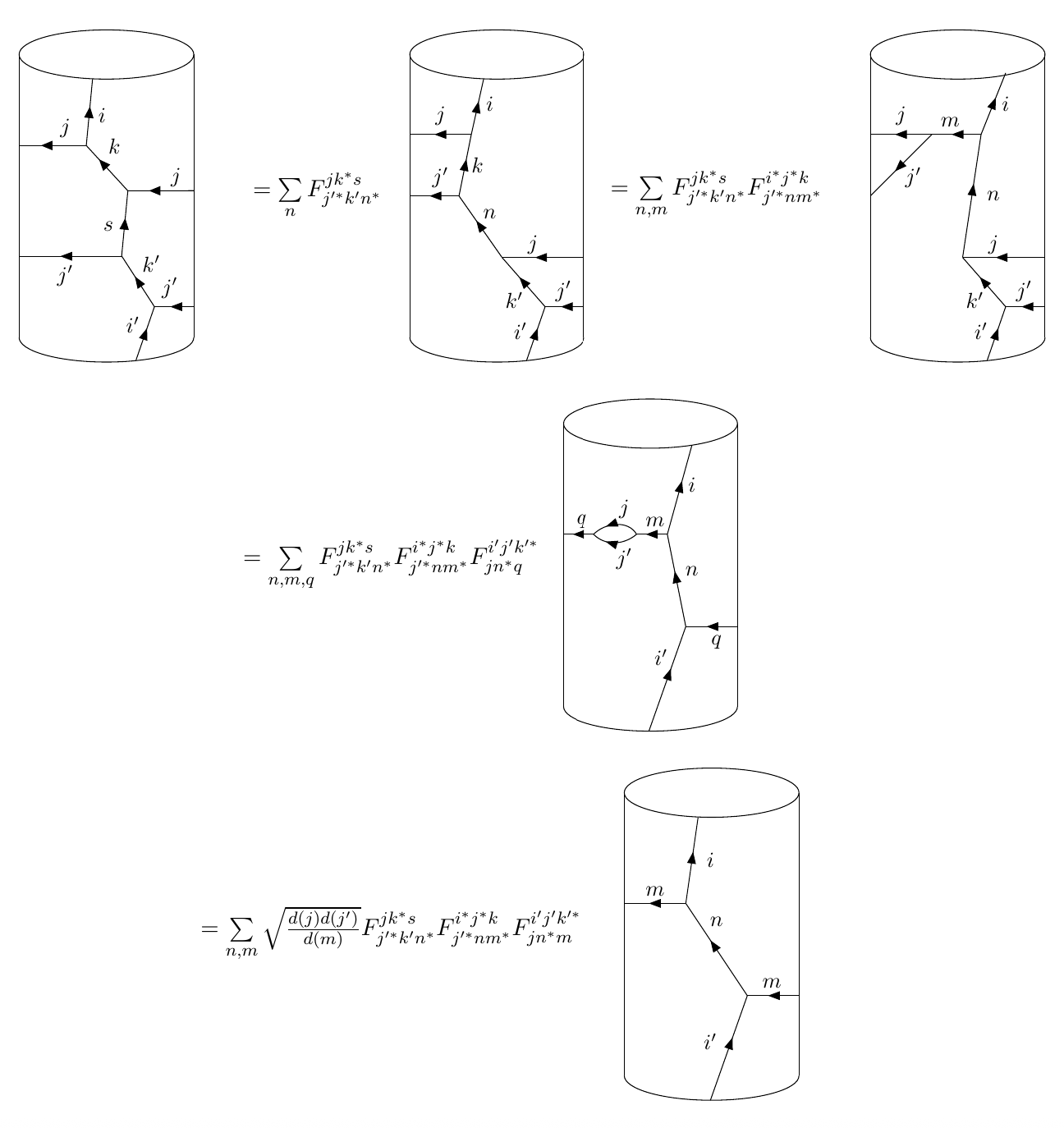}\,.
	\end{aligned}
\end{equation}
The fusion of the lasso maps has to be read, as usual for operators, from right to left. It corresponds to a network on the cylinder that has to be read from bottom to top.

\subsubsection{Minimal central idempotents for Fib}
\label{subsec:centralIdemp}
Let us first discuss the \emph{minimal central idempotents}, which are the analogues of the aforementioned projectors $\hat P_{g}^{\pi_g}$. Since in this case $s = i$ always, we will use the shorthand notation
\begin{equation}
	\cL_{i}^{jk} \colonequals \cL_{i \leftarrow i}^{jk}\,.
\end{equation}
For Fib there are 2 Hilbert spaces, so $i$ can be either `1' and `$\tau$'.

\paragraph{Untwisted Hilbert space\\}
In the untwisted Hilbert space $\cH_1$ we have two lasso operators:
\begin{equation}
	\cL_{1}^{11} \equalscolon \id_1, \qquad \cL_{1}^{\tau \tau} \equalscolon \tau\,.
\end{equation}
The tube algebra is therefore just the fusion algebra of the defect $\tau$ itself. This immediately implies that $\tau$ must have eigenvalue either $\xi$ or $-1/\xi$, and a simple calculation then shows that:
\begin{equation}
	\label{projectors_untwistedFib}
\begin{split}
	\hat{P}_1^{(1)} &= \frac{1}{1 + \xi^2} \left( \id_1 +\xi \tau  \right), \\
	\hat{P}_1^{(2)} &= \frac{1}{1 + 1/\xi^2} \left( \id_1 -\xi^{-1} \tau  \right)\,,
\end{split}
\end{equation}
are the orthogonal projectors onto the corresponding subspaces of $\cH_1$. In particular, we know that the vacuum $|\Omega\rangle$ has eigenvalue $d(\tau) = \xi$, so $\hat P_1^{(1)} |\Omega\rangle = |\Omega\rangle$ and $\hat P_1^{(2)} |\Omega\rangle = 0$.

\paragraph{Twisted Hilbert space\\}
Let us now show that the twisted Hilbert space splits into three orthogonal subspaces. First we note that there are three operators:
\begin{equation}
	\cL_{\tau}^{1 \tau} \equalscolon \id_\tau, \qquad \cL_{\tau}^{\tau 1} \equalscolon T_\tau, \qquad \cL_{\tau}^{\tau \tau}\, .
\end{equation}
On the cylinder, they correspond to the following networks:
\begin{center}
	\includegraphics[scale=0.7]{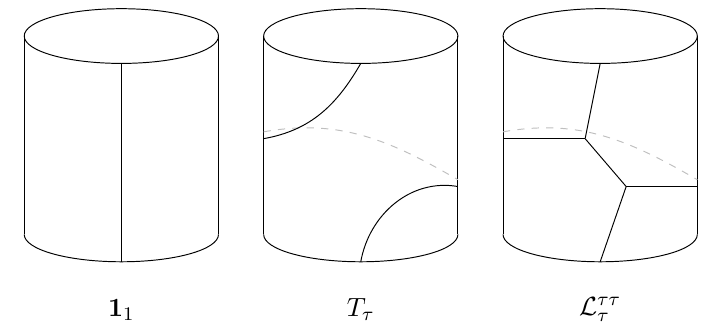}\, .
\end{center}
This figure also makes clear that $T_\tau$ is just a $ 2\pi$ rotation (or Dehn twist) of the cylinder\footnote{We have used the letter $T$ in anticipation of the following section, where we identify both ends of the cylinder to form a torus and then $T_\tau$ will correspond to the modular T transformation in the presence of the vertical $\tau$ line.}, so:
\begin{equation}
	T_\tau = e^{2 \pi i P_\tau}\,,
\end{equation}
with $P_\tau$ the generator of an infinitesimal rotation of the cylinder in the $\tau$-twisted sector. In CFTs $P_\tau$ also corresponds to a rotation on the plane, so its eigenvalues are just the spins of the operators on which a $\tau$ line can end. Knowing the eigenvalue of $T_\tau$ is therefore equivalent to knowing the spin of such operators modulo an integer.

The tube algebra for the lasso operators acting within $\cH_\tau$ is then generated by:
\begin{align}
	\label{algebraFib2}
	\mathcal{L}_\tau^{\tau \tau} \times \mathcal{L}_\tau^{\tau \tau}  &=  - \xi^{-5/2} \mathcal{L}_\tau^{\tau \tau} - \xi^{-1} \id_\tau+ T_\tau, \\
	\label{algebraFib3}
	T_\tau \times T_\tau  &=  \phantom{-}\xi^{-1/2} \mathcal{L}_\tau^{\tau \tau} + \xi^{-1} \id_\tau, \\
	\label{algebraFib4}
	\mathcal{L}_\tau^{\tau \tau} \times  T_\tau  &=  - \xi^{-1} \mathcal{L}_\tau^{\tau \tau} + \xi^{-1/2} \id_\tau \, .
\end{align}
These relations for Fib of course follow from the general equation \eqref{algebra_Lasso}, although it is a satisfying exercise to derive them graphically using the simple F moves of equation \eqref{FmovesFibgraphical} with the coefficients of equation \eqref{eq:FFib}.

Next we note that it is possible to diagonalize all three operators at once, even though the above produces three equations for two unknown eigenvalues (of course $\id_\tau$ always has eigenvalue $+1$). The possible eigenvalues are:
\begin{center}
\renewcommand{\arraystretch}{2}
\begin{tabular}{c|ccc}
		& $\id_\tau$ & $T_\tau$ & $\cL_\tau^{\tau \tau}$ \\
		\hline
	(1) & 1 & 1 & $\displaystyle \frac{\sqrt{\xi}}{1 + \xi}$\\
	(2) & 1 & $e^{4 \pi i /5}$ & $\displaystyle \frac{\sqrt{\xi}}{1 +e^{4 \pi i /5} \xi}$ \\
	(3) & 1 & $e^{-4 \pi i / 5}$ & $\displaystyle  \frac{\sqrt{\xi}}{1 +e^{-4 \pi i /5} \xi}$
\end{tabular}
\end{center}
As promised, we find three possible superselection sectors, two of which have fractional spins $\pm 2/5 + \Z$. The corresponding three projectors are easily found to be:
\begin{align}
	\label{ProjectorsTwistedFib}
	\hat{P}_\tau^{(1)} &= \frac{1}{1 + \xi^2} \left( \xi \id_\tau +\xi T_\tau + \frac{1}{\sqrt{\xi}} \cL_\tau^{\tau \tau} \right) , \nn \\ 
	\hat{P}_\tau^{(2)} &= \frac{1}{1 + \xi^2} \left(\id_\tau +(\xi e^{-3\pi i /5} + e^{3\pi i/5})T_\tau + e^{3\pi i /5} \sqrt{\xi} \, \cL_\tau^{\tau \tau} \right) , \\
	\hat{P}_\tau^{(3)} &=\frac{1}{1 + \xi^2} \left(\id_\tau +(\xi e^{3\pi i /5} + e^{-3\pi i/5})T_\tau + e^{-3\pi i /5} \sqrt{\xi} \, \cL_\tau^{\tau \tau} \right) \nn \, .
\end{align}

The construction of the minimal central idempotents is easily generalizable. Suppose one is in a given Hilbert space $\cH_i$, with networks $\cL_{i}^{j k}$. Suppose also that in a given representation $(\Gamma)$ they take values $\rho^{(\Gamma)}_i\left( \cL_{i}^{j k}\right)$, which have trace $\chi^{(\Gamma)}_i\left( \cL_{i}^{j k}\right)$. As for a group, the structure of the projector is then
\begin{align}
	\label{projectorsInGeneral}
	\hat P_i^{(\Gamma)} = N_i^{(\Gamma)}\sum_{j,k} \bar \chi^{(\Gamma)}_i\left( \cL_{i}^{j k}\right) \cL_{i}^{j k}\, ,
\end{align}
where $N_i^{(\Gamma)}$ is a constant to be fixed requiring that the projectors square to themselves.\footnote{The lecture notes \cite{etingof2009introduction} provide an excellent reference for these and many other standard results in the representation theory of finite-dimensional algebras.}

In the following, we will denote as $\cH_i^{(\Gamma)}$ the subspace of $\cH_i$ selected by the projector $\hat P_i^{(\Gamma)}$. 

\subsubsection{Further lassos} 
\label{subsec:Lasso}
Let us now consider what happens when one examines the lasso maps $\cL_{i \leftarrow s}^{jk}$ for $s \neq i$. For Fib the only two such maps
are
\begin{equation}
	\label{nontrivialLassomapsFib}
	\mathcal{L}_{\tau \leftarrow 1}^{\tau \tau}, \qquad \mathcal{L}_{1 \leftarrow \tau }^{\tau \tau}\,.
\end{equation}
These lassos map states in $\cH_1$ to states in $\cH_\tau$ with the same energy and spin. They can however have non-trivial kernels and co-kernels and a little computation is needed to understand their action in detail.

First, we note that:
\begin{align}
	\mathcal{L}_{\tau \leftarrow 1}^{\tau \tau} \times \hat P_1^{(1)}  &= 0, \nn \\
	 \mathcal{L}_{\tau \leftarrow 1}^{\tau \tau} \times \hat P_1^{(2)}  &=  \mathcal{L}_{\tau \leftarrow 1}^{\tau \tau}  \,,
\end{align}
and therefore $\mathcal{L}_{\tau \leftarrow 1}^{\tau \tau}$ can only act  non-trivially on the states in the eigenspace (2) of the untwisted sector. Similarly, $\mathcal{L}_{1 \leftarrow \tau}^{\tau \tau}$ can only act non-trivially on the eigenspace (1) of $\cH_\tau$ because:
\begin{align}
	\mathcal{L}_{1 \leftarrow \tau}^{\tau \tau} \times \hat P_\tau^{(1)} &= \mathcal{L}_{1 \leftarrow \tau}^{\tau \tau} , \nn  \\
	\mathcal{L}_{1 \leftarrow \tau}^{\tau \tau} \times \hat P_\tau^{(2)} &=  0,  \\
	\mathcal{L}_{1 \leftarrow \tau}^{\tau \tau} \times \hat P_\tau^{(3)} &=  0 \, . \nn
\end{align}
The above equations do not guarantee that the action indeed \emph{is} non-trivial in the given subspaces. However, we can also compute
\begin{align}
	\label{Lasso1tau}
	\mathcal{L}_{1 \leftarrow \tau}^{\tau \tau} \times \mathcal{L}_{\tau \leftarrow 1}^{\tau \tau}   &= \frac{1 + \xi^2}{\xi^{3/2}} \hat P_1^{(2)} ,\\
	\label{Lassotau1}
	\mathcal{L}_{\tau \leftarrow 1}^{\tau \tau} \times \mathcal{L}_{ 1 \leftarrow \tau}^{\tau \tau}  &= \frac{1 + \xi^2}{\xi^{3/2}} \hat P_\tau^{(1)},
\end{align}
which shows that the two lasso operators under consideration are each other's inverse in the given subspaces. An invertible operator must have a trivial kernel, so the Hilbert spaces that $\hat P_\tau^{(1)}$ and $\hat P_1^{(2)}$ project on must have identical spectra.

\subsubsection{Reflection positivity}
\label{subsubsect:reflectionpositivity}
Given an element of the tube algebra, so a network of lines on the cylinder, we can define a conjugate network by reflecting the cylinder vertically and changing every line $i$ to its conjugate $i^*$, which is the same as reversing the orientation. For a simple network this operation can be combined with an F move to obtain
\begin{equation}
	(\cL^{jk}_{s \leftarrow i})^\dagger = \sum_n F^{jsk^*}_{j^*i^*n} \, \cL^{jn}_{i \leftarrow s}\, , 
\end{equation}
which we can represent graphically as:
\begin{center}
	\includegraphics[scale = 0.8]{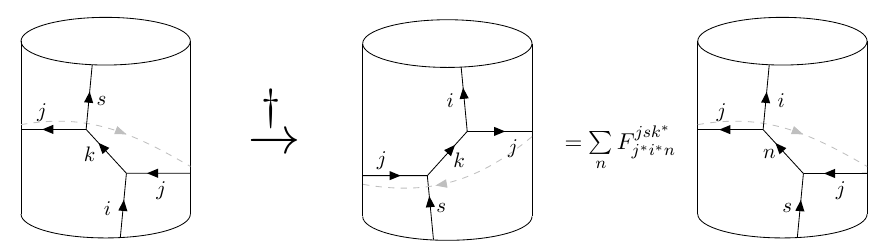}\,.
\end{center}
In a reflection positive theory the combined operation
\begin{equation}
	(\cL^{jk}_{s \leftarrow i})^\dagger \cL^{jk}_{s \leftarrow i}: \cH_i \to \cH_i
\end{equation}
should be positive semidefinite.

We note that unitarity also imposes constraints on the quantum dimensions. Consider the vacuum expectation value of a line $i$ on the cylinder. After compactifying it to a sphere, we see that this vev is related to that of $i^*$ by a simple 180-degree rotation of the sphere and therefore
\begin{equation}
	d(i) = d(i^*)\,.
\end{equation}
By rotating the sphere 90 degrees we see that this expectation value is also an inner product of a state in $\cH_{i,i^*}$, the Hilbert space with both $i$ and $i^*$ inserted, so in a unitary theory we must have
\begin{equation}
	d(i) \geq 0.
\end{equation}
Finally, since the quantum dimensions obey the fusion rules and $d(1) = 1$, we find that
\begin{equation}
	d(i)^2 = d(i^*) d(i) = 1 + \ldots
\end{equation}
where the dots are not negative. Therefore
\begin{equation}
	d(i) \geq 1
\end{equation}
in a reflection positive theory.\footnote{Slightly different derivations of these same results exist, see for example \cite{Bhardwaj:2017xup,Chang:2018iay}.}

\paragraph{Other consequences on the lasso maps\\}
In the Fib case the two lasso maps $\cL_{1 \leftarrow \tau}^{\tau \tau}$ and $\cL_{\tau \leftarrow 1}^{\tau \tau}$ are hermitian conjugates one of the other. Moreover, being simple linear maps from $\cH_\tau^{(1)}$ to $\cH_1^{(2)}$ and vice versa they are represented as simple complex numbers, which we denote as $\rho_{1, \tau}^{(2, 1)}\left(\cL_{1 \leftarrow \tau}^{\tau \tau} \right)$ and $\rho_{\tau, 1}^{(1, 2)}\left(\cL_{\tau \leftarrow 1}^{\tau \tau} \right)$. The phase of these numbers depends on a choice of orthonormal basis, but their magnitude can be determined by hermitian conjugation and equations  \eqref{Lasso1tau} and \eqref{Lassotau1}, which yield
\begin{align}
	\rho_{\tau,1}^{(1, 2)}(\mathcal{L}_{\tau \leftarrow 1}^{\tau \tau}) = ( \rho_{1,\tau}^{(2 ,1)}(\mathcal{L}_{1 \leftarrow \tau}^{\tau \tau}) )^*\, , \quad \left| \rho_{1,\tau}^{(2, 1)}(\mathcal{L}_{1 \leftarrow \tau}^{\tau \tau}) \right|^2 =  \frac{1 + \xi^2}{\xi^{3/2}}\, .
\end{align}

More generally, consider a lasso map $\cL_{i \leftarrow s}^{j k}$ and suppose there exist two Hilbert spaces $\cH_i^{(\Gamma)}$ and $\cH_s^{(\Lambda)}$, of dimension respectively given by $\text{dim}(\Gamma)$ and $\text{dim}(\Lambda)$, that are intertwined. Then, the maps
\begin{align}
	\cL_{i \leftarrow s}^{(\Gamma \leftarrow \Lambda), jk} &\equalscolon \hat P_i^{(\Gamma)} \times \cL_{i \leftarrow s}^{j k} \times \hat P_s^{(\Lambda)}: \cH_{s}^{(\Lambda)} \to \cH_{i}^{(\Gamma)} 
\end{align}
and its hermitian conjugate one can be viewed respectively as a $\text{dim}(\Lambda) \times \text{dim}(\Gamma)$ matrix $\rho_{i,s}^{(\Gamma,\Lambda)}\left(\cL_{i \leftarrow s}^{jk}\right)$ and its conjugate transpose matrix. These matrices are constrained by the tube algebra up to multiplication with a $\dim(\Gamma)$ unitary matrix on the left and an independent $\dim(\Lambda)$ unitary on the right.

\subsubsection{From the tube algebra to the Drinfeld center}
\label{sec:Drinfeld center}
We have seen that the lasso maps $\cL_{i}^{jk}$ solving the tube algebra \eqref{algebra_Lasso} are represented in terms of matrices $\rho^{(\Gamma)}_i \left(\cL_{i}^{jk}\right)$ acting on the twisted Hilbert space $\cH_i$. However, as we saw in subsection \ref{subsec:Lasso} the tube algebra contains more information than just these representation matrices. In particular, it also determines the maps that intertwine the projectors associated to different Hilbert spaces.

The existence of such intertwiners implies that certain representations living in different $\cH_i$ are not independent. Instead, they are linked together as parts of a single coherent structure. For example, suppose we have the representations $(\Gamma)$ in $\cH_i$, $(\Lambda)$ in $\cH_j$, $(\Omega)$ in $\cH_k$, and so on, which are related by intertwiners. We might associate to them a formal (not necessarily simple) object $X$ in the category $\cC$, defined as
\begin{equation}
	\label{defX}
	X = \text{dim}(\Gamma) i + \text{dim}(\Lambda) j + \text{dim}(\Omega) k + \dots \, , 
\end{equation}
where the dots stand for all other representations that intertwine with these. 

Taking Fib as an example, this method associates the object $X = 1$ to the representation of $\hat P_1^{(1)}$, $X = \tau$ to both $\hat P_\tau^{(2)}$ and $\hat P_\tau^{(3)}$ and $X = 1 + \tau$ to the set $\{\hat P_1^{(2)},\hat P_\tau^{(1)}\}$.

Notice that the object $X$ only keeps track of the type and dimension of the representations. It does not contain any information about the explicit representation matrices and therefore does not entirely characterize a representation. It may therefore happen that two different sets $\{(\Gamma,i), (\Lambda,j), (\Omega,k), \dots\}$ and $\{(\Gamma',i), (\Lambda',j), (\Omega',k), \dots\}$ give rise to the same object $X$. Indeed, already for Fib the object $X = \tau$ corresponds to two different representations.

All the information about the representations of a fusion category $\cC$ can nevertheless be packaged into a new fusion category, known as the Drinfeld center $\cZ(\cC)$.\footnote{We will only give the briefest of reviews here. The interested reader is encouraged to consult \emph{e.g.} \cite{Lin:2022dhv,Lootens:2022avn,Bhardwaj:2023wzd,Bhardwaj:2023ayw} for reviews, applications, the connection to three-dimensional TFTs, and references to the original literature.} A simple object of $\cZ(\cC)$ is a pair $(X,e_X)$, where $X$ is an object of $\cC$ as in equation \eqref{defX}, and $e_X$, called the \emph{half-braiding}, contains the information about the representation matrices and the intertwiners.\footnote{The fusion rules in the Drinfeld center correspond to the decomposition of tensor products of irreducible representations, which is an operation that we will not consider here, but that will be important when studying four-point functions.}

A convenient way to think about the half-braiding $e_X$ is that it distinguishes between the different realizations of the same formal object $X$. Therefore for Fib we expect one half-braiding associated to $X = 1$, two half-braidings associated to $X = \tau$ and one half-braiding associated to $X = 1 + \tau$. 

Concretely, a half-braiding is given by a collection of linear maps
\begin{equation}
	e_X(i): X \otimes i \to i \otimes X\, ,
\end{equation}
defined for every simple object $i$ in $\cC$ (and extended to arbitrary objects by linearity and fusion). 
On the plane, we can represent the half-braiding $e_X(i)$ graphically as
\begin{center}
	\begin{tikzpicture}
	\draw [postaction={decorate},
decoration={markings, mark=at position 0.7 with {
		\arrow{Triangle[width=4pt,length=5.5pt]}
}}](1,-1) -- (-1,1) ;

	\draw [thick,postaction={decorate},
decoration={markings, mark=at position 0.6 with {
		\arrow{Triangle[width=4pt,length=5.5pt]}
}}](-1,-1) -- (-0.05,-0.05) ;

	\draw [thick, postaction={decorate},
decoration={markings, mark=at position 0.6 with {
		\arrow{Triangle[width=4pt,length=5.5pt]}
}}](0.05,0.05) -- (1,1);

\node at (-1,0.6){$i$};
\node at (1,0.6){$X$};
\node at (-1,-0.6){$X$};
\node at (1,-0.6){$i$};
	\end{tikzpicture}
	\, . 
\end{center}
The half-braiding is natural, meaning it is independent of the choice of basis or decomposition, and its defining property is that it must be compatible with the fusion in $\cC$:
\begin{equation}
	\label{eq:fusionHALFBR}
	\begin{aligned}
	\begin{tikzpicture}
		\draw [postaction={decorate},
		decoration={markings, mark=at position 0.7 with {
				\arrow{Triangle[width=4pt,length=5.5pt]}
		}}](1,-1) -- (-1,1) ;
		
		\draw [thick,postaction={decorate},
		decoration={markings, mark=at position 0.7 with {
				\arrow{Triangle[width=4pt,length=5.5pt]}
		}}](-0.5,-0.5) -- (-0.05,-0.05) ;
		
		\draw [thick, postaction={decorate},
		decoration={markings, mark=at position 0.6 with {
				\arrow{Triangle[width=4pt,length=5.5pt]}
		}}](0.05,0.05) -- (1,1);
	
	\draw [yshift = -0.55cm, xshift = -0.55cm, postaction={decorate},
decoration={markings, mark=at position 0.7 with {
		\arrow{Triangle[width=4pt,length=5.5pt]}
}}](1,-1) -- (-1,1) ;

		\draw [thick, postaction={decorate},
decoration={markings, mark=at position 0.6 with {
		\arrow{Triangle[width=4pt,length=5.5pt]}
}}](-1.5,-1.5) -- (-0.55,-0.55);
		
		\node at (-1,0.6){$j$};
		\node at (1,0.6){$X$};
		\node at (1,-0.6){$j$};
			\node[yshift = -0.55cm, xshift = -0.55cm] at (-1,0.6){$i$};
				\node[yshift = -0.55cm, xshift = -0.55cm] at (1,-0.6){$i$};
				\node at (-1.5,-1.1){$X$};
				
				\node at (2,-0.2){$=$};
				
					\draw [xshift = 4cm, postaction={decorate},
				decoration={markings, mark=at position 0.7 with {
						\arrow{Triangle[width=4pt,length=5.5pt]}
				}}](1,-1) -- (-1,1) ;
				
				\draw [xshift = 4cm,thick,postaction={decorate},
				decoration={markings, mark=at position 0.6 with {
						\arrow{Triangle[width=4pt,length=5.5pt]}
				}}](-1,-1) -- (-0.05,-0.05) ;
				
				\draw [xshift = 4cm,thick, postaction={decorate},
				decoration={markings, mark=at position 0.6 with {
						\arrow{Triangle[width=4pt,length=5.5pt]}
				}}](0.05,0.05) -- (1,1);
				
				\node[xshift = 4cm] at (-1.1,0.5){\small $i \times j$};
				\node[xshift = 4cm] at (1,0.6){$X$};
				\node[xshift = 4cm] at (-1,-0.6){$X$};
				\node [xshift = 4cm]at (1.1,-0.5){\small $i \times j$};
	\end{tikzpicture}
	\, .
		\end{aligned}
\end{equation}
In this way, the half-braiding collects all the data that is lost when passing from the explicit tube algebra representation to the formal object $X$ alone. 

To work out the half-braidings from the constraint \eqref{eq:fusionHALFBR} we must pick a convenient basis. Let us consider the case when $X$ is a direct sum of simple lines of $\cC$ with unit coefficients. On the cylinder, $(X,e_X(j))$ can be expressed as
\begin{equation}
		\label{basisHalfBraidingsSum}
	\begin{aligned}
		\begin{tikzpicture}
			
			\draw[xshift = -8cm] (1.42,3.3) ellipse (1.42 and 0.4);
			\draw[xshift = -8cm]  (0,3.3) -- (0,0.4);
			\draw[xshift = -8cm]  (2.84,3.3) -- (2.84,0.4);
			\draw[xshift = -8cm]  (2.84,0.4) arc (0:-180:1.42 and 0.4);
			
			\draw[dashed, color =gray!50, xshift =  -8cm,postaction={decorate},
			decoration={markings, mark=at position 0.5 with {
					\arrowreversed{Triangle[width=4pt,length=5.5pt]}
			}}] (2.84,2) arc (0:180:1.42 and 0.4);

			\draw[xshift =  -8cm,postaction={decorate},
			decoration={markings, mark=at position 0.7 with {
					\arrow{Triangle[width=4pt,length=5.5pt]}
			}}]  (2.84,2) arc (0:-180:1.42 and 0.4);
			
			\draw [xshift =  -8cm,postaction={decorate},
			decoration={markings, mark=at position 0.6 with {
					\arrow{Triangle[width=4pt,length=5.5pt]}
			}}] (1.4,0) -- (1.61,1.53) ;
			
			\draw [xshift =  -8cm,postaction={decorate},
			decoration={markings, mark=at position 0.6 with {
					\arrow{Triangle[width=4pt,length=5.5pt]}
			}}] (1.63,1.64) -- (1.8,2.9);
			
			\node[xshift = -8cm] at (2.3,1.4){\small $j$};
			\node[xshift = -8cm]  at (1.3,0.5){\small $X$};
			\node[xshift = -8cm] at (2,2.3){\small $X$};
			
			\node at (-3,1.7) {$= \sum\limits_{\substack{i, s \in X , \\ k}} \sqrt{\frac{d(k)}{d(i) d(j)}} e_{i \leftarrow s}^{j k}$};
			\draw[xshift = -1cm]  (1.42,3.3) ellipse (1.42 and 0.4);
			\draw[xshift = -1cm] (0,3.3) -- (0,0.4);
			\draw[xshift = -1cm] (2.84,3.3) -- (2.84,0.4);
			\draw[xshift = -1cm] (2.84,0.4) arc (0:-180:1.42 and 0.4);
			\draw[xshift = -1cm,dashed,color =gray!50, postaction={decorate},
			decoration={markings, mark=at position 0.8 with {
					\arrow{Triangle[width=4pt,length=5.5pt]}
			}}] (0,1.9) to[out=10, in=150] (2.84, 1.2);
			
			\draw [xshift = -1cm,postaction={decorate},
			decoration={markings, mark=at position 0.6 with {
					\arrow{Triangle[width=4pt,length=5.5pt]}
			}}](1.3,1.9) -- (0,1.9)  ;
			
			\draw [xshift = -1cm,postaction={decorate},
			decoration={markings, mark=at position 0.6 with {
					\arrow{Triangle[width=4pt,length=5.5pt]}
			}}](1.9,1) -- (1.3,1.9) ;
			
			\draw [xshift = -1cm,postaction={decorate},
			decoration={markings, mark=at position 0.6 with {
					\arrow{Triangle[width=4pt,length=5.5pt]}
			}}] (2.84,1) -- (1.9,1)  ;
			
			\draw [xshift = -1cm,postaction={decorate},
			decoration={markings, mark=at position 0.6 with {
					\arrow{Triangle[width=4pt,length=5.5pt]}
			}}] (1.42,0) -- (1.9,1) ;
			
			\draw [xshift = -1cm,postaction={decorate},
			decoration={markings, mark=at position 0.6 with {
					\arrow{Triangle[width=4pt,length=5.5pt]}
			}}] (1.3,1.9) -- (1.42,2.9);
			
			\node[xshift = -1cm] at (1.7,2.3){\small $i$};
			\node[xshift = -1cm] at (0.8,2.2){\small $j$};
			\node[xshift = -1cm] at (1.3,1.4){\small $k$};
			\node[xshift = -1cm] at (2.3,1.3){\small $j$};
			\node[xshift = -1cm] at (1.3,0.5){\small $s$};
			
			\node at (4.5,1.7) {$= \sum\limits_{\substack{i, s \in X , \\ k }} \sqrt{\frac{d(k)}{d(i) d(j)}} e_{i \leftarrow s}^{j k} \cL_{i \leftarrow s}^{j k}\, ,$};
			
		\end{tikzpicture}
	\end{aligned}
\end{equation}
where now $e_{i\leftarrow s}^{jk}$ is a basis for the half-braiding $e_X(j)$. Equation \eqref{eq:fusionHALFBR} then translates, making use of the tube algebra \eqref{algebra_Lasso} for the lasso maps, into the following consistency conditions:
\begin{equation}
	\label{algebra_HBsum}
	\sum_{k,l} \sqrt{\frac{d(j) d(j')}{d(i)d(m)}} \, e_{i \leftarrow s}^{j k} e_{s \leftarrow i'}^{j' k'} F^{j k^* s}_{j' k'^* n^*} F^{i^* j^* k}_{j'^* n m^*} F^{ i' j' k'^*}_{j n^* m}   = \, \delta_{j' j m^*} e_{i \leftarrow i'}^{m n}  \, .
\end{equation} 
The solutions to equation \eqref{algebra_HBsum} are essentially the representation matrices emerging from the tube algebra. Suppose $X$ is as in \eqref{defX} with all dimensions equal to 1. Then
\begin{equation}
	e_{i \leftarrow i}^{j k} = \sqrt{\frac{d(i)}{d(j) d(k) }} \rho^{(\Gamma)}_i \left( \cL_i^{jk}\right) 
\end{equation}
for all the representations $(\Gamma)$ in $\cH_i$ associated to $X$. Analogously, if $i \neq s$
\begin{equation}
	e_{i \leftarrow s}^{j k} = \sqrt{\frac{d(i)}{d(j) d(k) }} \rho^{(\Gamma,\Lambda)}_{i \leftarrow s} \left( \cL_{i \leftarrow s}^{jk}\right) \, , 
\end{equation}
for all the representations $(\Gamma)$ in $\cH_i$ and $(\Lambda)$ in $\cH_s$ associated to $X$ such that their respective Hilbert spaces $\cH_i^{(\Gamma)}$ and $\cH_s^{(\Lambda)}$ are intertwined. 

Let us consider the Fib case. For $X = 1$ we have only one half-braiding, which in this basis reads 
\begin{equation}
	e_{1 \leftarrow 1}^{1 1} = \rho^{(1)}_1 \left( \id_1 \right) = 1 \, ,  \quad e_{1 \leftarrow 1}^{\tau \tau} = \frac{1}{\xi } \rho^{(1)}_1 \left( \tau \right) = 1 \, .
\end{equation}
For $X = \tau$ we have two possible half-braidings:
\begin{align}
	e_{\tau \leftarrow \tau}^{1 \tau} &= \rho^{(2)}_\tau \left( \id_\tau \right) = 1\, , &	e_{\tau \leftarrow \tau}^{\tau 1} = \rho^{(2)}_\tau \left( T_\tau \right) = e^{4 \pi i /5}\,,&& e_{\tau \leftarrow \tau}^{\tau \tau} &= \frac{1}{\sqrt{\xi}} \rho^{(2)}_\tau \left( \cL_\tau^{\tau \tau} \right) = e^{-3 \pi i /5}\, , \\ 
		e_{\tau \leftarrow \tau}^{1 \tau} &= \rho^{(3)}_\tau \left( \id_\tau \right) = 1\, , &	e_{\tau \leftarrow \tau}^{\tau 1} = \rho^{(3)}_\tau \left( T_\tau \right) = e^{- 4 \pi i /5}\, , &&	e_{\tau \leftarrow \tau}^{\tau \tau} &=  \frac{1}{\sqrt{\xi}} \rho^{(3)}_\tau \left( \cL_\tau^{\tau \tau} \right) = e^{3 \pi i /5}\, , 
\end{align}
while for $X = 1 + \tau$ we expect only one half-braiding, which contains the information about the lassos that map $\hat P_1^{(2)}$ in $\hat P_\tau^{(1)}$ and vice versa:
\begin{align}
	e_{1 \leftarrow 1}^{1 1} &= \rho^{(2)}_1 \left( \id_1 \right) = 1 \, , \, \,  &e_{1 \leftarrow 1}^{\tau \tau} &= \frac{1}{\xi} \rho^{(2)}_1 \left( \tau \right) = 	- \frac{1}{\xi^2} \, ,\nn  \\ e_{\tau \leftarrow \tau}^{1 \tau} &= \rho^{(1)}_\tau \left( \id_\tau \right) = 1\, , \, \,  &e_{\tau \leftarrow \tau}^{ \tau 1} &= \rho^{(1)}_\tau \left( T_\tau \right) = 1\, ,
 \, \,  \quad e_{\tau \leftarrow \tau}^{\tau \tau} =  \frac{1}{\sqrt{\xi}} \rho^{(1)}_\tau \left( \cL_\tau^{\tau \tau} \right) =   \frac{1}{\xi^2}  \, , \\ 
 e_{1 \leftarrow \tau}^{\tau \tau} &= \frac{1}{\xi}\rho^{(2,1)}_{1,\tau} \left( \cL_{1 \leftarrow \tau}^{\tau \tau} \right)\, , \, \,  &e_{\tau \leftarrow 1}^{\tau \tau} &= \frac{1}{\sqrt{\xi}} \rho^{(1,2)}_{\tau,1} \left(\cL_{\tau \leftarrow 1}^{\tau \tau}\right) =  \frac{1 + \xi^2}{\xi^{3}} (e_{1 \leftarrow \tau}^{\tau \tau} )^{-1} \, . \nn 
\end{align}

When the coefficients of the simple lines are not equal to one, the situation is not really more complicated. One can in fact easily generalize equation \eqref{basisHalfBraidingsSum} by associating a multiplicity to the lines $i$ and $s$, turning the half-braiding coefficients into matrices. In practice, it is the same as solving equation \eqref{algebra_HBsum} with a matrix product between the $e_{i \leftarrow s}^{j k}$'s. See for example \cite{Bhardwaj:2023ayw} for some worked out examples.

In all cases the equations \eqref{algebra_HBsum} are very similar to those of the tube algebra themselves that we wrote in equation \eqref{algebra_Lasso}. Up to some convention-dependent factors, determining the matrices $e_{i \leftarrow s}^{j k}$ turns out to be computationally almost identical to determining the representation matrices $\rho(\cL^{jk}_{i \leftarrow s})$. For our category of interest we decided to proceed with the latter computation, which has the advantage of having a more direct physical interpretation. We will briefly mention the connection to the Drinfeld center again in subsection \ref{subsec:drinfeldcenter_complicated}.

\subsection{Partition function and modular symmetry}
Consider a two-dimensional QFT on a cylinder which is a CFT. The usual torus partition function is obtained by identifying the two ends of the cylinder:
\begin{align}
	\label{partition function}
	Z(u, \bar u) = \Tr \left( q^{L_0 - c/24} \bar q^{\bar L_0 - c/24}\right)\, ,
\end{align}
where $q \colonequals e^{2 \pi i u}$ and, since the letter $\tau$ has already been used, we write $u$ to denote the modulus of the torus. 

Now endow the cylinder with the network of topological lines corresponding to the operator $\cL_i^{jk}$ in the twisted Hilbert space $\mathcal{H}_i$.  Since it commutes with the Virasoro symmetry, the resulting torus partition function in the presence of this network is
\begin{equation}
		\label{partition function with lines}
	Z_{i}^{jk}(u,\bar u) =  \Tr_{\mathcal{H}_i} \left( \, q^{L_0 - c/24} \bar q^{\bar L_0 - c/24}\, \cL_i^{jk}\,   \right) \, .
\end{equation}
We denote by $Z_i^{(\Gamma)}(u,\bar u)$ the torus partition function produced by inserting a linear combination of networks on the cylinder corresponding to the minimal central idempotent $\hat P_i^{(\Gamma)}$:
\begin{equation}
	Z_{i}^{(\Gamma)}(u,\bar u) =  \Tr_{\mathcal{H}_i} \left( \, q^{L_0 - c/24} \bar q^{\bar L_0 - c/24}\, \hat P_i^{(\Gamma)}\,   \right) \, .
\end{equation}
 Each $Z_i^{(\Gamma)}(u,\bar u)$ contains operators with the same spins modulo an integer, and has positive integer coefficients. They are therefore the physical partition functions of the theory.
 
 Now suppose there exist two lassos $\cL_{s \leftarrow i}^{jk}$ and $\cL_{i \leftarrow s}^{j' k'}$ that map all the states in the sector obtained by projecting with $\hat P_i^{(\Gamma)}$ to those obtained by projecting with $\hat P_s^{(\Gamma')}$ and vice versa. Since these sectors have identical spectra, the corresponding partition functions must coincide:
 \begin{equation}
 	\label{equivalencesZs}
 	Z_i^{(\Gamma)}(u,\bar u) = Z_s^{(\Gamma')}(u,\bar u)\, .
 \end{equation} 
 For example, our previous analysis shows that in a theory with a single Fib there are five partition functions of the form $Z_i^{(\Gamma)}$ but only four of them are independent. We will assemble them into a single vector and write:
 \begin{equation}
 	\label{vectorZ}
 	\vec{Z}(u,\bar u) = ( Z_1^{(1)}(u,\bar u), Z_1^{(2)}(u,\bar u), Z_\tau^{(2)}(u,\bar u),Z_\tau^{(3)}(u,\bar u) )\, .
 \end{equation}
 In terms of the Drinfeld center, $\vec{Z}(u,\bar u)$ is the vector that collects the partition functions $Z_{(X,e_X)}(u, \bar u)$ associated to the simple objects $(X, e_X) \in \cZ(\cC)$. For Fib indeed we could have labeled the partition functions in \eqref{vectorZ} as
 \begin{equation}
	\vec{Z}(u,\bar u) = ( Z_{(1,e_1)}(u,\bar u), Z_{(1+ \tau,e_{1 + \tau})}(u,\bar u), Z_{(\tau,e_\tau(1))}(u,\bar u),Z_{(\tau,e_\tau(2))}(u,\bar u) )\, ,
\end{equation}
where $e_\tau(1)$ and $e_\tau(2)$ label the two different half-braidings associated to $X = \tau$. 
 
 \subsubsection{Completeness}
 	\label{sec:completeness}
Although the $Z_i^{(\Gamma)}$’s represent the physical partition functions, the number of independent $Z_i^{jk}$’s in a given Hilbert space turns out to coincide with the number of $Z_i^{(\Gamma)}$’s. Moreover, these independent $Z_i^{jk}$’s are simply a linear combination of the $Z_i^{(\Gamma)}$’s.

To see this, consider first the simple case of a finite group $G$. The minimal central idempotents in a Hilbert space $\cH_g$ are the projectors onto the irreps of the stabilizer $C_g$ \eqref{Projgroup}. Complete reducibility ensures that the entire Hilbert space can be decomposed as
 \begin{align}
 	\cH_{g} = \bigoplus_{\pi_g} \hat \cH_{g}^{\pi_g} \otimes V_{\pi_g}\, ,
 \end{align}
 where $V_{\pi_g}$ denotes the representation space of $C_g$ associated with the irrep $\pi_g$. The simple network $\cL_{g}^{h}$, corresponding to the insertion of $h$ in $\cH_g$, acts as
 \begin{equation}
 	\sum_{\pi_g} \hat P_{g}^{\pi_g} \otimes \rho_{\pi_g}(h)\,,
 \end{equation} 
 where $\rho_{\pi_g}(h)$ is the representation of $\cL_g^h$ on the subspace $\hat \cH_{g}^{\pi_g} \otimes V_{\pi_g}$ that $\hat P_{g}^{\pi_g}$ projects on. Its trace is given by the character $\chi_{\pi_g}(h)$. The associated partition function is then
 \begin{align}
 	\label{completeness_eq}
 	Z_g^h(u, \bar u) &= \Tr_{\cH_{g}} \left(q^{L_0 - c/24} \bar q^{\bar L_0 - c/24}\, \cL_g^h \right)
 	= \sum_{\pi_g} \chi_{\pi_g}(h)\, Z_{g}^{\pi_g}(u, \bar u) \, .
 \end{align}
This is the result that allows us to express any of the $Z_g^{h}$ in terms of the $Z_g^{\pi_g}$. We have already established the converse operation, so there exists an invertible linear map between the two vectors of partition functions. 

The same completeness result holds in our setup: unitary representations on a QFT Hilbert space are completely reducible, and the irreducible representations of our finite-dimensional tube algebra are all finite-dimensional. See again \cite{etingof2009introduction} for more details and mathematical background.

\subsubsection{Modular invariance}
Under modular S and T transformations, the partition function $Z_{i}^{jk}$ of equation \eqref{partition function with lines} transforms as
\begin{center}
\includegraphics[scale=0.5]{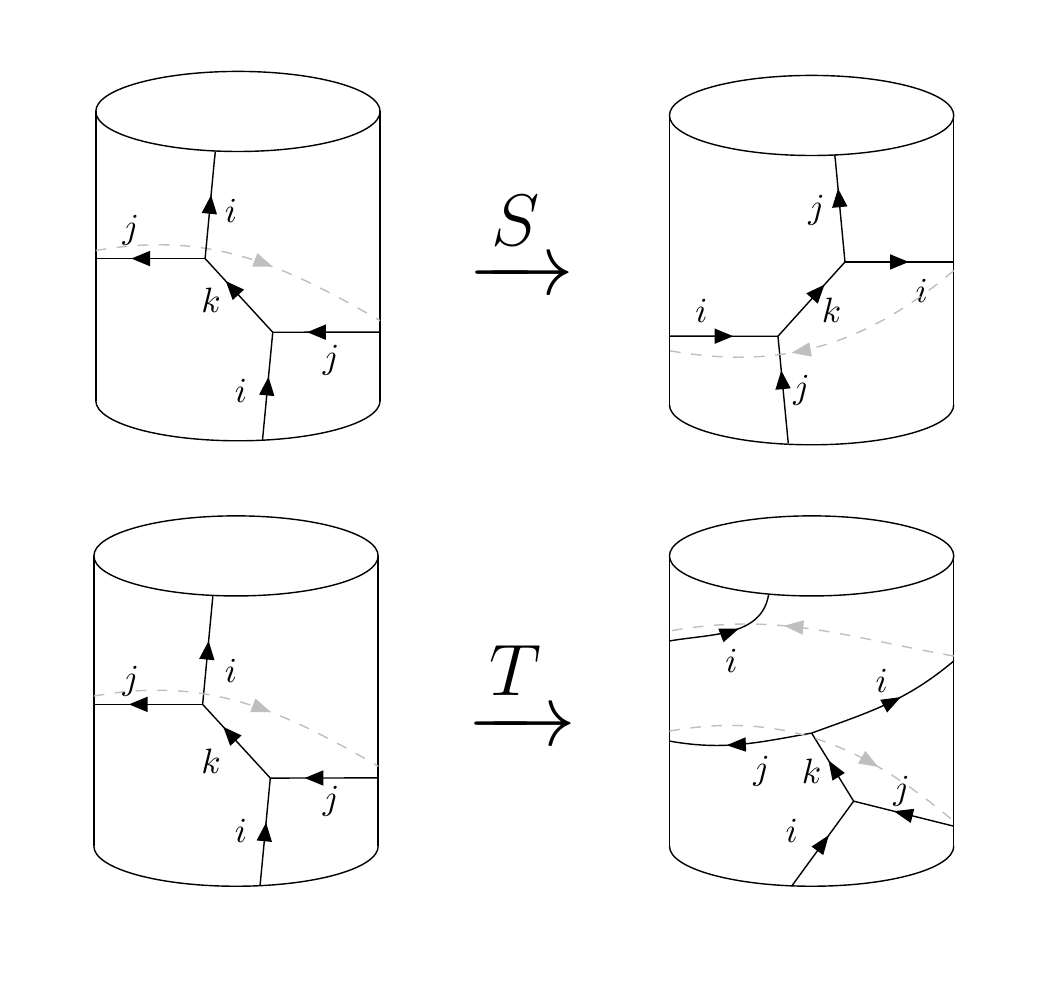}\,.
\end{center}
The transformed partition function can be rewritten as a linear combination of all the $Z_{i}^{jk}$ with a single F move, namely
\begin{align}
 	\label{Stransform}
 	Z_i^{j k}\left(-1/u, -1/\bar u\right)  &= \sum_m F^{i j k^*}_{i^* j^* m} Z_j^{i^* m}\left(u, \bar u\right) \, , \\ 
 	\label{Ttransform}
 	Z_i^{j k}\left(u + 1, \bar u + 1 \right) &=  \sum_m F^{i^* j ^* k}_{i j  m^*} Z_i^{m j}\left(u, \bar u\right)\, .
\end{align}

As a consequence of equations \eqref{Stransform} and \eqref{Ttransform}, the decomposition of the projectors as linear combinations of simple networks \eqref{projectorsInGeneral} and the completeness of the decomposition of a Hilbert space into irreps of the tube algebra, as expressed in equation \eqref{completeness_eq}, we obtain that the modular S and T transformations act covariantly on $\vec{Z}(u,\bar u)$:
\begin{align}
	\label{Smatrix}
	\vec{Z}\left(-1/u, -1/\bar u\right) & = \,  S \,	\vec{Z}\left(u, \bar u\right) \, , \\ 
	\label{Tmatrix}
	\vec{Z}\left(u + 1, \bar u + 1 \right) & =  \, T \,	\vec{Z}\left(u, \bar u\right) \, ,
\end{align}
where $S$ and $T$ are the modular matrices. The modular $T$ matrix is diagonal, and its eigenvalues correspond to those of the network $T_i$ for any sector $(\Gamma)$ in the Hilbert space $\cH_i$. 

As a simple example let us consider the Fib case. From \eqref{Ttransform}, \eqref{vectorZ}, \eqref{Tmatrix} and the expression of the minimal central idempotents for Fib in terms of simple networks in equations \eqref{projectors_untwistedFib} and \eqref{ProjectorsTwistedFib} we find that:
\begin{equation}
T = \text{diag} \left(e^{2 \pi i} \, , e^{2 \pi i} \, ,  e^{4 \pi i/5} \, , e^{-4 \pi i/5} \right)\, ,
\end{equation} 
and we can combine equations \eqref{equivalencesZs}, \eqref{vectorZ}, \eqref{Stransform}, \eqref{Smatrix}, \eqref{projectors_untwistedFib} and \eqref{ProjectorsTwistedFib} to obtain
\begin{align}
	\label{SmatrixFib}
	\begin{pmatrix}
		Z_1^{(1)} \\
		Z_1^{(2)} \\
		Z_\tau^{(2)} \\
		Z_\tau^{(3)}
	\end{pmatrix}\left(-1/u, -1/\bar u\right) =  
	\frac{1}{1 + \xi^2}	\begin{pmatrix}
		1 & \xi^2 & \xi & \xi \\
		\xi^2 & 1 & -\xi & -\xi  \\
		\xi & - \xi & -1 & \xi^2 \\
		\xi & - \xi &  \xi^2 & -1 \\
	\end{pmatrix}
	\begin{pmatrix}
	Z_1^{(1)} \\
	Z_1^{(2)} \\
	Z_\tau^{(2)} \\
	Z_\tau^{(3)}
	\end{pmatrix}\left(u,\bar u \right) \, .
\end{align}

This completes our review of tube algebras, their representations, lasso operations, and modular properties of torus partition functions. 

\section{Multiplying categories}
The category of interest is generated by two copies of the Fibonacci category plus the $S_2$ permutation symmetry which swaps the two copies
\begin{equation}
	(\text{Fib} \boxtimes \text{Fib}) \rtimes S_2 \, .
\end{equation}
To understand it, we first review the direct product of two categories and then discuss a method to study the case of a group acting on a category.

\subsection{Direct product of two categories}
Consider the direct (or Deligne) product of two fusion categories $\mathcal{C}_1$ and $ \mathcal{C}_2$
\begin{equation}
\mathcal{C}_1 \boxtimes \mathcal{C}_2 \, .
\end{equation}
When the full categorical data of each factor, \emph{i.e.} their simple objects, fusion rules, quantum dimensions, and F symbols, are known, the structure of the product category is likewise completely determined. Graphically the Deligne product structure means that the lines of $\cC_1$ and $\cC_2$ can cross each other freely. One could take the view that the product category is not a fusion category of itself (at least not in the way we defined it above), because of the existence of the quartic vertex
\begin{center}
	\begin{tikzpicture}
	\draw[thick,
	postaction={decorate},
	decoration={markings, mark=at position 0.8 with {\arrow{>}}}, color=red
	] (11.2,-0.3) -- (13.2,-0.3); \node at (13,-0.7) {$i_1$}; \node at (11.5,-0.7) {$i_1$};
	
	\draw[thick,
	postaction={decorate},
	decoration={markings, mark=at position 0.7 with {\arrow{>}}}, color=green!70!black!60
	] (12.3,-1) -- (12.3,0.5); \node at (12.6,0.3) {$i_2$};
\end{tikzpicture}
\, .
\end{center}
(Here $i_1$ is a simple object of $\cC_1$ and $i_2$ is a simple object of $\cC_2$. Their lines are respectively drawn in \textcolor{red}{red} and \textcolor{green!50!black!60}{green}.) However, it is straightforward to see that $\mathcal{C}_1 \boxtimes \mathcal{C}_2$ can actually be given the canonical structure of a fusion category, \emph{i.e.}, a finite set of simple lines and corresponding fusion rules. Indeed, its simple objects are given by all possible copies of simple objects $(i_1,i_2)$
\begin{equation}
	\begin{aligned}
		\begin{tikzpicture}
		\draw[thick,
		postaction={decorate},
		decoration={markings, mark=at position 0.5 with {\arrow{>}}}, color=red
		] (1,0) -- (2.5,0); \node at (1.5,0.3) {$i_1$};
		
		\draw[thick,
		postaction={decorate},
		decoration={markings, mark=at position 0.5 with {\arrow{>}}},color=green!70!black!60
		] (1,-0.5) -- (2.5,-0.5); \node at (1.5,-0.8) {$i_2$};
		
		\node at (3.5,-0.25) {$\equiv \, (i_1,i_2)$ };
	\end{tikzpicture}
	\end{aligned}
\, , 
\end{equation}
 and the fusion rules are 
\begin{equation}
(i_1,i_2) \times (j_1,j_2) = (i_1 \times_1 j_1,i_2 \times_2 j_2) \, ,
\end{equation}
where `$\times_{1}$' ($\times_{2}$) means the fusion in $\mathcal{C}_{1}$ ($\mathcal{C}_{2}$) and we may use linearity to expand the right-hand side into a sum of simple objects.
Given this structure, we can indeed resolve the quartic vertex as two cubic vertices involving simple lines: 
\begin{equation}
			\label{quartic}
				\begin{aligned}
	\begin{tikzpicture}
	\draw[thick,
	postaction={decorate},
	decoration={markings, mark=at position 0.8 with {\arrow{>}}}, color=red
	] (-1,1) -- (1,-1); \node at (-1.3,1.2) {$i_1$}; \node at (1.3,-1.2) {$i_1$};
	
	\draw[thick,
	postaction={decorate},
	decoration={markings, mark=at position 0.7 with {\arrow{>}}}, color=green!70!black!60
	] (-1,-1) -- (1,1); \node at (1.3,1) {$i_2$}; \node at (-1.2,-1.2) {$i_2$};
	
	\node at (2,0){$=$};

	\draw[thick,
	postaction={decorate},
	decoration={markings, mark=at position 0.5 with {\arrow{>}}}, color=red, xshift = 4cm
	] (-1,1) -- (-0.5,0.2); \node[xshift = 4cm] at (-1.3,1.3) {$i_1$};
	
	\draw[thick, color=red, xshift = 4cm,postaction={decorate},
	decoration={markings, mark=at position 0.5 with {\arrow{>}}}]  (-0.5,0.2)-- (1,0.2);
	\draw[thick, color=red, postaction={decorate},
	decoration={markings, mark=at position 0.3 with {\arrow{<}}},xshift = 4cm]  (1.7,-1) -- (1,0.2);

	\draw[thick,
	postaction={decorate},
	decoration={markings, mark=at position 0.5 with {\arrow{>}}}, color=green!70!black!60,xshift = 4cm
	] (-1,-1) -- (-0.5,-0.2); \node[xshift = 4cm] at (1.5,1.3) {$i_2$};
	\draw[thick, color=green!70!black!60, xshift = 4cm,postaction={decorate},
	decoration={markings, mark=at position 0.5 with {\arrow{>}}}]  (-0.5,-0.2) -- (1,-0.2);
	\draw[thick, color=green!70!black!60, xshift = 4cm,postaction={decorate},
	decoration={markings, mark=at position 0.3 with {\arrow{<}}}]  (1.7,1) -- (1,-0.2);
	\node[xshift = 4cm] at (1.8,-1.3) {$i_1$};
	\node[xshift = 4cm] at (-1.2,-1.2) {$i_2$};
	
	\node[xshift = 4cm] at (2.5,0){$=$};

	\draw[thick,
	postaction={decorate},
	decoration={markings, mark=at position 0.5 with {\arrow{>}}}, color=red, xshift = 8.5cm
	] (-1,1) -- (-0.5,0.2); \node[xshift = 8.5cm] at (-1.3,1.3) {\small$(i_1,e)$};
	\node[xshift = 8.5cm] at (1.8,-1.3) {\small$(i_1,e)$};
	
	\draw[thick, color=red, xshift = 8.5cm,postaction={decorate},
	decoration={markings, mark=at position 0.5 with {\arrow{>}}}]  (-0.5,0.2)-- (1,0.2);
	\draw[thick, color=red, postaction={decorate},
	decoration={markings, mark=at position 0.3 with {\arrow{<}}},xshift = 8.5cm]  (1.7,-1) -- (1,0.2);

	\draw[thick,
	postaction={decorate},
	decoration={markings, mark=at position 0.5 with {\arrow{>}}}, color=green!70!black!60,xshift = 8.5cm
	] (-1,-1) -- (-0.5,-0.2); \node[xshift = 8.5cm] at (1.5,1.3) {\small$(e,i_2)$}; \node[xshift = 8.5cm] at (-1.3,-1.3) {\small$(e,i_2)$};
	\draw[thick, color=green!70!black!60, xshift = 8.5cm,postaction={decorate},
	decoration={markings, mark=at position 0.5 with {\arrow{>}}}]  (-0.5,-0.2) -- (1,-0.2);
	\draw[thick, color=green!70!black!60, xshift = 8.5cm,postaction={decorate},
	decoration={markings, mark=at position 0.3 with {\arrow{<}}}]  (1.7,1) -- (1,-0.2);
	
	\node[xshift = 8.5cm] at (0.3,0.6) {\small$(i_1,i_2)$};
	\filldraw[xshift = 8.5cm, fill=gray!60!white!40, draw=black, thick] (-0.5,0) circle (0.3cm);
	\filldraw[xshift = 8.5cm, fill=gray!60!white!40, draw=black, thick] (1.2,0) circle (0.3cm);
\end{tikzpicture}
\end{aligned}
\, .
\end{equation}

Furthermore, the quantum dimensions are given by the product of the quantum dimensions of the objects
\begin{equation}
d(i_1, i_2) = d(i_1) d(i_2),
\end{equation}
and the F symbols factorize as:
\begin{equation}
F^{(i_1,i_2)(j_1,j_2)(m_1,m_2)}_{(k_1,k_2)(l_1,l_2)(n_1,n_2)} = F^{i_1 j_1 m_1}_{k_1 l_1 n_1} F^{i_2 j_2 m_2}_{k_2 l_2 n_2} \, .
\end{equation}
Likewise, the minimal central idempotents are given by the direct product of the minimal central idempotents of the two categories.

\subsection{Semidirect product of a group and a category}
Let now $\mathcal{C}$ be a fusion category and $G$ a group acting on it,  giving rise to the semidirect product 
\begin{equation}
	\mathcal{C} \rtimes G \, .
\end{equation}
 We treat $G$ as a fusion category and, as in the previous case, assume that all the categorical data of each factor are known.

The group elements can act on the lines of $\mathcal{C}$, which again naturally leads to quartic vertices. Given $i$ a simple object of $\mathcal{C}$ and $a$ an element of $G$, the action of $a$ on $i$ is realized through the following quartic vertex:
\begin{center}
	\begin{tikzpicture}
	\draw[thick,
	postaction={decorate},
	decoration={markings, mark=at position 0.8 with {\arrow{>}}}, color=red
	] (11.2,-0.3) -- (13.2,-0.3); \node at (13,-0.7) {$i$}; \node at (11.5,-0.7) {$\phi_a(i)$};
	
	\draw[thick,
	postaction={decorate},
	decoration={markings, mark=at position 0.7 with {\arrow{>}}}, color=blue
	] (12.3,-1) -- (12.3,0.5); \node at (12.5,-0.9) {$a$};
\end{tikzpicture}
\,.
\end{center}
(Here the group lines are colored \textcolor{blue}{blue} and the lines of the original category $\mathcal{C}$ are colored \textcolor{red}{red}.) This quartic vertex is not trivial, and for consistency with the semidirect product structure it must obey two fundamental properties:
\begin{equation}
	\begin{aligned}
	\begin{tikzpicture}
	
	\draw[thick,
	postaction={decorate},
	decoration={markings, mark=at position 0.7 with {\arrow{<}}}
	,color = red] (1,0) -- (-0.5,1); \node at (0.7,0.6) {$i$}; \node at (-0.5,1.2) {$\phi_{a}(i)$};
	\draw[thick,
	postaction={decorate},
	decoration={markings, mark=at position 0.7 with {\arrow{<}}},
	color = red ] (1,0) -- (-0.5,-1); \node at (-0.4,-1.2) {$\phi_{a}(j)$}; \node at (0.8,-0.6) {$j$};
	\draw[thick,
	postaction={decorate},
	decoration={markings, mark=at position 0.5 with {\arrow{>}}}, color = red
	] (1,0) -- (2.5,0); \node at (2.4,0.3) {$k$};
	\filldraw[color = red] (1,0) circle (2pt);
	
	\draw[thick,
	postaction={decorate},
	decoration={markings, mark=at position 0.5 with {\arrow{>}}}, color = blue
	] (0.2,-1.5) -- (0.2,1.5); \node at (-0.4,0.3) {$a$};
	
	\node at (3, 0) {$=$};
	
	\draw[thick,
	postaction={decorate},
	decoration={markings, mark=at position 0.5 with {\arrow{<}}}
	,color = red] (4.5,0) -- (3.5,1);  \node at (3.7,1.2) {$\phi_a(i)$};
	\draw[thick,
	postaction={decorate},
	decoration={markings, mark=at position 0.5 with {\arrow{<}}},
	color = red ] (4.5,0) -- (3.5,-1);  \node at (3.7,-1.2) {$\phi_a(j)$};
	\draw[thick,
	postaction={decorate},
	decoration={markings, mark=at position 0.5 with {\arrow{>}}}, color = red
	] (4.5,0) -- (6.5,0); \node at (6.4,0.3) {$k$}; \node at (5.3,0.3) {$\phi_{a}(k)$};
	\filldraw[color = red] (4.5,0) circle (2pt);
	
	\draw[thick,
	postaction={decorate},
	decoration={markings, mark=at position 0.7 with {\arrow{>}}}, color = blue
	] (6,-1.5) -- (6,1.5); \node at (6.2,1) {$a$};

\end{tikzpicture}
\end{aligned}
\, ,
\end{equation}

\begin{equation}
	\begin{aligned}
	\begin{tikzpicture}
	
	\draw[thick,
	postaction={decorate},
	decoration={markings, mark=at position 0.5 with {\arrow{<}}}
	,color = blue] (1,0) -- (-0.5,1);  \node at (0.5,0.8) {$a$};
	\draw[thick,
	postaction={decorate},
	decoration={markings, mark=at position 0.5 with {\arrow{<}}},
	color = blue ] (1,0) -- (-0.5,-1);  \node at (0.4,-0.9) {$b$};
	\draw[thick,
	postaction={decorate},
	decoration={markings, mark=at position 0.5 with {\arrow{>}}}, color = blue
	] (1,0) -- (2.5,0); \node at (2.4,0.3) {$c$};
	\filldraw[color = blue] (1,0) circle (2pt);
	
	\draw[thick,
	postaction={decorate},
	decoration={markings, mark=at position 0.5 with {\arrow{>}}}, color = red
	] (-0.2,-1.5) -- (-0.2,1.7); \node at (-0.7,0.1) {$\phi_b(i)$}; \node at (-0.4,-1.3) {$i$}; \node at (-1,1.7) {$\phi_a(\phi_b(i))$};
	
	\node at (3, 0) {$=$};
	
	\draw[thick,
	postaction={decorate},
	decoration={markings, mark=at position 0.5 with {\arrow{<}}}
	,color = blue] (4.5,0) -- (3.5,1);  \node at (3.7,1.2) {$a$};
	\draw[thick,
	postaction={decorate},
	decoration={markings, mark=at position 0.5 with {\arrow{<}}},
	color = blue ] (4.5,0) -- (3.5,-1);  \node at (3.7,-1.2) {$b$};
	\draw[thick,
	postaction={decorate},
	decoration={markings, mark=at position 0.5 with {\arrow{>}}}, color = blue
	] (4.5,0) -- (6.5,0);  \node at (5.3,0.3) {$c$};
	\filldraw[color = blue] (4.5,0) circle (2pt);
	
	\draw[thick,
	postaction={decorate},
	decoration={markings, mark=at position 0.7 with {\arrow{>}}}, color = red
	] (6,-1.5) -- (6,1.5); \node at (6.5,1) {$\phi_c(i)$}; \node at (6.2,-1.4) {$i$};

\end{tikzpicture}
\end{aligned}
\, .
\end{equation}
In equations, the first one corresponds to:
\begin{equation}
	i \times j \to k \qquad \Rightarrow \qquad \phi_a(i) \times \phi_a(j) \to \phi_a(k)\, ,
\end{equation}
while the second ones indicates that the group action should be consistent with the group multiplication:
\begin{equation}
 \phi_a(\phi_b(i)) = \phi_{ab}(i)\, .
\end{equation}
If a vertex is oriented in the opposite direction, we can always make a change of variables and revert it to our standard definition. Defining $i' \colonequals \phi_a(i)$ and using that $\phi_{a^*}(\phi_a(i)) = i$, we have
\begin{equation}
	\label{vertex2}
	\begin{aligned}
	\begin{tikzpicture}
	\draw[thick,
	postaction={decorate},
	decoration={markings, mark=at position 0.8 with {\arrow{>}}}, color=red
	] (11.2,-0.3) -- (13.2,-0.3); \node at (13,-0.7) {$\phi_a(i)$}; \node at (11.5,-0.7) {$i$};
	
	\draw[thick,
	postaction={decorate},
	decoration={markings, mark=at position 0.7 with {\arrow{<}}}, color=blue
	] (12.3,-1) -- (12.3,0.5); \node at (12.5,0.3) {$a$};
	
	\node at (14.3,-0.2){$=$};
	
	\draw[thick,
	postaction={decorate},
	decoration={markings, mark=at position 0.8 with {\arrow{>}}}, color=red, xshift = 4cm
	] (11.2,-0.3) -- (13.2,-0.3); \node[xshift = 4cm] at (13,-0.7) {$\phi_a(i)$}; \node[xshift = 4cm] at (11.5,-0.7) {$i$};
	
	\draw[thick,
	postaction={decorate},
	decoration={markings, mark=at position 0.7 with {\arrow{>}}}, color=blue, xshift = 4cm
	] (12.3,-1) -- (12.3,0.5); \node[xshift = 4cm] at (12.6,0.3) {$a^*$};
	
	\node at (18.3,-0.2){$=$};
	
	\draw[thick,
	postaction={decorate},
	decoration={markings, mark=at position 0.8 with {\arrow{>}}}, color=red, xshift = 8cm
	] (11.2,-0.3) -- (13.2,-0.3); \node[xshift = 8cm] at (13,-0.7) {$i'$}; \node[xshift = 8cm] at (11.5,-0.7) {$\phi_{a^*}(i')$};
	
	\draw[thick,
	postaction={decorate},
	decoration={markings, mark=at position 0.7 with {\arrow{>}}}, color=blue, xshift = 8cm
	] (12.3,-1) -- (12.3,0.5); \node[xshift = 8cm] at (12.6,0.3) {$a^*$};
\end{tikzpicture}\, ,
\end{aligned}
\end{equation}
where we recall that $a^* = a^{-1}$ for all $a \in G$. 

\subsubsection{Fusion category structure}
As above, we would like to define a fusion category without quartic vertices. Fortunately $\mathcal{C} \rtimes G$ can again be given such a  structure, in the same straightforward way as for $\cC_1 \boxtimes \cC_2$.

The simple objects of $\mathcal{C} \rtimes G$ are labeled in the same way as the ones of $\mathcal{C} \boxtimes G$ and have the same quantum dimensions.
Graphically, we \emph{define} the simple objects $(i,a)$ by placing the group line \emph{to the right} of the $\mathcal{C}$ line:
	\begin{equation}
		\begin{aligned}
	\begin{tikzpicture}
	\draw[thick,
	postaction={decorate},
	decoration={markings, mark=at position 0.5 with {\arrow{>}}}, color=red
	] (1,0) -- (2.5,0); \node at (1.5,0.3) {$i$};
	
	\draw[thick,
	postaction={decorate},
	decoration={markings, mark=at position 0.5 with {\arrow{>}}},color=blue
	] (1,-0.5) -- (2.5,-0.5); \node at (1.5,-0.7) {$a$};
	
	\node at (3.5,-0.25) {$\equiv \, (i,a)$ };
\end{tikzpicture}
\end{aligned}
\, .
	\end{equation}
If a blue line lies to the left of a red line, the configuration does not represent a simple object. However, we can always resolve such a network locally as
	\begin{equation}
		\label{wrongside}
		\begin{aligned}
		\begin{tikzpicture}
		\draw[thick,
		postaction={decorate},
		decoration={markings, mark=at position 0.7 with {\arrow{>}}}, color=blue
		] (8,-1) -- (8,1.2); \node at (8.3,-0.8) {$a$};
		
		\draw[thick,
		postaction={decorate},
		decoration={markings, mark=at position 0.7 with {\arrow{>}}}, color=red
		] (9,-1) -- (9,1.2); \node at (9.3,-0.8) {$i$};
		
		\node at (9.9,0){$=$};

		\draw[thick,
		postaction={decorate},
		decoration={markings, mark=at position 0.5 with {\arrow{>}}}, color=blue
		] (10.8,-1) -- (10.8,-0.25); \node at (11.1,-0.9) {$a$};
		
		\draw[thick,
		postaction={decorate},
		decoration={markings, mark=at position 0.6 with {\arrow{>}}}, color=red
		] (11.5,-1) -- (11.5,1.4); \node at (11.8,-0.8) {$i$}; \node at (11,0.2){$\phi_a(i)$}; \node at (11.8,1.2) {$i$};

		\draw[thick,
		postaction={decorate},
		decoration={markings, mark=at position 0.5 with {\arrow{>}}}, color=blue
		] (12,-0.25) -- (12,0.75); \node at (12.3,0.25) {$a$};

		\draw[thick,
		postaction={decorate},
		decoration={markings, mark=at position 0.5 with {\arrow{>}}}, color=blue
		] (10.8,-0.25) -- (12,-0.25); 
		
		\draw[thick,
		postaction={decorate},
		decoration={markings, mark=at position 0.5 with {\arrow{<}}}, color=blue
		] (10.8,0.75) -- (12,0.75); 
		
		\draw[thick,
		postaction={decorate},
		decoration={markings, mark=at position 0.5 with {\arrow{>}}}, color=blue
		] (10.8,0.75) -- (10.8,1.4); \node at (11.1,1.2) {$a$};

	\end{tikzpicture}\, .
	\end{aligned}
	\end{equation}
We can then absorb the two quartic vertices into the rest of the network, redefining some cubic vertices if needed. So without loss of generality we can always move the group lines to the right:
	\begin{equation}
		\label{equivalence}
		\begin{aligned}
		\begin{tikzpicture}
		
		\draw[thick,
		postaction={decorate},
		decoration={markings, mark=at position 0.7 with {\arrow{>}}}, color=blue
		] (8,-2) -- (8,1); \node at (8.3,-0.8) {$a$};
		
		\draw[thick,
		postaction={decorate},
		decoration={markings, mark=at position 0.7 with {\arrow{>}}}, color=red
		] (9.2,-2) -- (9.2,1); \node at (9.5,-0.8) {$i$};
		
		\fill[white!20!gray!80] (8.6,1) ellipse (0.8cm and 0.4cm);
		\fill[white!20!gray!80] (8.6,-2) ellipse (0.8cm and 0.4cm);
		
		\node at (10.5,-0.5){$\leftrightarrow$}; 
		
		\draw[thick,
		postaction={decorate},
		decoration={markings, mark=at position 0.7 with {\arrow{>}}}, color=red
		] (11.8,-2) -- (11.8,1); \node at (12.3,-0.8) {$\phi_a(i)$};
		
		\draw[thick,
		postaction={decorate},
		decoration={markings, mark=at position 0.7 with {\arrow{>}}}, color=blue
		] (13,-2) -- (13,1); \node at (13.3,-0.8) {$a$};
		
		\fill[white!20!gray!80] (12.4,1) ellipse (0.8cm and 0.4cm);
		\fill[white!20!gray!80] (12.4,-2) ellipse (0.8cm and 0.4cm);
		
	\end{tikzpicture}\, .
	\end{aligned}
	\end{equation}
The gray blobs here represent anything that can be attached to these lines.

\paragraph{Fusion rules \\}
Suppose we want to fuse two simple objects $(i, a)$ and $(j,b)$. Their fusion follows from the standard semidirect product structure:
\begin{equation}
	\label{junctionsemidirect}
	(i, a) \times (j,b) = (i \times \phi_a(j), ab)\, .
\end{equation}
 Making use of \eqref{vertex2}, we can represent it graphically as:
\begin{center}
			\begin{tikzpicture}
			\draw[thick,
			postaction={decorate},
			decoration={markings, mark=at position 0.5 with {\arrow{<}}}
			,color = red] (1,0) -- (-0.5,1); \node at (0.5,0.75) {$i$};
			\draw[thick,
			postaction={decorate},
			decoration={markings, mark=at position 0.5 with {\arrow{<}}},
			color = red ] (1,0) -- (-0.5,-1); \node at (-0.4,-0.6) {$j$};
			\draw[thick,
			postaction={decorate},
			decoration={markings, mark=at position 0.5 with {\arrow{>}}}, color = red
			] (1,0) -- (2.5,0); \node at (2.4,0.3) {$k$};
			\filldraw[color = red] (1,0) circle (2pt);
			\node at (1.4,0.3) {\small $\delta_{i\phi_a(j) k^*}$};
			
			\draw[thick,
			postaction={decorate},
			decoration={markings, mark=at position 0.5 with {\arrow{<}}}
			,color = blue] (1.1,-0.5) -- (-0.7,0.7); \node at (-0.7,0.2) {$a$};
			\draw[thick,
			postaction={decorate},
			decoration={markings, mark=at position 0.5 with {\arrow{<}}},
			color = blue ] (1.1,-0.5) -- (-0.3,-1.4); \node at (0.1,-1.4) {$b$};
			\draw[thick,
			postaction={decorate},
			decoration={markings, mark=at position 0.5 with {\arrow{>}}}, color = blue
			] (1.1,-0.5) -- (2.5,-0.5); \node at (2.4,-0.7) {$c$};
			\filldraw[color = blue] (1.1,-0.5) circle (2pt);
			\node at (1.4,-0.8) {\small $\delta_{a b c^*}$};

		\end{tikzpicture}\, .
\end{center}
Notice that the cubic vertices between the simple lines in $\cC \rtimes G$ are defined in terms of a single quartic vertex as well as two cubic vertices: one in $\mathcal{C}$ ($ i \times \phi_a(j) \to k$) and one in $G$ ($ab \to c$).

\paragraph{Resolution of quartic vertices\\}
The quartic vertices can be resolved into two cubic vertices in essentially the same obvious manner as for the product category:
\begin{equation}
			\label{quarticsemidirect}
				\begin{aligned}
	\begin{tikzpicture}
	\draw[thick,
	postaction={decorate},
	decoration={markings, mark=at position 0.8 with {\arrow{>}}}, color=red
	] (-1,1) -- (1,-1); \node at (-1.4,1.3) {$\phi_a(i)$}; \node at (1.2,-1.3) {$i$};
	
	\draw[thick,
	postaction={decorate},
	decoration={markings, mark=at position 0.7 with {\arrow{>}}}, color=blue
	] (-1,-1) -- (1,1); \node at (1.3,1.2) {$a$}; \node at (-1.2,-1.3) {$a$};
	
	\node at (2,0){$=$};

	\draw[thick,
	postaction={decorate},
	decoration={markings, mark=at position 0.5 with {\arrow{>}}}, color=red, xshift = 4cm
	] (-1,1) -- (-0.5,0.2); \node[xshift = 4cm] at (-1.3,1.3) {$\phi_a(i)$};
	
	\draw[thick, color=red, xshift = 4cm,postaction={decorate},
	decoration={markings, mark=at position 0.5 with {\arrow{>}}}]  (-0.5,0.2)-- (1,0.2);
	\draw[thick, color=red, postaction={decorate},
	decoration={markings, mark=at position 0.3 with {\arrow{<}}},xshift = 4cm]  (1.7,-1) -- (1,0.2);

	\draw[thick,
	postaction={decorate},
	decoration={markings, mark=at position 0.5 with {\arrow{>}}}, color=blue,xshift = 4cm
	] (-1,-1) -- (-0.5,-0.2); \node[xshift = 4cm] at (1.7,1.3) {$a$};
	\draw[thick, color=blue, xshift = 4cm,postaction={decorate},
	decoration={markings, mark=at position 0.5 with {\arrow{>}}}]  (-0.5,-0.2) -- (1,-0.2);
	\draw[thick, color=blue, xshift = 4cm,postaction={decorate},
	decoration={markings, mark=at position 0.3 with {\arrow{<}}}]  (1.7,1) -- (1,-0.2);
	\node[xshift = 4cm] at (1.7,-1.3) {$i$};
	\node[xshift = 4cm] at (-1.2,-1.3) {$a$};
	
	\node[xshift = 4cm] at (2.5,0){$=$};

	\draw[thick,
	postaction={decorate},
	decoration={markings, mark=at position 0.5 with {\arrow{>}}}, color=red, xshift = 8.5cm
	] (-1,1) -- (-0.5,0.2); \node[xshift = 8.5cm] at (-1.3,1.3) {\small$(\phi_a(i),e)$};
	\node[xshift = 8.5cm] at (2,-1.3) {\small$(i,e)$};
	
	\draw[thick, color=red, xshift = 8.5cm,postaction={decorate},
	decoration={markings, mark=at position 0.5 with {\arrow{>}}}]  (-0.5,0.2)-- (1.3,0.2);
	\draw[thick, color=red, postaction={decorate},
	decoration={markings, mark=at position 0.3 with {\arrow{<}}},xshift = 8.5cm]  (2,-1) -- (1.3,0.2);

	\draw[thick,
	postaction={decorate},
	decoration={markings, mark=at position 0.5 with {\arrow{>}}}, color=blue,xshift = 8.5cm
	] (-1,-1) -- (-0.5,-0.2); \node[xshift = 8.5cm] at (2,1.3) {\small$(e,a)$}; \node[xshift = 8.5cm] at (-1.3,-1.3) {\small$(e,a)$};
	\draw[thick, color=blue, xshift = 8.5cm,postaction={decorate},
	decoration={markings, mark=at position 0.5 with {\arrow{>}}}]  (-0.5,-0.2) -- (1.3,-0.2);
	\draw[thick, color=blue, xshift = 8.5cm,postaction={decorate},
	decoration={markings, mark=at position 0.3 with {\arrow{<}}}]  (2,1) -- (1.3,-0.2);
	
	\node[xshift = 8.5cm] at (0.4,0.6) {\small$(\phi_a(i),a)$};
	\filldraw[xshift = 8.5cm, fill=gray!60!white!40, draw=black, thick] (-0.5,0) circle (0.3cm);
	\filldraw[xshift = 8.5cm, fill=gray!60!white!40, draw=black, thick] (1.4,0) circle (0.3cm);
\end{tikzpicture}
\end{aligned}
\, .
\end{equation}

\paragraph{Dual objects \\}

Next we consider the dual object of $(i,a)$. To do so we need to find a simple object $(j,b)$ which can fuse with $(i,a)$ into the identity line, \emph{i.e.} we need to make sense of:
		\begin{center}
		\begin{tikzpicture}
		\draw[thick,
		postaction={decorate},
		decoration={markings, mark=at position 0.5 with {\arrow{<}}}
		,color = red] (1,0) -- (-0.5,1); \node at (0.5,0.75) {$i$};
		\draw[thick,
		postaction={decorate},
		decoration={markings, mark=at position 0.5 with {\arrow{<}}},
		color = red ] (1,0) -- (-0.5,-1); \node at (-0.4,-0.6) {$j$};
		\draw[thick, dashed, color = red
		] (1,0) -- (2.5,0);
		
		\draw[thick,
		postaction={decorate},
		decoration={markings, mark=at position 0.5 with {\arrow{<}}}
		,color = blue] (1.1,-0.5) -- (-0.7,0.7); \node at (-0.7,0.2) {$a$};
		\draw[thick,
		postaction={decorate},
		decoration={markings, mark=at position 0.5 with {\arrow{<}}},
		color = blue ] (1.1,-0.5) -- (-0.3,-1.4); \node at (0.1,-1.4) {$b$};
		\draw[thick,	dashed, color = blue
		] (1.1,-0.5) -- (2.5,-0.5); 
		\node at (1.4,-0.8) {\small $\delta_{a b 1}$};
		\node at (1.4,0.3) {\small $\delta_{i \phi_a(j) 1}$};
		\filldraw[color = blue] (1.1,-0.5) circle (2pt);
		\filldraw[color = red] (1,0) circle (2pt);
	\end{tikzpicture}\, .
		\end{center}
We claim that the solution is:
\begin{equation}
	\label{dual}
	(i,a)^* = (\phi_{a^*}(i^*), a^*)\, .
\end{equation} 
Indeed, thanks to \eqref{vertex2}, it is immediate to rewrite this vertex as
	\begin{equation}
		\begin{aligned}
	\begin{tikzpicture}
	\draw[thick,
	postaction={decorate},
	decoration={markings, mark=at position 0.5 with {\arrow{<}}}
	,color = red] (1,0) -- (-0.5,1); \node at (0.5,0.75) {$i$};
	\draw[thick,
	postaction={decorate},
	decoration={markings, mark=at position 0.5 with {\arrow{<}}},
	color = red ] (1,0) -- (-0.5,-1); \node at (-0.4,-0.6) {$j$};
	\draw[thick, dashed, color = red
	] (1,0) -- (2.5,0);
	
	\draw[thick,
	postaction={decorate},
	decoration={markings, mark=at position 0.5 with {\arrow{<}}}
	,color = blue] (1.1,-0.5) -- (-0.7,0.7); \node at (-0.7,0.2) {$a$};
	\draw[thick,
	postaction={decorate},
	decoration={markings, mark=at position 0.5 with {\arrow{<}}},
	color = blue ] (1.1,-0.5) -- (-0.3,-1.4); \node at (0.1,-1.4) {$b$};
	\draw[thick,	dashed, color = blue
	] (1.1,-0.5) -- (2.5,-0.5); 
	\filldraw[color = blue] (1.1,-0.5) circle (2pt);
	\filldraw[color = red] (1,0) circle (2pt);
	\node at (1.4,-0.8) {\small $\delta_{a b 1}$};
	\node at (1.4,0.3) {\small $\delta_{i \phi_a(j) 1}$};
	
	\node at (3.2,-0.2){$=$};
	
	\draw[thick,
	postaction={decorate},
	decoration={markings, mark=at position 0.5 with {\arrow{<}}}
	,color = red,xshift = 5cm] (1,0) -- (-0.5,1); \node[xshift = 5cm ] at (0.5,0.75) {$i$}; \node[xshift = 5cm ] at (0.45,0) {$i^*$};
	\draw[thick,
	postaction={decorate},
	decoration={markings, mark=at position 0.2 with {\arrow{<}}},
	color = red,xshift = 5cm ] (1,0) -- (-0.5,-1); \node[xshift = 5cm ] at (-1,-1.3) {$\phi_{a^*}(i^*)$};
	\draw[thick, dashed, color = red,xshift = 5cm
	] (1,0) -- (2.5,0);
	
	\draw[thick,
	postaction={decorate},
	decoration={markings, mark=at position 0.8 with {\arrow{>}}}
	,color = blue,xshift = 5cm] (0.8,-0.7) -- (-0.7,0.4); \node[xshift = 5cm ] at (-0.8,0) {$a^*$};
	\draw[thick,
	postaction={decorate},
	decoration={markings, mark=at position 0.5 with {\arrow{<}}},
	color = blue,xshift = 5cm ] (0.8,-0.7) -- (-0.3,-1.4); \node[xshift = 5cm ] at (0.1,-1.4) {$a^*$};
	\draw[thick,	dashed, color = blue,xshift = 5cm
	] (0.8,-0.7) -- (2.5,-0.7); 
	
	\filldraw[color = blue,xshift = 5cm] (0.8,-0.7)circle (2pt);
	\filldraw[color = red,xshift = 5cm] (1,0) circle (2pt);
	
	\node[xshift = 5cm] at (1.4,-1) {\small $\delta_{a a^* 1}$};
	\node[xshift = 5cm] at (1.4,0.3) {\small $\delta_{i i^* 1}$};
\end{tikzpicture}
		\end{aligned}
\, .
	\end{equation}

\paragraph{F symbols\\}
It is now straightforward to compute the F symbols of $\mathcal{C} \rtimes G$. They are defined essentially as in a standard fusion category:
\begin{equation}
\begin{aligned}
			\resizebox{0.9\textwidth}{!}{
		\begin{tikzpicture}
		\draw[thick,
		postaction={decorate},
		decoration={markings, mark=at position 0.7 with {\arrow{<}}}, color = red
		] (1,0) -- (-0.5,1); \node at (0,1.05) {$i$};
		\draw[thick,
		postaction={decorate},
		decoration={markings, mark=at position 0.5 with {\arrow{<}}}, color = red
		] (1,0) -- (-0.5,-1); \node at (-0.5,-0.75) {$j$};
		\draw[thick,
		postaction={decorate},
		decoration={markings, mark=at position 0.5 with {\arrow{<}}}, color = red
		] (1,0) -- (2.5,0); \node at (1.7,-0.4) {$m$};
		\filldraw[color = red] (1,0) circle (2pt);
		\draw[thick,
		postaction={decorate},
		decoration={markings, mark=at position 0.5 with {\arrow{<}}}, color = red
		] (2.5,0) -- (4,1); \node at (4,0.75) {$l$};
		\draw[thick,
		postaction={decorate},
		decoration={markings, mark=at position 0.8 with {\arrow{<}}}, color = red
		] (2.5,0) -- (4,-1); \node at (3.8,-1.2) {$k$};
		\filldraw[color = red] (2.5,0) circle (2pt);
		
		\draw[thick,
		postaction={decorate},
		decoration={markings, mark=at position 0.5 with {\arrow{<}}}, color = blue
		] (-0.3,0.5) -- (2.5,0.5); \node at (1.7,0.8) {$e$};
		\filldraw[color = blue] (-0.3,0.5) circle (2pt);
		\filldraw[color = blue] (2.5,0.5) circle (2pt);
		
		\draw[thick,
		postaction={decorate},
		decoration={markings, mark=at position 0.5 with {\arrow{>}}}, color = blue
		] (-0.7,0.8) -- (-0.3,0.5); \node at (-1,0.5) {$a$};

		\draw[thick,
		postaction={decorate},
		decoration={markings, mark=at position 0.5 with {\arrow{<}}}, color = blue
		] (2.5,0.5) -- (4.3,-0.5); \node at (4.5,-0.5) {$c$};
		
		\draw[thick,
		postaction={decorate},
		decoration={markings, mark=at position 0.5 with {\arrow{<}}}, color = blue
		] (2.5,0.5) -- (4.3,1.6); \node at (4.5,1.5) {$d$};
		\draw[thick,
		postaction={decorate},
		decoration={markings, mark=at position 0.5 with {\arrow{>}}}, color = blue
		]  (0.2,-1.2) -- (1,-0.5); \node at (1,-1) {$b$};
		
		\draw[thick,
		postaction={decorate},
		decoration={markings, mark=at position 0.5 with {\arrow{>}}}, color = blue
		]  (1,-0.5) -- (-0.3,0.5);

		\node at (7,0) {$= \, \sum \limits_{(n,f) \in \mathcal{C} \rtimes G} F^{(i,a)(j,b)(m,e)}_{(k,c)(l,d)(n,f)}$};
		
		\draw[thick,
		postaction={decorate},
		decoration={markings, mark=at position 0.5 with {\arrow{>}}}, color = red
		] (9,1.5) -- (10.5,0.5); \node at (9,1.7) {$i$};
		\draw[thick,
		postaction={decorate},
		decoration={markings, mark=at position 0.5 with {\arrow{>}}}, color = red
		] (12,1.5) -- (10.5,0.5); \node at (12,1.7) {$l$};
		\draw[thick,
		postaction={decorate},
		decoration={markings, mark=at position 0.5 with {\arrow{<}}}, color = red
		] (10.5,0.5) -- (10.5,-0.5); \node at (10.2,0) {$n$};
		\filldraw[color = red] (10.5,0.5) circle (2pt);
		\filldraw[color = red] (10.5,-0.5) circle (2pt);
		\draw[thick,
		postaction={decorate},
		decoration={markings, mark=at position 0.5 with {\arrow{>}}}, , color = red
		] (9,-1.5) -- (10.5,-0.5); \node at (8.9,-1.4) {$j$};
		\draw[thick,
		postaction={decorate},
		decoration={markings, mark=at position 0.5 with {\arrow{>}}}, color = red
		] (12,-1.5) -- (10.5,-0.5); \node at (12,-1.7) {$k$};
		
		\draw[thick,
		postaction={decorate},
		decoration={markings, mark=at position 0.5 with {\arrow{<}}}, color = blue
		] (11,1.3) -- (11,-0.5); \node at (11.3,0) {$f$};
		\filldraw[color = blue] (11,1.3) circle (2pt);
		\filldraw[color = blue] (11,-0.5) circle (2pt);
		
		\draw[thick,
		postaction={decorate},
		decoration={markings, mark=at position 0.5 with {\arrow{>}}}, , color = blue
		] (9.2,-1.8) -- (11,-0.5); \node at (10,-1.6) {$b$};
		
		\draw[thick,
		postaction={decorate},
		decoration={markings, mark=at position 0.5 with {\arrow{>}}}, color = blue
		] (12.5,-1.5) -- (11,-0.5); \node at (12.5,-1.7) {$c$};
		
		\draw[thick,
		postaction={decorate},
		decoration={markings, mark=at position 0.5 with {\arrow{>}}}, color = blue
		] (12,2) -- (11,1.3); \node at (11.5,2) {$d$};
		
		\draw[thick,
		postaction={decorate},
		decoration={markings, mark=at position 0.5 with {\arrow{>}}}, color = blue
		] (9,1) -- (9.8,0.4); \node at (8.9,1.2) {$a$};
		
		\draw[thick,
		postaction={decorate},
		decoration={markings, mark=at position 0.5 with {\arrow{>}}}, color = blue
		] (9.8,0.4) -- (11,1.3); 
	\end{tikzpicture}
}
\end{aligned}
\, ,
\end{equation}
where we already resolved the cubic vertices in terms of more elementary vertices as before.

To explicitly compute them one can perform a sequence of F moves for the category $\mathcal{C}$ and the group $G$ while being careful about the action of the quartic vertices. The graphical calculus reads:
\begin{equation}
\begin{aligned}
	\resizebox{0.9\textwidth}{!}{
		\begin{tikzpicture}
		\draw[thick,
		postaction={decorate},
		decoration={markings, mark=at position 0.7 with {\arrow{<}}}, color = red
		] (1,0) -- (-0.5,1); \node at (-0.2,1.2) {$i$};
		\draw[thick,
		postaction={decorate},
		decoration={markings, mark=at position 0.5 with {\arrow{<}}}, color = red
		] (1,0) -- (-0.5,-1); \node at (-0.5,-0.75) {$j$};
		\draw[thick,
		postaction={decorate},
		decoration={markings, mark=at position 0.5 with {\arrow{<}}}, color = red
		] (1,0) -- (2.5,0); \node at (1.7,-0.4) {$m$};
		\filldraw[color = red] (1,0) circle (2pt);
		\draw[thick,
		postaction={decorate},
		decoration={markings, mark=at position 0.5 with {\arrow{<}}}, color = red
		] (2.5,0) -- (4,1); \node at (4,0.75) {$l$};
		\draw[thick,
		postaction={decorate},
		decoration={markings, mark=at position 0.8 with {\arrow{<}}}, color = red
		] (2.5,0) -- (4,-1); \node at (3.8,-1.2) {$k$};
		\filldraw[color = red] (2.5,0) circle (2pt);
		
		\draw[thick,
		postaction={decorate},
		decoration={markings, mark=at position 0.5 with {\arrow{<}}}, color = blue
		] (-0.3,0.5) -- (2.5,0.5); \node at (1.7,0.8) {$e$};
		\filldraw[color = blue] (-0.3,0.5) circle (2pt);
		\filldraw[color = blue] (2.5,0.5) circle (2pt);
		
		\draw[thick,
		postaction={decorate},
		decoration={markings, mark=at position 0.5 with {\arrow{>}}}, color = blue
		] (-0.7,0.8) -- (-0.3,0.5); \node at (-1,0.5) {$a$};

		\draw[thick,
		postaction={decorate},
		decoration={markings, mark=at position 0.5 with {\arrow{<}}}, color = blue
		] (2.5,0.5) -- (4.3,-0.5); \node at (4.5,-0.5) {$c$};
		
		\draw[thick,
		postaction={decorate},
		decoration={markings, mark=at position 0.5 with {\arrow{<}}}, color = blue
		] (2.5,0.5) -- (4.3,1.6); \node at (4.5,1.5) {$d$};
		\draw[thick,
		postaction={decorate},
		decoration={markings, mark=at position 0.5 with {\arrow{>}}}, color = blue
		]  (0.2,-1.2) -- (1,-0.5); \node at (1,-1) {$b$};
		
		\draw[thick,
		postaction={decorate},
		decoration={markings, mark=at position 0.5 with {\arrow{>}}}, color = blue
		]  (1,-0.5) -- (-0.3,0.5);

		\node at (5,0) {$ = $};
		
		\draw[thick,
		postaction={decorate},
		decoration={markings, mark=at position 0.7 with {\arrow{<}}}, color = red,
		xshift = 6cm] (1,0) -- (-0.5,1); \node[xshift = 6cm] at (-0.2,1.2) {$i$};
		\draw[thick,
		postaction={decorate},
		decoration={markings, mark=at position 0.5 with {\arrow{<}}}, color = red,  xshift = 6cm] (1,0) -- (-0.5,-1); \node[xshift = 6cm] at (-0.5,-0.75) {$j$};
		\draw[thick,
		postaction={decorate},
		decoration={markings, mark=at position 0.5 with {\arrow{<}}}, color = red,xshift = 6cm] (1,0) -- (2.5,0); \node[xshift = 6cm] at (1.7,-0.4) {$m$};
		
		\draw[thick,
		postaction={decorate},
		decoration={markings, mark=at position 0.5 with {\arrow{<}}}, color = red, xshift = 6cm] (2.5,0) -- (4,1); \node[xshift = 6cm] at (4,0.75) {$l$};
		\draw[thick,
		postaction={decorate},
		decoration={markings, mark=at position 0.8 with {\arrow{<}}}, color = red, xshift = 6cm] (2.5,0) -- (4,-1); \node[xshift = 6cm] at (3.8,-1.2) {$k$};
		\filldraw[color = red, xshift = 6cm] (2.5,0) circle (2pt);
		\filldraw[color = red, xshift = 6cm] (1,0) circle (2pt);
		
		\draw[thick,
		postaction={decorate},
		decoration={markings, mark=at position 0.5 with {\arrow{<}}}, color = blue, xshift = 6cm
		] (2.5,0.5) -- (4.3,-0.5); \node[xshift = 6cm] at (4.5,-0.5) {$c$};
		
		\draw[thick,
		postaction={decorate},
		decoration={markings, mark=at position 0.5 with {\arrow{<}}}, color = blue, xshift = 6cm
		] (2.5,0.5) -- (4.3,1.6); \node[xshift = 6cm] at (4.5,1.5) {$d$};
		\draw[thick,
		postaction={decorate},
		decoration={markings, mark=at position 0.5 with {\arrow{>}}}, color = blue,xshift = 6cm
		]  (0.2,-1.2) -- (1,-0.5); \node[xshift = 6cm] at (1,-1) {$b$};
		
		\draw[thick,
		postaction={decorate},
		decoration={markings, mark=at position 0.8 with {\arrow{>}}}, color = blue,xshift = 6cm
		]  (1,-0.5) -- (0.3,0);

		\draw[thick,
		postaction={decorate},color = blue,xshift = 6cm
		]  (0.3,0) -- (0.9,0.5); 
		
		\draw[thick,
		postaction={decorate},
		decoration={markings, mark=at position 0.5 with {\arrow{<}}}, color = blue, xshift = 6cm
		] (0.9,0.5) -- (2.5,0.5); \node[xshift = 6cm] at (1.7,0.8) {$e$};
		\filldraw[color = blue,xshift = 6cm] (0.9,0.5) circle (2pt);
		\filldraw[color = blue,xshift = 6cm] (2.5,0.5) circle (2pt);
		
		\draw[thick,
		postaction={decorate},
		decoration={markings, mark=at position 0.5 with {\arrow{>}}}, color = blue, xshift = 6cm
		] (-0.7,0.8) -- (-0.3,0.5); \node[xshift = 6cm] at (-1,0.5) {$a$};
		
		\draw[thick,
		postaction={decorate},
		decoration={markings, mark=at position 0.2 with {\arrow{>}}}, color = blue, xshift = 6cm]  (-0.3,0.5) -- (0.9,0.5);
		
		\node at (-2,-5) {$ = \sum \limits_{f} F^{abe}_{cdf}$};

		\draw[thick,
		postaction={decorate},
		decoration={markings, mark=at position 0.8 with {\arrow{<}}}, color = red,
		yshift = -5cm] (1,0) -- (-0.5,1); \node[yshift = - 5cm] at (-0.2,1.2) {$i$};
		\draw[thick,
		postaction={decorate},
		decoration={markings, mark=at position 0.5 with {\arrow{<}}}, color = red,  yshift = -5cm] (1,0) -- (-0.5,-1); \node[yshift = -5cm] at (-0.5,-0.75) {$j$};
		\draw[thick,
		postaction={decorate},
		decoration={markings, mark=at position 0.5 with {\arrow{<}}}, color = red,yshift = -5cm] (1,0) -- (2.5,0); \node[yshift = -5cm] at (1.7,-0.4) {$m$};
		
		\draw[thick,
		postaction={decorate},
		decoration={markings, mark=at position 0.5 with {\arrow{<}}}, color = red, yshift = -5cm] (2.5,0) -- (4,1); \node[yshift = -5cm] at (4,0.75) {$l$};
		\draw[thick,
		postaction={decorate},
		decoration={markings, mark=at position 0.8 with {\arrow{<}}}, color = red, yshift = -5cm] (2.5,0) -- (4,-1); \node[yshift = -5cm] at (3.8,-1.2) {$k$};
		\filldraw[color = red, yshift = -5cm] (2.5,0) circle (2pt);
		\filldraw[color = red, yshift = -5cm] (1,0) circle (2pt);

		\draw[thick,
		postaction={decorate},
		decoration={markings, mark=at position 0.5 with {\arrow{<}}}, color = blue, yshift = -5cm
		] (2.5,0.5) -- (4.3,-0.5); \node[yshift = -5cm] at (4.5,-0.5) {$c$};
		
		\draw[thick,
		postaction={decorate},
		decoration={markings, mark=at position 0.5 with {\arrow{<}}}, color = blue, yshift = -5cm
		] (1.7,1.6) -- (4.3,1.6); \node[yshift = -5cm] at (4.5,1.5) {$d$};
		\draw[thick,
		postaction={decorate},
		decoration={markings, mark=at position 0.5 with {\arrow{>}}}, color = blue,yshift = -5cm
		]  (0.2,-1.2) -- (1,-0.5); \node[yshift = -5cm] at (1,-1) {$b$};
		
		\draw[thick,
		postaction={decorate},
		decoration={markings, mark=at position 0.8 with {\arrow{>}}}, color = blue,yshift = -5cm
		]  (1,-0.5) -- (0.3,0);

		\draw[thick,
		postaction={decorate},color = blue,yshift = -5cm
		]  (0.3,0) -- (0.9,0.5); 
		
		\draw[thick,
		postaction={decorate},
		decoration={markings, mark=at position 0.5 with {\arrow{>}}}, color = blue, yshift = -5cm
		] (0.9,0.5) -- (2,0.5) ; 
		
		\draw[thick,
		postaction={decorate},
		decoration={markings, mark=at position 0.5 with {\arrow{<}}}, color = blue, yshift = -5cm
		] (2,0.5) -- (2.5,0.5); 
		
		\draw[thick,
		postaction={decorate},
		decoration={markings, mark=at position 0.5 with {\arrow{>}}}, color = blue,yshift = -5cm] (1.7,0.5) -- (1.7,1.6); \node[yshift = -5cm] at (2,1) {$f$};
		
		\draw[thick,
		postaction={decorate},
		decoration={markings, mark=at position 0.5 with {\arrow{>}}}, color = blue, yshift = -5cm
		] (-0.7,0.8) -- (-0.3,0.5); \node[yshift = -5cm] at (0.8,1.4) {$a$};

		\draw[thick,
		postaction={decorate},
		decoration={markings, mark=at position 0.5 with {\arrow{>}}}, color = blue, yshift = -5cm
		]  (-0.3,0.5) --(1.7,1.6) ; 
		
		\filldraw[color = blue,yshift = -5cm] (1.7,1.6) circle (2pt);
		\filldraw[color = blue,yshift = -5cm] (1.7,0.5) circle (2pt);
		
		\node at (7,-5) {$ = \sum \limits_{n, f} F^{\phi_e(i) \phi_{b^*}(j) m}_{k \phi_c(l) n} F^{abe}_{cdf}$};

		\draw[thick,
		postaction={decorate},
		decoration={markings, mark=at position 0.8 with {\arrow{<}}}, color = red,
		yshift = -5cm, xshift = 10cm] (1,0) -- (-0.5,1); \node[yshift = - 5cm,xshift = 10cm] at (-0.2,1.2) {$i$};
		
		\draw[thick,
		postaction={decorate},
		decoration={markings, mark=at position 0.8 with {\arrow{<}}}, color = red,yshift = -5cm,xshift = 10cm] (2,0) -- (2.5,0); \node[yshift = -5cm,xshift = 10cm] at (2.4,-0.5) {$n$};
		
		\draw[thick,
		postaction={decorate},
		decoration={markings, mark=at position 0.6 with {\arrow{<}}}, color = red,yshift = -5cm,xshift = 10cm] (2,0) -- (1.0,0);
		
		\draw[thick,
		postaction={decorate},
		decoration={markings, mark=at position 0.5 with {\arrow{<}}}, color = red, yshift = -5cm,xshift = 10cm] (2.5,0) -- (4,1); \node[yshift = -5cm,xshift = 10cm] at (4,0.75) {$l$};
		\draw[thick,
		postaction={decorate},
		decoration={markings, mark=at position 0.8 with {\arrow{<}}}, color = red, yshift = -5cm,xshift = 10cm] (2,-1) -- (4,-2); \node[yshift = -5cm,xshift = 10cm] at (4,-1.5) {$k$};
		
		\draw[thick,
		postaction={decorate},
		decoration={markings, mark=at position 0.5 with {\arrow{>}}}, color = red, yshift = -5cm,xshift = 10cm] (2,-1) -- (2,0); 
		
		\draw[thick,
		postaction={decorate},
		decoration={markings, mark=at position 0.8 with {\arrow{<}}}, color = red,  yshift = -5cm,xshift = 10cm] (2,-1) -- (-0.5,-1.5); \node[yshift = -5cm,xshift = 10cm] at (-0.5,-1.2) {$j$};

		\filldraw[color = red, yshift = -5cm,xshift = 10cm] (2,0) circle (2pt);
		\filldraw[color = red, yshift = -5cm,xshift = 10cm] (2,-1) circle (2pt);

		\draw[thick,
		postaction={decorate},
		decoration={markings, mark=at position 0.5 with {\arrow{<}}}, color = blue, yshift = -5cm,xshift = 10cm
		] (2.5,0.5) -- (4.3,-0.5); \node[yshift = -5cm,xshift = 10cm] at (4.5,-0.5) {$c$};
		
		\draw[thick,
		postaction={decorate},
		decoration={markings, mark=at position 0.5 with {\arrow{<}}}, color = blue, yshift = -5cm,xshift = 10cm
		] (1.7,1.6) -- (4.3,1.6); \node[yshift = -5cm,xshift = 10cm] at (4.5,1.5) {$d$};
		\draw[thick,
		postaction={decorate},
		decoration={markings, mark=at position 0.5 with {\arrow{>}}}, color = blue,yshift = -5cm,xshift = 10cm
		]  (-0.4,-2) -- (1,-1.7); \node[yshift = -5cm,xshift = 10cm] at (-0.1,-2.3) {$b$};
		
		\draw[thick,
		postaction={decorate},
		decoration={markings, mark=at position 0.8 with {\arrow{>}}}, color = blue,yshift = -5cm,xshift = 10cm
		]  (1,-1.7) -- (0.3,0);

		\draw[thick,
		postaction={decorate},color = blue,yshift = -5cm,xshift = 10cm
		]  (0.3,0) -- (0.9,0.5); 
		
		\draw[thick,
		postaction={decorate},
		decoration={markings, mark=at position 0.5 with {\arrow{>}}}, color = blue, yshift = -5cm,xshift = 10cm
		] (0.9,0.5) -- (2,0.5) ; 
		
		\draw[thick,
		postaction={decorate},
		decoration={markings, mark=at position 0.5 with {\arrow{<}}}, color = blue, yshift = -5cm,xshift = 10cm
		] (2,0.5) -- (2.5,0.5); 
		
		\draw[thick,
		postaction={decorate},
		decoration={markings, mark=at position 0.5 with {\arrow{>}}}, color = blue,yshift = -5cm,xshift = 10cm] (1.7,0.5) -- (1.7,1.6); \node[yshift = -5cm,xshift = 10cm] at (2,1) {$f$};
		
		\draw[thick,
		postaction={decorate},
		decoration={markings, mark=at position 0.5 with {\arrow{>}}}, color = blue, yshift = -5cm,xshift = 10cm
		] (-0.7,0.8) -- (-0.3,0.5); \node[yshift = -5cm,xshift = 10cm] at (0.8,1.4) {$a$};

		\draw[thick,
		postaction={decorate},
		decoration={markings, mark=at position 0.5 with {\arrow{>}}}, color = blue, yshift = -5cm,xshift = 10cm
		]  (-0.3,0.5) --(1.7,1.6) ; 
		
		\filldraw[color = blue,yshift = -5cm,xshift = 10cm] (1.7,1.6) circle (2pt);
		\filldraw[color = blue,yshift = -5cm,xshift = 10cm] (1.7,0.5) circle (2pt);
		
		
		\node at (-1,-10) {$ = \sum \limits_{n, f} F^{\phi_e(i) \phi_{b^*}(j) m}_{k \phi_c(l) n} F^{abe}_{cdf}$};

		\draw[thick,
		postaction={decorate},
		decoration={markings, mark=at position 0.5 with {\arrow{>}}}, color = red,
		yshift = -10cm,xshift = -8cm] (9,1.5) -- (10.5,0.5); \node[	yshift = -10cm,xshift = -8cm] at (9,1.7) {$i$};
		\draw[thick,
		postaction={decorate},
		decoration={markings, mark=at position 0.5 with {\arrow{>}}}, color = red, 	yshift = -10cm,xshift = -8cm
		] (12,1.5) -- (10.5,0.5); \node[yshift = -10cm,xshift = -8cm] at (12,1.7) {$l$};
		\draw[thick,
		postaction={decorate},
		decoration={markings, mark=at position 0.5 with {\arrow{<}}}, color = red,	yshift = -10cm,xshift = -8cm
		] (10.5,0.5) -- (10.5,-0.5); \node[	yshift = -10cm,xshift = -8cm] at (9.8,0) {$\phi_b(n)$};
		\filldraw[color = red,	yshift = -10cm,xshift = -8cm] (10.5,0.5) circle (2pt);
		\filldraw[color = red,	yshift = -10cm,xshift = -8cm] (10.5,-0.5) circle (2pt);
		\draw[thick,
		postaction={decorate},
		decoration={markings, mark=at position 0.5 with {\arrow{>}}}, , color = red,	yshift = -10cm,xshift = -8cm
		] (9,-1.5) -- (10.5,-0.5); \node[yshift = -10cm,xshift = -8cm] at (8.9,-1.4) {$j$};
		\draw[thick,
		postaction={decorate},
		decoration={markings, mark=at position 0.5 with {\arrow{>}}}, color = red,	yshift = -10cm,xshift = -8cm
		] (12,-1.5) -- (10.5,-0.5); \node[	yshift = -10cm,xshift = -8cm] at (12,-1.7) {$k$};
		
		\draw[thick,
		postaction={decorate},
		decoration={markings, mark=at position 0.5 with {\arrow{<}}}, color = blue,	yshift = -10cm,xshift = -8cm
		] (11,1.3) -- (11,-0.5); \node[	yshift = -10cm,xshift = -8cm] at (11.3,0) {$f$};
		\filldraw[color = blue,	yshift = -10cm,xshift = -8cm] (11,1.3) circle (2pt);
		\filldraw[color = blue,	yshift = -10cm,xshift = -8cm] (11,-0.5) circle (2pt);
		
		\draw[thick,
		postaction={decorate},
		decoration={markings, mark=at position 0.5 with {\arrow{>}}}, , color = blue,	yshift = -10cm,xshift = -8cm	] (9.2,-1.8) -- (11,-0.5); \node[yshift = -10cm,xshift = -8cm] at (10,-1.6) {$b$};
		
		\draw[thick,
		postaction={decorate},
		decoration={markings, mark=at position 0.5 with {\arrow{>}}}, color = blue,	yshift = -10cm,xshift = -8cm
		] (12.5,-1.5) -- (11,-0.5); \node[yshift = -10cm,xshift = -8cm] at (12.5,-1.7) {$c$};
		
		\draw[thick,
		postaction={decorate},
		decoration={markings, mark=at position 0.5 with {\arrow{>}}}, color = blue,	yshift = -10cm,xshift = -8cm
		] (12,2) -- (11,1.3); \node[yshift = -10cm,xshift = -8cm] at (11.5,2) {$d$};
		
		\draw[thick,
		postaction={decorate},
		decoration={markings, mark=at position 0.5 with {\arrow{>}}}, color = blue,	yshift = -10cm,xshift = -8cm
		] (9,1) -- (9.8,0.4); \node[yshift = -10cm,xshift = -8cm] at (8.9,1.2) {$a$};
		
		\draw[thick,
		postaction={decorate},
		decoration={markings, mark=at position 0.5 with {\arrow{>}}}, color = blue,	yshift = -10cm,xshift = -8cm
		] (9.8,0.4) -- (11,1.3);

		
		\node at (7,-10) {$ = \sum \limits_{n, f} F^{\phi_e(i) \phi_{b^*}(j) m}_{k \phi_c(l) \phi_{b^*}(n)} F^{abe}_{cdf}$};

		\draw[thick,
		postaction={decorate},
		decoration={markings, mark=at position 0.5 with {\arrow{>}}}, color = red,
		yshift = -10cm] (9,1.5) -- (10.5,0.5); \node[	yshift = -10cm] at (9,1.7) {$i$};
		\draw[thick,
		postaction={decorate},
		decoration={markings, mark=at position 0.5 with {\arrow{>}}}, color = red, 	yshift = -10cm
		] (12,1.5) -- (10.5,0.5); \node[yshift = -10cm] at (12,1.7) {$l$};
		\draw[thick,
		postaction={decorate},
		decoration={markings, mark=at position 0.5 with {\arrow{<}}}, color = red,	yshift = -10cm
		] (10.5,0.5) -- (10.5,-0.5); \node[	yshift = -10cm] at (10.2,0) {$n$};
		\filldraw[color = red,	yshift = -10cm] (10.5,0.5) circle (2pt);
		\filldraw[color = red,	yshift = -10cm] (10.5,-0.5) circle (2pt);
		\draw[thick,
		postaction={decorate},
		decoration={markings, mark=at position 0.5 with {\arrow{>}}}, , color = red,	yshift = -10cm
		] (9,-1.5) -- (10.5,-0.5); \node[yshift = -10cm] at (8.9,-1.4) {$j$};
		\draw[thick,
		postaction={decorate},
		decoration={markings, mark=at position 0.5 with {\arrow{>}}}, color = red,	yshift = -10cm
		] (12,-1.5) -- (10.5,-0.5); \node[	yshift = -10cm] at (12,-1.7) {$k$};
		
		\draw[thick,
		postaction={decorate},
		decoration={markings, mark=at position 0.5 with {\arrow{<}}}, color = blue,	yshift = -10cm
		] (11,1.3) -- (11,-0.5); \node[	yshift = -10cm] at (11.3,0) {$f$};
		\filldraw[color = blue,	yshift = -10cm] (11,1.3) circle (2pt);
		\filldraw[color = blue,	yshift = -10cm] (11,-0.5) circle (2pt);
		
		\draw[thick,
		postaction={decorate},
		decoration={markings, mark=at position 0.5 with {\arrow{>}}}, , color = blue,	yshift = -10cm	] (9.2,-1.8) -- (11,-0.5); \node[yshift = -10cm] at (10,-1.6) {$b$};
		
		\draw[thick,
		postaction={decorate},
		decoration={markings, mark=at position 0.5 with {\arrow{>}}}, color = blue,	yshift = -10cm
		] (12.5,-1.5) -- (11,-0.5); \node[yshift = -10cm] at (12.5,-1.7) {$c$};
		
		\draw[thick,
		postaction={decorate},
		decoration={markings, mark=at position 0.5 with {\arrow{>}}}, color = blue,	yshift = -10cm
		] (12,2) -- (11,1.3); \node[yshift = -10cm] at (11.5,2) {$d$};
		
		\draw[thick,
		postaction={decorate},
		decoration={markings, mark=at position 0.5 with {\arrow{>}}}, color = blue,	yshift = -10cm
		] (9,1) -- (9.8,0.4); \node[yshift = -10cm] at (8.9,1.2) {$a$};
		
		\draw[thick,
		postaction={decorate},
		decoration={markings, mark=at position 0.5 with {\arrow{>}}}, color = blue,	yshift = -10cm
		] (9.8,0.4) -- (11,1.3); 
	\end{tikzpicture}
	\, .
}
\end{aligned}
\end{equation}
The end result is then:
\begin{align}
	\label{FsymbolsCG}
	F^{(i,a)(j,b)(m,e)}_{(k,c)(l,d)(n,f)} = \textcolor{red}{F^{\phi_e(i) \phi_{b^*}(j) m}_{k \phi_c(l) \phi_{b^*}(n)}}  \textcolor{blue}{ F^{abe}_{cdf}} \, .
\end{align}

\subsection{$(\text{Fib} \boxtimes \text{Fib}) \rtimes S_2$}
We are now ready to analyze in detail our category of interest. Given simple objects $i, \, j$ in Fib  and $a$ in $S_2$, we label the simple objects of $(\text{Fib} \boxtimes \text{Fib}) \rtimes S_2$ as $(i, j, a)$. Their quantum dimensions are simply $d(i)d(j)$.

The category contains eight simple objects. Denoting by $p$ the non-trivial element of $S_2$, they are:
\begin{align}
	&c_1 \colonequals (1,1,1), &&c_3 \colonequals  (\tau,1,1), &&c_5\colonequals   (1,\tau,1), &&c_7 \colonequals (\tau,\tau,1),\\
	&c_2 \colonequals  (1,1,p), &&c_4 \colonequals (\tau,1,p), &&c_6 \colonequals (1,\tau,p), &&c_8 \colonequals (\tau,\tau,p)\, .
\end{align}
 For notational simplicity, these eight lines are labeled by $c_1, \dots, c_8$. Almost all of them are self-dual, except $c_4$ and $c_6$: from equation \eqref{dual}, it follows that $c_4^* = c_6$ and vice versa. The simple object $c_1$ is the identity of the fusion category and we will refer to it as `1' whenever the meaning is clear. Analogously, $c_2$ corresponds to a simple $p$ line and it will sometimes be denoted just by `$p$'.
 
In the graphical calculus we represent the two non-trivial lines of the two Fibonacci categories by \textcolor{red}{red} and \textcolor{green!50!black!60}{green} lines. The permutation line, which exchanges red and green lines, is instead represented by a \textcolor{blue}{blue} one. We can therefore draw:
\begin{center}
	\label{fig:simpleobj}
	\resizebox{1\textwidth}{!}{
	\begin{tikzpicture}
		
		\draw[xshift = -6cm] (1.42,3.3) ellipse (1.42 and 0.4);
		\draw[xshift = -6cm]  (0,3.3) -- (0,0.4);
		\draw[xshift = -6cm]  (2.84,3.3) -- (2.84,0.4);
		\draw[xshift = -6cm]  (2.84,0.4) arc (0:-180:1.42 and 0.4);
		\draw[xshift = -6cm]  (2.84,0.4) arc (0:-180:1.42 and 0.4);

		\draw[xshift = -2cm] (1.42,3.3) ellipse (1.42 and 0.4);
		\draw[xshift = -2cm]  (0,3.3) -- (0,0.4);
		\draw[xshift = -2cm]  (2.84,3.3) -- (2.84,0.4);
		\draw[xshift = -2cm]  (2.84,0.4) arc (0:-180:1.42 and 0.4);
		\draw[xshift = -2cm]  (2.84,0.4) arc (0:-180:1.42 and 0.4);
		
		\draw[xshift = -2cm, color =  blue ] (1.42,0.01) -- (1.42, 2.9);
		
		\draw[xshift = 2cm] (1.42,3.3) ellipse (1.42 and 0.4);
		\draw[xshift = 2cm]  (0,3.3) -- (0,0.4);
		\draw[xshift = 2cm]  (2.84,3.3) -- (2.84,0.4);
		\draw[xshift = 2cm]  (2.84,0.4) arc (0:-180:1.42 and 0.4);
		\draw[xshift = 2cm]  (2.84,0.4) arc (0:-180:1.42 and 0.4);
		
		\draw[xshift = 2cm, color =  red ] (1.42,0.01) -- (1.42, 2.9);

		\draw[xshift = 6cm] (1.42,3.3) ellipse (1.42 and 0.4);
		\draw[xshift = 6cm]  (0,3.3) -- (0,0.4);
		\draw[xshift = 6cm]  (2.84,3.3) -- (2.84,0.4);
		\draw[xshift = 6cm]  (2.84,0.4) arc (0:-180:1.42 and 0.4);
		\draw[xshift = 6cm]  (2.84,0.4) arc (0:-180:1.42 and 0.4);
		
		\draw[xshift = 6cm, color =  red] (1.2,0.01) -- (1.2, 2.9);
		\draw[xshift = 6cm, color =  blue ] (1.6,0.01) -- (1.6, 2.9);
		
		\node[xshift = -6cm]  at (1.42,-0.5){$c_1$};
		\node[xshift = -2cm]  at (1.42,-0.5){$c_2$};
		\node[xshift = 2cm]  at (1.42,-0.5){$c_3$};
		\node[xshift = 6cm]  at (1.42,-0.5){$c_4$};
		
		\draw[xshift = -6cm, yshift = -5 cm] (1.42,3.3) ellipse (1.42 and 0.4);
		\draw[xshift = -6cm,yshift = -5 cm]  (0,3.3) -- (0,0.4);
		\draw[xshift = -6cm,yshift = -5 cm]  (2.84,3.3) -- (2.84,0.4);
		\draw[xshift = -6cm,yshift = -5 cm]  (2.84,0.4) arc (0:-180:1.42 and 0.4);
		\draw[xshift = -6cm,yshift = -5 cm]  (2.84,0.4) arc (0:-180:1.42 and 0.4);
		
		\draw[xshift = -6cm, yshift = -5 cm,color =  green!70!black!60 ] (1.42,0.01) -- (1.42, 2.9);

		\draw[xshift = -2cm,yshift = -5 cm] (1.42,3.3) ellipse (1.42 and 0.4);
		\draw[xshift = -2cm,yshift = -5 cm]  (0,3.3) -- (0,0.4);
		\draw[xshift = -2cm,yshift = -5 cm]  (2.84,3.3) -- (2.84,0.4);
		\draw[xshift = -2cm,yshift = -5 cm]  (2.84,0.4) arc (0:-180:1.42 and 0.4);
		\draw[xshift = -2cm,yshift = -5 cm]  (2.84,0.4) arc (0:-180:1.42 and 0.4);
		
		\draw[xshift = -2cm, yshift = -5 cm, color =   green!70!black!60] (1.2,0.01) -- (1.2, 2.9);
		\draw[xshift = -2cm, yshift = -5 cm, color =  blue ] (1.6,0.01) -- (1.6, 2.9);
		
		\draw[xshift = 2cm,yshift = -5 cm] (1.42,3.3) ellipse (1.42 and 0.4);
		\draw[xshift = 2cm,yshift = -5 cm]  (0,3.3) -- (0,0.4);
		\draw[xshift = 2cm,yshift = -5 cm]  (2.84,3.3) -- (2.84,0.4);
		\draw[xshift = 2cm,yshift = -5 cm]  (2.84,0.4) arc (0:-180:1.42 and 0.4);
		\draw[xshift = 2cm,yshift = -5 cm]  (2.84,0.4) arc (0:-180:1.42 and 0.4);
		
		\draw[xshift = 2cm, yshift = -5 cm, color =   red] (1.2,0.01) -- (1.2, 2.9);
		\draw[xshift = 2cm, yshift = -5 cm, color = green!70!black!60 ] (1.6,0.01) -- (1.6, 2.9);

		\draw[xshift = 6cm,yshift = -5 cm] (1.42,3.3) ellipse (1.42 and 0.4);
		\draw[xshift = 6cm,yshift = -5 cm]  (0,3.3) -- (0,0.4);
		\draw[xshift = 6cm,yshift = -5 cm]  (2.84,3.3) -- (2.84,0.4);
		\draw[xshift = 6cm,yshift = -5 cm]  (2.84,0.4) arc (0:-180:1.42 and 0.4);
		\draw[xshift = 6cm,yshift = -5 cm]  (2.84,0.4) arc (0:-180:1.42 and 0.4);
		
		\draw[xshift = 6cm, yshift = -5 cm, color =  red] (1.1,0.01) -- (1.1, 2.9);
		\draw[xshift = 6cm, yshift = -5 cm, color = green!70!black!60 ] (1.5,0.01) -- (1.5, 2.9);
		\draw[xshift = 6cm, yshift = -5 cm, color =  blue ] (1.9,0.01) -- (1.9, 2.9);
		
		\node[xshift = -6cm,yshift = -5 cm]  at (1.42,-0.5){$c_5$};
		\node[xshift = -2cm,yshift = -5 cm]  at (1.42,-0.5){$c_6$};
		\node[xshift = 2cm,yshift = -5 cm]  at (1.42,-0.5){$c_7$};
		\node[xshift = 6cm,yshift = -5 cm]  at (1.42,-0.5){$c_8$};
		
	\end{tikzpicture}\,.
}
\end{center}
If the cylinder supports a QFT, each simple line of the category, when inserted vertically on it, gives rise to a twisted Hilbert space.

 \section{Minimal central Idempotents of two copies of Fib}
 \label{sec:irreps}
 We will now discuss the minimal central idempotents of our category, building on the analysis from subsection \ref{subsec:centralIdemp}. In fact, the graphical calculus of the previous section allows us to write the equations for the tube algebra \eqref{algebra_Lasso}.

 To check if we constructed all the minimal central idempotents in a given Hilbert space, we will match their dimension with the dimension of the subalgebra as follows. Suppose we are in the $i$-th Hilbert space, which has $|\mathcal{L}_i|$ possible simple networks. As for a group, $|\mathcal{L}_i|$ is the dimension of the regular representation of the algebra in $\cH_i$, which in turn contains all the possible irreps weighted by their dimensions. Therefore, given the projector $\hat P_i^{(\chi)}$ which projects onto a $d_i^{(\chi)}$-dimensional irrep $(\chi)$, we have that
 \begin{equation}
 |\mathcal{L}_i| = \sum_\chi \left(d_i^{(\chi)}\right)^2\, .
 \end{equation}
 
 In this section, every Hilbert space will be treated independently and only later on will we discuss the lasso maps that relate sectors of different Hilbert spaces. 
 
 \subsection{Untwisted Hilbert space}
 In the untwisted Hilbert space ($\cH_{c_1}$ or simply $\cH_1$ in our notation) the simple networks are just horizontal lines inserted on the cylinder.
 The one-dimensional (diagonal) solutions of the algebra  \eqref{algebra_Lasso} are
 \begin{center}
 	\renewcommand{\arraystretch}{2}
 	\begin{tabular}{c|cccccccc}
 		& $\id_{1}$ & $c_2$ & $c_3$ & $c_4$ & $c_5$ & $c_6$ & $c_7$ & $c_8$\\
 		\hline
 		(1) & $1$ & $1$ & $\xi$ & $\xi$ & $\xi$ & $\xi$ & $\xi^2$ & $\xi^2$\\
 		(2) & $1$ & $-1$ & $\xi$ & $-\xi$ & $\xi$ & $-\xi$ & $\xi^2$ & $-\xi^2$\\
 		(3) & $1$ & $1$ & $-1/\xi$ & $-1/\xi$ & $-1/\xi$ & $-1/\xi$ & $1/\xi^2$ & $1/\xi^2$\\
 		(4) & $1$ & $-1$ & $-1/\xi$ & $1/\xi$ & $-1/\xi$ & $1/\xi$ & $1/\xi^2$ & $-1/\xi^2$
 	\end{tabular}
 \end{center}
 The derivation is straightforward. First, $c_2$ is just a $p$ line which can take values $\pm 1$, and $c_3 $ and $c_5$ are simple $\tau$ lines (red and green, respectively) which can take values $\xi$ or $-1/\xi$. Note however that the eigenvalues of $c_3$ and $c_5$ must be identical because they are related by the action of $p$, which here acts diagonally. This produces the four possibilities, with the eigenvalues of the other lines determined by consistency. For example $c_7$, being constructed with a red and a green line, is given by  $c_3 c_5$ or $c_4 =  (\tau,1,p)$ is given by $c_3 c_2$.
 
 There is also a two-dimensional (off-diagonal) solution, which corresponds to 
 \begin{equation}
 	c_2 = p = \begin{pmatrix}
 	0 & 1 \\
 	1 & 0
 	\end{pmatrix}\, .
 \end{equation}
The two $\tau$ lines $c_3$ and $c_5$ are now differently represented: when one has eigenvalue $\xi$, the other has eigenvalue  $-1/\xi$. 
The full representation is given by
\begin{align}
&	c_1 = \mathbb{1}_2 \, ,\,\,
 &&c_3 = \begin{pmatrix}
		\xi & 0 \\
		0 & -1/\xi
	\end{pmatrix}\, ,
	\,\, &&c_5 = \begin{pmatrix}
	-1/\xi & 0 \\
	0 & \xi
	\end{pmatrix}\, ,
\,\, && c_ 7 = c_3 c_5 = - \mathbb{1}_2\, , \\
	&	c_2  = \begin{pmatrix}
		0 & 1 \\
		1 & 0
	\end{pmatrix}\, , \,\,
	&& c_4 = \begin{pmatrix}
		0 & -1/\xi \\
		\xi & 0
	\end{pmatrix} \, , \,\,
  &&c_6 = \begin{pmatrix}
		0 & \xi \\
		-1/\xi & 0
	\end{pmatrix} \, ,
	 \qquad &&c_8 = - \begin{pmatrix}
	  	0 & 1 \\
	  	1 & 0
	  \end{pmatrix}\, .
\end{align}
 
 The five minimal central idempotents in the untwisted Hilbert space are then found to be
 \begin{align}
 	\hat P_1^{(1)} & = \frac{1}{10 \xi^2} \left( \id_{1} + c_2 + \xi(c_3 + c_4 + c_5 + c_6) + \xi^2 (c_7 + c_8) \right)\, , \nn \\
 	\hat P_1^{(2)} & = \frac{1}{10 \xi^2} \left( \id_{1} - c_2 + \xi(c_3 - c_4 + c_5 - c_6) + \xi^2 (c_7 - c_8) \right)\, , \nn \\
 		\hat P_1^{(3)} & = \frac{\xi^2}{10} \left( \id_{1} + c_2 -1/\xi(c_3 + c_4 + c_5 + c_6) + 1/\xi^2 (c_7 + c_8) \right)\, , \\
 	\hat P_1^{(4)} & = \frac{\xi^2}{10} \left( \id_{1} - c_2 -1/ \xi(c_3 - c_4 + c_5 - c_6) + 1/\xi^2 (c_7 - c_8) \right)\, , \nn \\
 		\hat P_1^{(5)} & = \frac{1}{5} \left( 2 \, \id_{1} + c_3 + c_5 - 2 \,  c_7  \right)\, . \nn
 \end{align}
Note how the first two projectors differ by the coefficient of the $p = c_2$ line, while the second pair differs from the first by the replacement $\xi \to -1/\xi$. The fifth projector corresponds to the two-dimensional irrep. 
 
 \subsection{Twisted p (or $c_2$) Hilbert space}
 Let us move on to the Hilbert space generated by the permutation line $p = c_2$. Here there are four possible networks we can construct:
 \begin{center}
 \begin{tikzpicture}
 	
 	\draw[xshift = 1.5cm] (1.42,3.3) ellipse (1.42 and 0.4);
 	\draw[xshift = 1.5cm]  (0,3.3) -- (0,0.4);
 	\draw[xshift = 1.5cm]  (2.84,3.3) -- (2.84,0.4);
 	\draw[xshift = 1.5cm]  (2.84,0.4) arc (0:-180:1.42 and 0.4);
 	\draw[xshift = 1.5cm, color = blue]  (1.42,2.9) -- (1.42,0);

 	\draw[xshift = 5cm] (1.42,3.3) ellipse (1.42 and 0.4);
 	\draw[xshift = 5cm]  (0,3.3) -- (0,0.4);
 	\draw[xshift = 5cm]  (2.84,3.3) -- (2.84,0.4);
 	\draw[xshift = 5cm]  (2.84,0.4) arc (0:-180:1.42 and 0.4);

 	\draw[dashed,color =gray!50,xshift = 5cm ] (0,1.9) to[out=10, in=150] (2.84, 1.2);
 	
 	\draw[xshift = 5cm, color = blue ] (0,1.8) to[out=10, in=240] (1.42, 2.9);
 	\draw[xshift = 5cm, color = blue ] (1.42,0) to[out=80, in=170] (2.84, 1.1);
 	
 	\draw[xshift = 8.5cm] (1.42,3.3) ellipse (1.42 and 0.4);
 	\draw[xshift = 8.5cm]  (0,3.3) -- (0,0.4);
 	\draw[xshift = 8.5cm]  (2.84,3.3) -- (2.84,0.4);
 	\draw[xshift = 8.5cm]  (2.84,0.4) arc (0:-180:1.42 and 0.4);
 	\draw[xshift = 8.5cm]  (2.84,0.4) arc (0:-180:1.42 and 0.4);
 	
 	\draw[xshift = 8.5cm, color = blue]  (1.42,2.9) -- (1.42,0);
 	
 	\draw[xshift = 8.5cm,green!70!black!60]  (2.84,2.3) to[out=220, in=20] (1.42, 1.7);
 	\draw[xshift = 8.5cm,red]  (0,1.8) to[out=0, in=200] (1.42, 1.7);
 	\draw[xshift = 8.5cm,green!70!black!60]  (0,2.3) to[out=-45, in=160] (1.42, 1.2);
 	\draw[xshift = 8.5cm,red]  (2.84,1.8) to[out=220, in=20] (1.42, 1.2);

 	\draw[dashed, color =gray!50, xshift = 8.5cm]  (2.84,2.3) arc (0:180:1.42 and 0.4);
 	\draw[dashed, color =gray!50, xshift = 8.5cm]  (2.84,1.8) arc (0:180:1.42 and 0.4);
 	
 	\draw[xshift = 12cm] (1.42,3.3) ellipse (1.42 and 0.4);
 	\draw[xshift = 12cm]  (0,3.3) -- (0,0.4);
 	\draw[xshift = 12cm]  (2.84,3.3) -- (2.84,0.4);
 	\draw[xshift = 12cm]  (2.84,0.4) arc (0:-180:1.42 and 0.4);
 	\draw[xshift = 12cm]  (2.84,0.4) arc (0:-180:1.42 and 0.4);
 	
 	\draw[xshift = 12cm,green!70!black!60]  (2.84,1.5) to[out=220, in=20] (1.75, 0.9);
 	\draw[xshift = 12cm,red]  (0,0.9) to[out=0, in=200] (1.75, 0.9);
 	\draw[xshift = 12cm,green!70!black!60]  (0,1.5) to[out=-45, in=160] (1.5, 0.4);
 	\draw[xshift = 12cm,red]  (2.84,0.9) to[out=220, in=20] (1.5, 0.4);

 	\draw[dashed, color =gray!50, xshift = 12cm]  (2.84,1.5) arc (0:180:1.42 and 0.4);
 	\draw[dashed, color =gray!50, xshift = 12cm]  (2.84,1) arc (0:180:1.42 and 0.4);
 	
 	\draw[xshift = 12cm, color = blue ] (0,1.9) to[out=10, in=240] (1.42, 2.9);
 	\draw[xshift = 12cm, color = blue ] (1.42,0) to[out=80, in=180] (2.84, 1.6);

 	\node [xshift = 1.5cm] at (1.42,-0.7){$\id_p$};
 	\node [xshift = 5cm] at (1.42,-0.7){$T_p$};
 	\node [xshift = 8.5cm] at (1.42,-0.7){$\mathcal{L}_{p}^{c_7 c_8}$};
 	\node [xshift = 12cm] at (1.42,-0.7){$\mathcal{L}_{p}^{c_8 c_7}$};
 	
 \end{tikzpicture}\,.
 \end{center}
Observe that the last two networks are related by a $T$ transformation.
 
The algebra is straightforwardly written down, and its solutions are all one-dimensional. They are given by:
 \begin{center}
 	\renewcommand{\arraystretch}{2}
 	\begin{tabular}{c|cccc}
 		& $\id_{p}$ & $T_p$ & $\cL_{p}^{c_7 c_8}$ & $\cL_{p}^{c_8 c_7}$ \\
 		\hline
 		(1) & $1$ & $-1$ & $-1/\xi$ & $1/\xi$ \\
 		(2) & $1$ & $1$ & $-1/\xi$ & $-1/\xi$ \\
 		(3) & $1$ & $-1$ & $\xi$ & $-\xi$  \\
 		(4) & $1$ & $1$ & $\xi$ & $\xi$
 	\end{tabular}
 \end{center}
As can be seen from the eigenvalues of $T_p$, these solutions correspond to sectors with either integer or half-integer spin. Their minimal central idempotents are
 \begin{align}
 	\hat P_p^{(1)} & = \frac{1 + \xi^2}{10} \left( \id_{p} - T_p -1/\xi(\cL_{p}^{c_7 c_8} - \cL_{p}^{c_8 c_7}) \right)\, , \nn \\
 	\hat P_p^{(2)} & = \frac{1 + \xi^2}{10} \left( \id_{p} + T_p - 1/\xi(\cL_{p}^{c_7 c_8} +  \cL_{p}^{c_8 c_7}) \right)\, , \\
 		\hat P_p^{(3)} & = \frac{1 + \xi^2}{10 \, \xi^2} \left( \id_{p} - T_p + \xi(\cL_{p}^{c_7 c_8} -  \cL_{p}^{c_8 c_7}) \right)\, , \nn \\
 \hat P_p^{(4)} & = \frac{1 + \xi^2}{10 \, \xi^2} \left( \id_{p} + T_p + \xi(\cL_{p}^{c_7 c_8} +  \cL_{p}^{c_8 c_7}) \right)\,. \nn
 	\end{align}
 	
\subsection{Twisted $c_3$ Hilbert space}
Now consider the twisted Hilbert space of $c_3 = (\tau,1,1)$. It admits seven simple networks: the vertical $c_3$ line $\id_{c_3}$, the modular T transformation of it $T_{c_3}$, and the networks $\cL_{c_3}^{c_3 c_3}$, $\cL_{c_3}^{c_5 c_7}$, $\cL_{c_3}^{c_7 c_5}$, $\cL_{c_3}^{c_7 c_7}$ and $\cL_{c_3}^{c_8 c_8}$. The first six networks do not contain a permutation line and can therefore be represented by red lines in the twisted sector together with green lines in the untwisted sector:
 \begin{center}
	\begin{tikzpicture}

	\draw[xshift = 1.5cm] (1.42,3.3) ellipse (1.42 and 0.4);
	\draw[xshift = 1.5cm]  (0,3.3) -- (0,0.4);
	\draw[xshift = 1.5cm]  (2.84,3.3) -- (2.84,0.4);
	\draw[xshift = 1.5cm]  (2.84,0.4) arc (0:-180:1.42 and 0.4);
	\draw[xshift = 1.5cm,red]  (1.42,2.9) -- (1.42,0);

	\draw[xshift = 5cm] (1.42,3.3) ellipse (1.42 and 0.4);
	\draw[xshift = 5cm]  (0,3.3) -- (0,0.4);
	\draw[xshift = 5cm]  (2.84,3.3) -- (2.84,0.4);
	\draw[xshift = 5cm]  (2.84,0.4) arc (0:-180:1.42 and 0.4);

	\draw[dashed,color =gray!50,xshift = 5cm ] (0,1.9) to[out=10, in=150] (2.84, 1.2);
	
	\draw[xshift = 5cm,red ] (0,1.8) to[out=10, in=240] (1.42, 2.9);
	\draw[xshift = 5cm,red ] (1.42,0) to[out=80, in=170] (2.84, 1.1);

	\draw[xshift = 8.5cm] (1.42,3.3) ellipse (1.42 and 0.4);
	\draw[xshift = 8.5cm]  (0,3.3) -- (0,0.4);
	\draw[xshift = 8.5cm]  (2.84,3.3) -- (2.84,0.4);
	\draw[xshift = 8.5cm]  (2.84,0.4) arc (0:-180:1.42 and 0.4);

	\draw[dashed,color =gray!50,xshift = 8.5cm ] (0,1.9) to[out=10, in=150] (2.84, 1.2);
	
	\draw[xshift = 8.5cm,red ] (0,1.8) -- (1.2, 1.8);
	\draw[xshift = 8.5cm ,red] (1.2,1.8) -- (1.42, 2.9);
	\draw[xshift = 8.5cm,red ] (1.42,0) --(1.8, 1.1);
	\draw[xshift = 8.5cm ,red] (1.8,1.1)-- (2.84, 1.1);
	\draw[xshift = 8.5cm ,red] (1.8,1.1)-- (1.2,1.8);

	\node [xshift = 1.5cm] at (1.42,-1){$ \id_{c_3}$};
	\node [xshift = 5cm] at (1.42,-1){$T_{c_3}$};
	\node [xshift = 8.5cm] at (1.42,-1){$\mathcal{L}_{c_3}^{c_3 c_3}$};

	\draw[xshift = 1.5cm,yshift = -6cm] (1.42,3.3) ellipse (1.42 and 0.4);
	\draw[xshift = 1.5cm,yshift = -6cm] (0,3.3) -- (0,0.4);
	\draw[xshift = 1.5cm,yshift = -6cm]  (2.84,3.3) -- (2.84,0.4);
	\draw[xshift = 1.5cm,yshift = -6cm]  (2.84,0.4) arc (0:-180:1.42 and 0.4);
	\draw[xshift = 1.5cm,yshift = -6cm,red]  (1.42,2.9) -- (1.42,0);

	\draw[dashed, color =gray!50, xshift = 1.5cm,yshift = -6cm]  (2.84,2) arc (0:180:1.42 and 0.4);

	\draw[xshift = 1.5cm,yshift = -6cm,green!70!black!60]  (2.84,2) arc (0:-180:1.42 and 0.4);

	\draw[xshift = 5cm,yshift = -6cm] (1.42,3.3) ellipse (1.42 and 0.4);
	\draw[xshift = 5cm,yshift = -6cm]  (0,3.3) -- (0,0.4);
	\draw[xshift = 5cm,yshift = -6cm]  (2.84,3.3) -- (2.84,0.4);
	\draw[xshift = 5cm,yshift = -6cm]  (2.84,0.4) arc (0:-180:1.42 and 0.4);

	\draw[dashed,color =gray!50,xshift = 5cm,yshift = -6cm ] (0,1.9) to[out=10, in=150] (2.84, 1.2);
	
	\draw[xshift = 5cm,red,yshift = -6cm ] (0,1.8) to[out=10, in=240] (1.42, 2.9);
	\draw[xshift = 5cm,red,yshift = -6cm ] (1.42,0) to[out=80, in=170] (2.84, 1.1);
	
	\draw[dashed, color =gray!50, xshift = 5cm,yshift = -6.4cm]  (2.84,2) arc (0:180:1.42 and 0.4);

	\draw[xshift = 5cm,yshift = -6.4cm,green!70!black!60]  (2.84,2) arc (0:-180:1.42 and 0.4);

	\draw[xshift = 8.5cm,yshift = -6cm] (1.42,3.3) ellipse (1.42 and 0.4);
	\draw[xshift = 8.5cm,yshift = -6cm]  (0,3.3) -- (0,0.4);
	\draw[xshift = 8.5cm,yshift = -6cm]  (2.84,3.3) -- (2.84,0.4);
	\draw[xshift = 8.5cm,yshift = -6cm]  (2.84,0.4) arc (0:-180:1.42 and 0.4);

	\draw[dashed,color =gray!50,xshift = 8.5cm,yshift = -6cm ] (0,1.9) to[out=10, in=150] (2.84, 1.2);
	
	\draw[xshift = 8.5cm,red,yshift = -6cm ] (0,1.8) -- (1.2, 1.8);
	\draw[xshift = 8.5cm ,red,yshift = -6cm] (1.2,1.8) -- (1.42, 2.9);
	\draw[xshift = 8.5cm,red ,yshift = -6cm] (1.42,0) --(1.8, 1.1);
	\draw[xshift = 8.5cm ,red,yshift = -6cm] (1.8,1.1)-- (2.84, 1.1);
	\draw[xshift = 8.5cm ,red,yshift = -6cm] (1.8,1.1)-- (1.2,1.8);

	\draw[dashed, color =gray!50, xshift = 8.5 cm,yshift = -6.3cm]  (2.84,2) arc (0:180:1.42 and 0.4);

	\draw[xshift = 8.5 cm,yshift = -6.3cm,green!70!black!60]  (2.84,2) arc (0:-180:1.42 and 0.4);

	\node [xshift = 1.5cm,yshift = -6cm] at (1.42,-1){$\mathcal{L}_{c_3}^{c_5 c_7}$};
	\node [xshift = 5cm,yshift = -6cm] at (1.42,-1){$\mathcal{L}_{c_3}^{c_7 c_5}$};
	\node [xshift = 8.5cm,yshift = -6cm] at (1.42,-1){$\mathcal{L}_{c_3}^{c_7 c_7}$};
	
\end{tikzpicture}\,.
\end{center}
The seventh one $\cL_{c_3}^{c_8 c_8}$ contains a permutation line and mixes red and green lines:\footnote{This is the way the network should be drawn in terms of the lines and vertices in $\cC \rtimes G$. Of course, we could pull the green line through the blue line, and find a red lasso operation, an isolated horizontal blue line and then another red lasso operation. Such deformations can be useful to get an intuitive idea of the action of the network.}
 \begin{center}
\begin{tikzpicture}
	\draw (1.42,3.3) ellipse (1.42 and 0.4);
	\draw (0,3.3) -- (0,0.4);
	\draw  (2.84,3.3) -- (2.84,0.4);
	\draw  (2.84,0.4) arc (0:-180:1.42 and 0.4);

	\draw[dashed,color =gray!50 ] (0,1.9) to[out=10, in=150] (2.84, 1.2);
	
	\draw[red] (0,1.9) -- (0.5, 1.9);
	\draw[red] (1.35,2.35) -- (1.42, 2.9);
	\draw[red] (1.42,0) --(1.8, 1.1);
	\draw[red] (1.8,1.1)-- (2.84, 1.1);
	\draw[red] (1.8,1.1)-- (1.35, 1.7);
	\draw[green!70!black!60] (1.35,2.35) -- (1.3,2.1);
	
	\draw[green!70!black!60] (0,2.1) -- (0.5, 2.1);
	\draw[green!70!black!60] (1.3,2.1) -- (0.5, 2.1);
	\draw[red] (0.5,1.9) -- (1.2, 1.9);
	\draw[red] (1.2,1.9) -- (1.35, 1.7);
	\draw[green!70!black!60] (1.6,1.7)-- (1.3, 2.1);
	
	\draw[green!70!black!60] (1.6,1.7)-- (1.9, 1.3);
	\draw[green!70!black!60] (1.9,1.3)-- (2.84, 1.3);
	\draw[blue] (2,1.5)-- (2.84, 1.5);
	\draw[blue] (2,1.5)-- (1.35,2.35);
	\draw[blue] (0,2.35)-- (1.35,2.35);
\end{tikzpicture}\,.
\end{center}

There are seven different one-dimensional solutions of the corresponding algebra. For the first five $\cL_{c_3}^{c_8 c_8} = 0$, so the permutation line plays no role and correspondingly they are simply the solutions of Fib in the twisted $\tau$ Hilbert space for the first copy, multiplied by the solutions of Fib in the untwisted Hilbert space for the second copy. Their projectors are direct products:\footnote{To not overload our notation with sub- and superscripts, we leave it implicit that the first projector corresponds to the \textcolor{red}{red} Fib and the second to the \textcolor{green!50!black!60}{green} Fib.}
 \begin{align}
 	\label{trivialPc3}
	\hat P_{c_3}^{(1)} & =  \hat P_\tau^{(1)} \boxtimes \hat P_1^{(1)} \, , \nn \\
	\hat P_{c_3}^{(2)} & =  \hat P_\tau^{(2)} \boxtimes \hat P_1^{(1)} \, , \nn \\
	\hat P_{c_3}^{(3)} & =  \hat P_\tau^{(2)} \boxtimes \hat P_1^{(2)} \, ,  \\
		\hat P_{c_3}^{(4)} & =  \hat P_\tau^{(3)} \boxtimes \hat P_1^{(1)} \, , \nn  \\
			\hat P_{c_3}^{(5)} & =  \hat P_\tau^{(3)} \boxtimes \hat P_1^{(2)} \, . \nn 
\end{align}
We find two more non-trivial solutions where $\cL_{c_3}^{c_8 c_8}$ is not zero. They are:
\begin{center}
	 	\renewcommand{\arraystretch}{2}
	\begin{tabular}{c|ccccccc}
	& $\id_{c_3}$ & $T_{c_3}$ & $\cL_{c_3}^{c_3 c_3}$ & $\cL_{c_3}^{c_5 c_7}$ & $\cL_{c_3}^{c_7 c_5}$ & $\cL_{c_3}^{c_7 c_7}$ & $\cL_{c_3}^{c_8 c_8}$ \\ 
	\hline
	(6) & $1$ & $1$ & $1/\xi^{3/2}$ & $-1/\xi$ & $-1/\xi$ & $-1/\xi^{5/2}$ & $(\xi^2 + 1)/\xi^{3/2}$ \\
	(7) & $1$ & $1$ & $1/\xi^{3/2}$ & $-1/\xi$ & $-1/\xi$ & $-1/\xi^{5/2}$& $-(\xi^2 + 1)/\xi^{3/2}$ \\
\end{tabular}
\end{center}
The two projectors that arise from these solutions can be written as
 \begin{align}
	\hat P_{c_3}^{(6)} & = \frac{1}{2}\hat P_\tau^{(1)} \boxtimes \hat P_1^{(2)} + \frac{1}{2} \, \xi^{3/2} / (\xi^2 + 1)\cL_{c_3}^{c_8 c_8} \, ,  \\
	\hat P_{c_3}^{(7)} & =  \frac{1}{2} \hat P_\tau^{(1)} \boxtimes \hat P_1^{(2)} - \frac{1}{2} \, \xi^{3/2} / (\xi^2 + 1)\cL_{c_3}^{c_8 c_8}  \, .  
\end{align}
Here we find the projector $\hat P_\tau^{(1)} \boxtimes \hat P_1^{(2)}$ which was missing from the list of five given above. Indeed, this projector is special: for each of the two Fibs the Hilbert spaces of $\hat P_\tau^{(1)}$ and $\hat P_1^{(2)}$ are isomorphic, so here the permutation line can have a non-trivial action.

\subsection{Twisted $c_4$ Hilbert space}
In the twisted $c_4 = (\tau,1,p)$ Hilbert space there are six simple networks. The one that carries information about the spin of each sector is $\cL_{c_4}^{c_6 1} \equalscolon T_{c_4}$, while $\cL_{c_4}^{c_4 c_7} = T^{-1}_{c_4}$ and $\cL_{c_4}^{c_7 c_6} = T^{2}_{c_4}$. The other two possible networks are $\cL_{c_4}^{c_8 c_7}$ and its modular T transformation $\cL_{c_4}^{c_7 c_8}$:

\vspace{0.5cm}
\resizebox{1\textwidth}{!}{
\begin{tikzpicture}
	\draw(1.42,3.3) ellipse (1.42 and 0.4);
	\draw  (0,3.3) -- (0,0.4);
	\draw  (2.84,3.3) -- (2.84,0.4);
	\draw (2.84,0.4) arc (0:-180:1.42 and 0.4);

	\draw[dashed,color =gray!50 ] (0,1.9) to[out=10, in=150] (2.84, 1.2);

	\draw[red] (0.7,2.08) to[out=45, in=260] (1.12, 2.9);
	\draw[green!70!black!60]  (0,1.5) to[out=30, in=225] (0.7,2.08);
	\draw[blue ] (0,1.8) to[out=10, in=240] (1.42, 2.9);
	
	\draw[red] (1.12,0) to[out=70, in=190] (2, 0.9 );
	\draw[green!70!black!60]  (2,0.9) to[out=10, in=160] (2.84,0.8);
	\draw[blue] (1.42,0) to[out=80, in=170] (2.84, 1.1);

	\draw[xshift = 3.5cm](1.42,3.3) ellipse (1.42 and 0.4);
	\draw[xshift = 3.5cm]  (0,3.3) -- (0,0.4);
	\draw[xshift = 3.5cm]  (2.84,3.3) -- (2.84,0.4);
	\draw[xshift = 3.5cm] (2.84,0.4) arc (0:-180:1.42 and 0.4);

	\draw[dashed,color =gray!50,xshift = 3.5cm] (0,1.9) to[out=10, in=150] (2.84, 1.2);

	\draw[red,xshift = 3.5cm] (0.7,2.08) to[out=80, in=200] (1.12, 2.9);
	\draw[red,xshift = 3.5cm]  (0,1.5)to[out=40, in=40] (1.12,0);
	\draw[blue,xshift = 3.5cm ] (0,1.8) to[out=10, in=240] (1.42, 2.9);
	\draw[green!70!black!60,xshift = 3.5cm] (2,0.9) to[out=160, in=260] (0.7,2.08);
	
	\draw[xshift = 3.5cm,red]  (2,0.9) to[out=-10, in=180] (2.84,0.8);
	\draw[blue,xshift = 3.5cm] (1.42,0) to[out=80, in=170] (2.84, 1.1);
	
	\draw[xshift = 7cm](1.42,3.3) ellipse (1.42 and 0.4);
	\draw[xshift = 7cm]  (0,3.3) -- (0,0.4);
	\draw[xshift = 7cm]  (2.84,3.3) -- (2.84,0.4);
	\draw[xshift = 7cm] (2.84,0.4) arc (0:-180:1.42 and 0.4);

	\draw[dashed,color =gray!50,xshift = 7cm] (0,2.1) to[out=10, in=150] (2.84, 1.7);
	\draw[dashed,color =gray!50,xshift = 7cm] (0,1.6) to[out=10, in=150] (2.84, 1);
	
	\draw[red,xshift = 7cm ] (0,2) to[out=10, in=240] (1.12, 2.9);

	\draw[red,xshift = 7cm] (0.92,0) to[out=80, in=210] (1.42, 0.9);
	\draw[green!70!black!60,xshift = 7cm] (1.42,0.9) to[out=20, in=170]  (2.84, 1);
	
	\draw[xshift = 7cm,blue]  (1.42,2.9) -- (1.42,0);
	
	\draw[xshift = 7cm,red]  (2.84,1.6) arc (0:-90:1.42 and 0.4);
	\draw[xshift = 7cm,green!70!black!60]  (1.42,1.2) arc (-90:-180:1.42 and 0.4);

	\draw[xshift = 10.5cm](1.42,3.3) ellipse (1.42 and 0.4);
	\draw[xshift = 10.5cm]  (0,3.3) -- (0,0.4);
	\draw[xshift = 10.5cm]  (2.84,3.3) -- (2.84,0.4);
	\draw[xshift = 10.5cm] (2.84,0.4) arc (0:-180:1.42 and 0.4);

	\draw[dashed,color =gray!50,xshift = 10.5cm ] (0,1.9) to[out=10, in=150] (2.84, 1.2);

	\draw[red,xshift = 10.5cm] (0.7,2.08) -- (1.12, 2.9);

	\draw[green!70!black!60,xshift = 10.5cm]  (1.87,0.8) -- (1.7,0.8);
	\draw[green!70!black!60,xshift = 10.5cm]  (0.7,2.08) -- (0.5,1.7);
	\draw[green!70!black!60,xshift = 10.5cm]  (0.5,1.7) -- (0,1.6);
	\draw[green!70!black!60,xshift = 10.5cm]  (0.5,1.7) -- (1.7,0.8);
	\draw[red,xshift = 10.5cm]  (0.5,1.4) -- (0,1.4);
	\draw[red,xshift = 10.5cm]  (0.5,1.4) -- (1.53,0.6);
	\draw[red,xshift = 10.5cm]  (1.12,0) -- (1.53,0.6);
	\draw[red,xshift = 10.5cm]  (1.53,0.6) -- (1.68,0.6);
	\draw[red,xshift = 10.5cm]  (2.84,0.6)  to[out=180, in=0] (1.87,0.8);

	\draw[green!70!black!60,xshift = 10.5cm]  (1.68,0.6)  to[out=0, in=180] (2.84,0.8);

	\draw[blue,xshift = 10.5cm] (1.42,0) to[out=80, in=170] (2.84, 1.1);
	\draw[blue,xshift = 10.5cm ] (0,1.8) to[out=10, in=240] (1.42, 2.9);
	
	\draw[xshift = 14cm](1.42,3.3) ellipse (1.42 and 0.4);
	\draw[xshift = 14cm]  (0,3.3) -- (0,0.4);
	\draw[xshift = 14cm]  (2.84,3.3) -- (2.84,0.4);
	\draw[xshift = 14cm] (2.84,0.4) arc (0:-180:1.42 and 0.4);

	\draw[xshift = 14cm,blue]  (1.42,0) -- (1.42,2.9);
	\draw[red,xshift = 14cm] (0.92,0.03) to[out=90, in=180] (1.42, 0.5 );
	\draw[green!70!black!60,xshift = 14cm]  (1.42,0.5)to[out = 0, in = 180]   (2.84,1.1);
	
	\draw[red,xshift = 14cm]  (1.42,1.1) to[out = 0, in = 180]  (2.84,0.8);
	\draw[green!70!black!60,xshift = 14cm]  (1.42,1.1) -- (1.3,1.1);
	
	\draw[green!70!black!60,xshift = 14cm]  (1.3,1.8) -- (1.3,1.1);
	\draw[red,xshift = 14cm]  (1.1,1.8) -- (1.1,0.4);

	\draw[green!70!black!60,xshift = 14cm]  (0,2.4) -- (0.8,2.4);
	\draw[red,xshift = 14cm]   (0,2.2) -- (0.8,2.2);
	
	\draw[green!70!black!60,xshift = 14cm]  (1.3,1.8) -- (1.3,2.4);
	\draw[red,xshift = 14cm]  (1.1,1.8) -- (1.1,2.2);
	
	\draw[green!70!black!60,xshift = 14cm]  (0.8,2.4) -- (1.3,2.4);
	\draw[red,xshift = 14cm]  (0.8,2.2) -- (1.1,2.2);

	\draw[red,xshift = 14cm]  (0.6,2.2) -- (1.2,2.9);

	\node at (1.42,-0.7){$T_{c_4}$};
	\node[xshift = 3.5cm] at (1.42,-0.7){$T^{-1}_{c_4}$};
	\node[xshift = 7cm] at (1.42,-0.7){$T^{2}_{c_4}$};
	\node[xshift = 10.5cm] at (1.42,-0.7){$\mathcal{L}_{c_4}^{c_8 c_7}$};
	\node[xshift = 14cm] at (1.42,-0.7){$\mathcal{L}_{c_4}^{c_7 c_8}$};

\end{tikzpicture}\,.
}
Proceeding similarly to the other cases, we find the solutions of the algebra to be
\begin{center}
	\renewcommand{\arraystretch}{2}
	\begin{tabular}{c|cccccc}
		& $\id_{c_4}$ & $T_{c_4}$ & $T^{-1}_{c_4}$ & $T^2_{c_4}$ & $\cL_{c_4}^{c_8 c_7}$ & $\cL_{c_4}^{c_7 c_8}$ \\ 
		\hline
		(1) & $1$ & $-1$ & $-1$ & $1$ & $-1/\xi^{3/2}$ & $1/\xi^{3/2}$ \\
		(2) & $1$ & $1$ & $1$ & $1$ & $1/\xi^{3/2}$ & $1/\xi^{3/2}$  \\
		(3)  & $1$ & $e^{2\pi i/5}$ & $e^{-2\pi i/5}$ &  $e^{4\pi i/5}$ & $-\xi^{1/2}$ & $e^{-3\pi i/5} \xi^{1/2}$\\
		(4)  & $1$ & $e^{-3\pi i/5}$ & $e^{3\pi i/5}$ &  $e^{4\pi i/5}$ & $ \xi^{1/2}$ & $e^{-3\pi i/5} \xi^{1/2}$ \\
		(5) & $1$ & $e^{-2\pi i/5}$ & $e^{2\pi i/5}$ &  $e^{-4\pi i/5}$ & $- \xi^{1/2}$ & $e^{3\pi i/5}\xi^{1/2}$ \\
		(6)  & $1$ & $e^{3\pi i/5}$ & $e^{-3\pi i/5}$ &  $e^{-4\pi i/5}$ & $ \xi^{1/2}$ & $e^{3\pi i/5} \xi^{1/2}$ \\
	\end{tabular}
\end{center}
We therefore have six one-dimensional projectors, namely
 \begin{align}
	\hat P_{c_4}^{(1)}  = \frac{\xi}{2(1+ \xi^2)} & \left( \id_{c_4} - T_{c_4} - T^{-1}_{c_4} + T^2_{c_4} -\frac{1}{\xi^{3/2}}\left( \cL_{c_4}^{c_8 c_7} - \cL_{c_4}^{c_7 c_8} \right) \right) \, ,  \nn \\
  \hat P_{c_4}^{(2)}  =  \frac{\xi}{2(1+ \xi^2)} & \left( \id_{c_4} + T_{c_4} + T^{-1}_{c_4} + T^2_{c_4} + \frac{1}{\xi^{3/2}}\left( \cL_{c_4}^{c_8 c_7} + \cL_{c_4}^{c_7 c_8} \right) \right) \, ,  \nn \\
  \hat P_{c_4}^{(3)}  = \frac{1}{2(1+ \xi^2)} & \left( \id_{c_4} +  e^{-2\pi i/5} T_{c_4} +e^{2\pi i/5} T^{-1}_{c_4} + e^{-4\pi i/5} T^2_{c_4} -\xi^{1/2} \left( \cL_{c_4}^{c_8 c_7} + e^{-2\pi i/5} \cL_{c_4}^{c_7 c_8} \right) \right) \, ,  \\
  \hat P_{c_4}^{(4)}   = \frac{1}{2(1+ \xi^2)} & \left( \id_{c_4} + e^{3\pi i/5}T_{c_4}+  e^{-3\pi i/5}T^{-1}_{c_4} + e^{-4\pi i/5} T^2_{c_4} + \xi^{1/2} \left( \cL_{c_4}^{c_8 c_7} +e^{3\pi i/5} \cL_{c_4}^{c_7 c_8} \right) \right) \, , \nn \\
  \hat P_{c_4}^{(5)} =  \frac{1}{2(1+ \xi^2)} & \left( \id_{c_4} + e^{2\pi i/5}T_{c_4} +  e^{-2\pi i/5}T^{-1}_{c_4} +e^{4\pi i/5} T^2_{c_4} -\xi^{1/2}  \left( \cL_{c_4}^{c_8 c_7} + e^{2\pi i/5} \cL_{c_4}^{c_7 c_8} \right) \right) \, , \nn  \\
  \hat P_{c_4}^{(6)}  =  \frac{1}{2(1+ \xi^2)} & \left( \id_{c_4} +  e^{-3\pi i/5} T_{c_4} + e^{3\pi i/5} T^{-1}_{c_4} + e^{4\pi i/5}T^2_{c_4} + \xi^{1/2}  \left( \cL_{c_4}^{c_8 c_7} +  e^{-3\pi i/5} \cL_{c_4}^{c_7 c_8} \right) \right) \, . \nn
\end{align}

\subsection{Twisted $c_7$ Hilbert space}
In this Hilbert space we have 18 possible simple networks. We will find six one-dimensional representations and three two-dimensional representation. To illustrate our analysis, consider the network $\cL_{c_7}^{c_3 c_5}$, given by
\begin{center}
	\begin{tikzpicture}
	\draw[xshift = 7cm] (1.42,3.3) ellipse (1.42 and 0.4);
	\draw[xshift = 7cm]  (0,3.3) -- (0,0.4);
	\draw[xshift = 7cm]  (2.84,3.3) -- (2.84,0.4);
	\draw[xshift = 7cm]  (2.84,0.4) arc (0:-180:1.42 and 0.4);
	\draw[xshift = 7cm]  (2.84,0.4) arc (0:-180:1.42 and 0.4);
	
	\draw[dashed,color =gray!50,xshift = 7cm ] (0,1.9) to[out=10, in=150] (2.84, 1.2);
	
	\draw[ color = red,xshift = 7cm ] (0,1.8) to[out=20, in=260] (1, 2.92);
	\draw[color = red,xshift = 7cm ] (1.7,0.01) to[out=80, in=170] (2.84, 1.1);
	
	\draw[color =green!70!black!60 ,xshift = 7cm] (1.35,0.01) -- (1.35, 2.9);

	\node [xshift = 7cm] at (4,1.5){$= $ \textcolor{red}{$T_{\tau}$} \textcolor{green!50!black!60}{$\id_\tau$}\, .};
	
\end{tikzpicture}
\end{center}
Analogously, $\cL_{c_7}^{c_5 c_3} = $  \textcolor{red}{$\id_\tau$}\textcolor{green!50!black!60}{$T_\tau$}, and $T_{c_7} = $ \textcolor{red}{$T_\tau$}\textcolor{green!50!black!60}{$T_\tau$}. More generally, the idea is that networks without a permutation line can be interpreted simply as a separate red network times a green network, each acting in the twisted Hilbert space of the corresponding copy of Fib.

For the one-dimensional solutions it so happens that the representations of both copies coincide. In particular, the \textcolor{green!50!black!60}{$T_\tau$} is always represented with the same value as \textcolor{red}{$T_\tau$}. Let us denote these values as $t = e^{2 \pi i s}$, with $s = \{0, 2/5, -2/5\}$. We then recall that the one-dimensional representations of a single Fib (within the twisted Hilbert space) are entirely determined by $t$, leading directly to the following representation:
\begin{center}
	\renewcommand{\arraystretch}{2}
	\begin{tabular}{c|ccccccccc}
		& $\id_{c_7}$ & $\cL_{c_7}^{c_3 c_5}$ & $\cL_{c_7}^{c_5 c_3}$ &  $T_{c_7}$ & $\cL_{c_7}^{c_3 c_7}$ & $\cL_{c_7}^{c_5 c_7}$ & $\cL_{c_7}^{c_7 c_3}$ & $\cL_{c_7}^{c_7 c_5}$ & $\cL_{c_7}^{c_7 c_7}$ \\ 
		\hline
($s$) & $1$ & $t$ & $ = \cL_{c_7}^{c_3 c_5}$ & $t^2$ & $\xi^{1/2}/(1 + t \, \xi)$ &  $ = \cL_{c_7}^{c_3 c_7}$ &  $t\, \xi^{1/2}/(1 + t \, \xi)$ & $ = \cL_{c_7}^{c_7 c_3}$ & $\xi/(1 + t \, \xi)^2$
	\end{tabular}
\end{center}
where the ``$= \cL^{\cdot \cdot}_{\cdot}$'' simply means the networks are represented identically.

Now consider networks that contain a permutation line. For example, the insertion of a simple horizontal permutation line is realized by the network $\cL_{c_7}^{p c_8}$:
\begin{center}
	\begin{tikzpicture}

	\draw[xshift = -0.5cm] (1.42,3.3) ellipse (1.42 and 0.4);
	\draw[xshift = -0.5cm]  (0,3.3) -- (0,0.4);
	\draw[xshift = -0.5cm]  (2.84,3.3) -- (2.84,0.4);
	\draw[xshift = -0.5cm]  (2.84,0.4) arc (0:-180:1.42 and 0.4);
	\draw[xshift = -0.5cm]  (2.84,0.4) arc (0:-180:1.42 and 0.4);
	\draw[dashed, color =gray!50, xshift =-0.5cm]  (2.84,2) arc (0:180:1.42 and 0.4);
	
	\draw[xshift = -0.5cm, color =  green!70!black!60 ] (1.5,0.01) -- (1.5, 1.6);
	
	\draw[xshift = -0.5cm, color = red] (1,0.01) -- (1, 1.6);
	
	\draw[xshift = -0.5cm, color = red ] (1,2.9) to[out = 270, in = 90]  (1.5, 1.6);
	
	\draw[xshift = -0.5cm, color =  green!70!black!60] (1.5,2.9) to[out = 270, in = 90] (1, 1.6);

	\draw[xshift = -0.5cm,blue]  (2.84,2) arc (0:-180:1.42 and 0.4);

\end{tikzpicture} \, .
\end{center}
Its eigenvalues $q$ can be $+1$ and $-1$. The representation of the remaining simple networks is again completely fixed once we specify $q$, giving the two possibilities:
\begin{center}
	\renewcommand{\arraystretch}{2}
	\begin{tabular}{c|cccccccc}
		& $\cL_{c_7}^{c_4 c_6}$ &  $\cL_{c_7}^{c_6 c_4}$ & $\cL_{c_7}^{c_4 c_8}$ & $\cL_{c_7}^{c_6 c_8}$ & $\cL_{c_7}^{c_8 p}$ & $\cL_{c_7}^{c_8 c_4}$ & $\cL_{c_7}^{c_8 c_6}$ & $\cL_{c_7}^{c_8 c_8}$\\
		\hline
	$(q)$& $= q \, \cL_{c_7}^{c_3 c_5}$ & $= q\, \cL_{c_7}^{c_3 c_5}$ & $ = q\,\cL_{c_7}^{c_3 c_7}$ & $= q\,\cL_{c_7}^{c_3 c_7}$ & $= q\,T_{c_7}$ & $= q\,\cL_{c_7}^{c_7 c_3}$ & $= q\,\cL_{c_7}^{c_7 c_3}$ & $= q\,\cL_{c_7}^{c_7 c_7}$
	\end{tabular}
\end{center}
In words, this table says that the networks containing a permutation line have eigenvalues equal to $\pm$ those of the corresponding networks without a permutation line.

Combining all possible values of the pair $(s,q)$ we are led to six one-dimensional representations, as promised. The corresponding projectors are 
 \begin{align}
	&\hat P_{c_7}^{(s,q)} = \,  N^{(s)} \left(\id_{c_7} + q \cL_{c_7}^{p c_8} + t^{-2} \left( T_{c_7} + q \cL_{c_7}^{c_8 p} \right) + t^{-1} \left(\cL_{c_7}^{c_3 c_5} + \cL_{c_7}^{c_5 c_3}  + q  \left(\cL_{c_7}^{c_4 c_6} + \cL_{c_7}^{c_6 c_4}\right) \right)\right) \nonumber \\ & + N^{(s)} \left( \frac{\xi^{1/2}}{1 + t^{-1} \xi}\left(\cL_{c_7}^{c_3 c_7} + \cL_{c_7}^{c_5 c_7}  + q  \left(\cL_{c_7}^{c_4 c_8} + \cL_{c_7}^{c_6 c_8}\right) \right) +   \frac{\xi}{(1 + t^{-1} \xi)^2}\left(\cL_{c_7}^{c_7 c_7} + q \cL_{c_7}^{c_8 c_8}\right) \right) \nonumber \\ &  + \frac{t \, \xi^{1/2}}{1 + t^{-1} \xi} N^{(s)} \left(  \left( \cL_{c_7}^{c_7 c_3} + \cL_{c_7}^{c_7 c_5} \right) + q \left( \cL_{c_7}^{c_8 c_4} + \cL_{c_7}^{c_8 c_6} \right) \right) \, ,
\end{align}
with normalization $N^{(0)} = 1/10$ and $N^{(\pm 2/5)} = 1/(10 \, \xi^2)$. 

Because of the spin values $s$ and the parity symmetry labeled by $q$, it is straightforward to see that these projectors can be understood as arising partially from two copies of the twisted $\tau$ sector of Fib:
\begin{align}
	\label{projc7}
	\hat P_{c_7}^{(0,1)} + \hat P_{c_7}^{(0,-1)} & = \hat P_{\tau}^{(1)} \boxtimes \hat P_{\tau}^{(1)}\, , \nn \\ 
		\hat P_{c_7}^{(2/5,1)} + \hat P_{c_7}^{(2/5,-1)} & = \hat P_{\tau}^{(2)} \boxtimes \hat P_{\tau}^{(2)}\, , \\ 
			\hat P_{c_7}^{(-2/5,1)} + \hat P_{c_7}^{(-2/5,-1)} & = \hat P_{\tau}^{(3)} \boxtimes \hat P_{\tau}^{(3)}\, . \nn 
\end{align}
From the direct product of two copies of the Fibonacci category, however, we see that the combinations appearing in \eqref{projc7} do not exhaust all possible tensor products. In particular, the mixed combinations $\hat P_{\tau}^{(1)} \boxtimes \hat P_{\tau}^{(2)}$, $\hat P_{\tau}^{(2)} \boxtimes \hat P_{\tau}^{(3)}$, and $\hat P_{\tau}^{(1)} \boxtimes \hat P_{\tau}^{(3)}$, together with their images under the permutation $p$, are still missing.  Labeling the eigenvalues of the network $T_\tau$ in the two Fib copies as (\textcolor{red}{$t$}, \textcolor{green!50!black!60}{$t$}), these additional combinations appear when the two eigenvalues differ, i.e. when \textcolor{red}{$t$} $\neq$ \textcolor{green!50!black!60}{$t$}, in which case the permutation line $p$ -- represented in this Hilbert space by the network $\cL_{c_7}^{p c_8}$ -- acts off-diagonally.

Due to the non-trivial action of the permutation on these mixed Fib projectors, we therefore expect three additional two-dimensional projectors in this Hilbert space, which are indeed found to be
\begin{align}
	\hat P^{(7)}_{c_7} &= \hat P_{\tau}^{(1)} \boxtimes \hat P_{\tau}^{(2)} + \hat P_{\tau}^{(2)} \boxtimes \hat P_{\tau}^{(1)}\, , \nn \\
		\hat P^{(8)}_{c_7} &= \hat P_{\tau}^{(1)} \boxtimes \hat P_{\tau}^{(3)} + \hat P_{\tau}^{(3)} \boxtimes \hat P_{\tau}^{(1)}\, , \\
			\hat P^{(9)}_{c_7} &= \hat P_{\tau}^{(2)} \boxtimes \hat P_{\tau}^{(3)} + \hat P_{\tau}^{(3)} \boxtimes \hat P_{\tau}^{(2)}\, . \nn
\end{align}

\subsection{Twisted $c_8$ Hilbert space}
The twisted $c_8$ Hilbert space contains 14 simple networks and has six one-dimensional solutions to the algebra. Analogously to the $c_7$ case, these solutions can be organized into two subsectors, now distinguished by the eigenvalues ($q = \pm 1$) of the network $\cL_{c_8}^{p c_7}$:
\begin{center}
	\begin{tikzpicture}

	\draw[xshift = 5cm] (1.42,3.3) ellipse (1.42 and 0.4);
	\draw[xshift = 5cm]  (0,3.3) -- (0,0.4);
	\draw[xshift = 5cm]  (2.84,3.3) -- (2.84,0.4);
	\draw[xshift = 5cm]  (2.84,0.4) arc (0:-180:1.42 and 0.4);

	\draw[dashed,color =gray!50,xshift = 5cm ] (0,1.9) to[out=10, in=150] (2.84, 1.2);
	
	\draw[xshift = 5cm, color =  green!70!black!60 ] (1.5,0.01) -- (1.5, 2.15);
	
	\draw[xshift = 5cm, color = red] (1,0.01) -- (1, 1.9);
	
	\draw[xshift = 5cm, color = red ] (1,2.9) to[out = 270, in = 90]  (1.5, 2.15);
	
	\draw[xshift = 5cm, color =  green!70!black!60] (1.5,2.9) to[out = 270, in = 90] (1, 1.9);
	
	\draw[xshift = 5cm, blue] (0,1.8) to[out=0, in=260] (2, 2.93);
	\draw[xshift = 5cm, blue ] (2,0.03) to[out=80, in=170] (2.84, 1.1);
	
\end{tikzpicture}\,.
\end{center}
These two subsectors have the following eigenvalues in common
\begin{center}
	\renewcommand{\arraystretch}{2}
	\begin{tabular}{c|ccccccc}
		& $\id_{c_8}$ &  $\cL_{c_8}^{c_3 c_8}$ &  $\cL_{c_8}^{c_5 c_8}$ & $\cL_{c_8}^{c_7 c_2}$ & $\cL_{c_8}^{c_7 c_4}$ & $\cL_{c_8}^{c_7 c_6}$ & $\cL_{c_8}^{c_7 c_8}$ \\
		\hline
	(1)	& 1 & $\xi^{1/2}$ & $ = \cL_{c_8}^{c_3 c_8}$ & $1$ & $\xi^{1/2}$ & $ = \cL_{c_8}^{c_7 c_4}$ & $-1$ \\
	(2)	& 1 & $- \xi^{-1/2}$ & $ = \cL_{c_8}^{c_3 c_8}$ & $e^{-3 \pi i /5}$ & $e^{2 \pi i /5} \xi^{-1/2} $ & $ = \cL_{c_8}^{c_7 c_4} $ & $-1 + e^{2 \pi i /5}$ \\
		(3)	& 1 & $- \xi^{-1/2}$ & $ = \cL_{c_8}^{c_3 c_8}$ & $e^{3 \pi i /5}$ & $e^{-2 \pi i /5} \xi^{-1/2} $ & $ = \cL_{c_8}^{c_7 c_4}$ & $-1 + e^{-2 \pi i /5}$ \\
	\end{tabular}
\end{center}
and differ because of the following ones 
\begin{center}
	\renewcommand{\arraystretch}{2}
	\begin{tabular}{c|ccccccc}
		& $\cL_{c_8}^{p c_7}$ & $T_{c_8}$ &  $\cL_{c_8}^{c_4 c_7}$ &  $\cL_{c_8}^{c_6 c_7}$ & $\cL_{c_8}^{c_8 c_3}$ & $\cL_{c_8}^{c_8 c_5}$ & $\cL_{c_8}^{c_8 c_7}$  \\
		\hline
	$(q)$  & $q$ & $= q \, \cL_{c_8}^{c_7 c_2}$ & $= q \, \cL_{c_8}^{c_3 c_8}$ & $= q \, \cL_{c_8}^{c_3 c_8}$ & $= q \, \cL_{c_8}^{c_7 c_4}$ & $= q \, \cL_{c_8}^{c_7 c_4}$ & $= q \, \cL_{c_8}^{c_7 c_8}$ 
	\end{tabular}
\end{center}
Once again, since $\cL_{c_8}^{p c_7}$ acts diagonally, the networks with a single red line have the same  eigenvalues as the one with a single green line ($c_3 \leftrightarrow c_5$ and $c_4 \leftrightarrow c_6$). 

The projectors we can construct from these solutions of the algebra are
\begin{align}
	\hat P_{c_8}^{(1,q)} & = \frac{1}{2 \xi^2 \left(1 + \xi^2\right) }\left( \left(\id_{c_8} + q \cL_{c_8}^{p c_7} \right) + \left(\cL_{c_8}^{c_7 c_2} + q T_{c_8}\right)\right) \nonumber \\ & + \frac{1}{2 \xi^2 \left(1 + \xi^2\right) } \left( - \left( \cL_{c_8}^{c_7 c_8} + q \cL_{c_8}^{c_8 c_7}\right) + \xi^{1/2} \left(\cL_{c_8}^{c_3 c_8} + \cL_{c_8}^{c_5 c_8}  + q \left( \cL_{c_8}^{c_4 c_7} + \cL_{c_8}^{c_6 c_7} \right)\right) \right) \nonumber \\ & + \frac{\xi^{1/2}}{2 \xi^2 \left(1 + \xi^2\right) }  \left( \cL_{c_8}^{c_7 c_4} + \cL_{c_8}^{c_7 c_6}  + q \left(\cL_{c_8}^{c_8 c_3} + \cL_{c_8}^{c_8 c_5} \right)\right) \, , \nn \\ 
	\hat P_{c_8}^{(2,q)} & = \frac{1}{2 \xi \left(1 + \xi^2\right) } \left( \left(\id_{c_8} + q \cL_{c_8}^{p c_7} \right) + e^{3 \pi i /5} \left(\cL_{c_8}^{c_7 c_2} + q T_{c_8}\right)\right) \nonumber \\ & + \left( (-1 + e^{-2 \pi i /5}) \left( \cL_{c_8}^{c_7 c_8} + q \cL_{c_8}^{c_8 c_7}\right) - \xi^{-1/2} \left(\cL_{c_8}^{c_3 c_8} + \cL_{c_8}^{c_5 c_8}  + q \left( \cL_{c_8}^{c_4 c_7} + \cL_{c_8}^{c_6 c_7} \right)\right) \right) \nonumber \\ & -  \frac{\xi^{-1/2} e^{-2 \pi i /5}}{2 \xi \left(1 + \xi^2\right) }  \left( \cL_{c_8}^{c_7 c_4} + \cL_{c_8}^{c_7 c_6}  + q \left(\cL_{c_8}^{c_8 c_3} + \cL_{c_8}^{c_8 c_5} \right)\right)\, , \\ 
	\hat P_{c_8}^{(3,q)} & = \frac{1}{2 \xi \left(1 + \xi^2\right) } \left( \left(\id_{c_8} + q \cL_{c_8}^{p c_7} \right) + e^{-3 \pi i /5} \left(\cL_{c_8}^{c_7 c_2} + q T_{c_8}\right)\right) \nonumber \\ & + \left( (-1 + e^{2 \pi i /5}) \left( \cL_{c_8}^{c_7 c_8} + q \cL_{c_8}^{c_8 c_7}\right) - \xi^{-1/2} \left(\cL_{c_8}^{c_3 c_8} + \cL_{c_8}^{c_5 c_8}  + q \left( \cL_{c_8}^{c_4 c_7} + \cL_{c_8}^{c_6 c_7} \right)\right) \right) \nonumber \\ & -  \frac{\xi^{-1/2} e^{2 \pi i /5}}{2 \xi \left(1 + \xi^2\right) }  \left( \cL_{c_8}^{c_7 c_4} + \cL_{c_8}^{c_7 c_6}  + q \left(\cL_{c_8}^{c_8 c_3} + \cL_{c_8}^{c_8 c_5} \right)\right)\, . \nn
\end{align}
These six one-dimensional solutions are not the only ones in this Hilbert space. Indeed, there again exists a case where the permutation line (here represented by $\cL_{c_8}^{p c_7}$) acts off-diagonally on red and green lines, leading to two additional two-dimensional projectors associated with integer and half-integer spin sectors:
\begin{equation}
	T_{c_8} = x \mathbb{1}_2\, , \qquad  x = \pm 1 \, .
	\end{equation}
These two-dimensional projectors can be written as
\begin{align}
	\hat P_{c_8}^{(\text{2d}, x)} = \, & \frac{1}{1 + \xi^2} \left(\id_{c_8} + x T_{c_8}  + \frac{1}{2\xi^{3/2}} \left(  \cL_{c_8}^{c_3 c_8} + \cL_{c_8}^{c_5 c_8} + x \left(  \cL_{c_8}^{c_8 c_3} + \cL_{c_8}^{c_8 c_5} \right) \right)\right) \nonumber \\
	& -  \frac{1}{2 \xi^{1/2}(1 + \xi^2)} \left( \cL_{c_8}^{c_7 c_4} + \cL_{c_8}^{c_7 c_6} + x  \left(\cL_{c_8}^{c_4 c_7} + \cL_{c_8}^{c_6 c_7} \right)  \right)  \\
	& +  \frac{1}{\xi(1 + \xi^2)} \left(\cL_{c_8}^{c_7 c_8} + x \cL_{c_8}^{c_8 c_7} \right)\, .\nonumber
\end{align}

\section{Further lassos}
So far we have only considered the (many) simple networks and projectors onto various irreps within the same Hilbert space. The goal of this section is to extend our analysis to the maps between different Hilbert spaces, in order to identify the properly independent sectors of the theory. These maps correspond to morphisms between Hilbert spaces and, as we explained in subsection \ref{subsec:Lasso}, it is important to precisely determine their kernels and co-kernels.

First of all, the permutation line $p$ implies immediate lassos between the Hilbert spaces generated by a single $\tau$ line, so between $c_3$ and $c_5$, and between those generated by a $\tau$ and a permutation line, so between $c_4$ and $c_6$. These are isomorphisms and therefore:
\begin{align}
	\label{simpleisomorphisms}
	\cH_{c_3} \sim \cH_{c_5} \,,  \qquad	\cH_{c_4} \sim \cH_{c_6} \, .
\end{align}
Here and below, we use a physical definition of ``isomorphism'' meaning that we have in mind a map that commutes with the Hamiltonian and intertwines the projectors onto the different subspaces. This implies an equality between partition functions of the form
\begin{equation}
	\Tr_{\cH_i} \p{ \hat P_i^{(\chi)} q^{L_0 - c /24} \bar q^{\bar L_0 - c/24} }\,
\end{equation}
for different values of the pair $(i,\chi)$. For example, consider the projectors introduced in equation \eqref{trivialPc3} within the $c_3$ Hilbert space, written as $	\hat P_\tau^{(a)} \boxtimes 	\hat P_1^{(b)}$.  The $S_2$ symmetry indicates that these are isomorphic to $ \hat P_1^{(b)}  \boxtimes \hat P_\tau^{(a)}$,  which are precisely the projectors in the $c_5$ Hilbert space. 

The existence of the isomorphism \eqref{simpleisomorphisms} is of course why we did not discuss the $c_5$ and $c_6$ Hilbert spaces in the previous section: the solutions of the algebra and the projectors are entirely analogous to the ones of the $c_3$ and $c_4$ Hilbert spaces under the simultaneous exchange $c_3 \leftrightarrow c_5$ and $c_4 \leftrightarrow c_6$.

\subsection{Hilbert spaces without a permutation line}
\label{subsec:lassoe}

\subsubsection{$c_1 \leftrightarrow c_3$}
\label{subsubsec:strangelasso}
We have four simple maps that go from the untwisted Hilbert space to the $c_3$ twisted Hilbert space, and four that go the other way. They are
\begin{align}
&	(1 \to c_3): \,  \cL_{c_3 \leftarrow 1}^{c_3 c_3}\, ,  \cL_{c_3 \leftarrow 1}^{c_6 c_6}\, ,  \cL_{c_3 \leftarrow 1}^{c_7 c_7}\, ,  \cL_{c_3 \leftarrow 1}^{c_8 c_8}\, , \nn  \\
&	(c_3 \to 1) : \,  \cL_{1 \leftarrow c_3}^{c_3 c_3}\, ,  \cL_{1 \leftarrow c_3}^{c_4 c_4}\, ,  \cL_{1 \leftarrow c_3}^{c_7 c_7}\, ,  \cL_{1 \leftarrow c_3}^{c_8 c_8}\, . \nn
\end{align}
From them, we can build four independent linear combinations, namely
\begin{align}
	 &   \cL_{c_3 \leftarrow 1}^{(1)} = \frac{\xi ^{3/4}}{\left(\xi ^2+1\right)^{3/2}} \left( \cL_{c_3 \leftarrow 1}^{c_3 c_3} + \xi  \cL_{c_3 \leftarrow 1}^{c_7 c_7}\right) \, , \nn \\
	&  \cL_{c_3 \leftarrow 1}^{(2)} =\frac{\xi ^{3/4}}{\left(\xi ^2+1\right)^{3/2}} \left(\cL_{c_3 \leftarrow 1}^{c_6 c_6} + \xi  \cL_{c_3 \leftarrow 1}^{c_8 c_8}\right) \, , \\
	&  \cL_{c_3 \leftarrow 1}^{(3)} =\frac{\xi ^{11/4}}{2 \left(\xi ^2+1\right)^{3/2}} \left(\cL_{c_3 \leftarrow 1}^{c_3 c_3} + \cL_{c_3 \leftarrow 1}^{c_6 c_6} -\frac{1}{\xi} \left(  \cL_{c_3 \leftarrow 1}^{c_7 c_7} +  \cL_{c_3 \leftarrow 1}^{c_8 c_8}\right)\right) \, , \nn \\ 
		&  \cL_{c_3 \leftarrow 1}^{(4)} = \frac{\xi ^{11/4}}{2 \left(\xi ^2+1\right)^{3/2}} \left(\cL_{c_3 \leftarrow 1}^{c_3 c_3} - \cL_{c_3 \leftarrow 1}^{c_6 c_6} -\frac{1}{\xi} \left(  \cL_{c_3 \leftarrow 1}^{c_7 c_7} -  \cL_{c_3 \leftarrow 1}^{c_8 c_8}\right)\right) \, . \nn
\end{align}
Their conjugate maps, in the sense described in subsection \ref{subsubsect:reflectionpositivity}, are then the four combinations:
\begin{align}
	&   \cL_{1 \leftarrow c_3}^{(1)} =   \frac{\xi ^{3/4}}{\left(\xi ^2+1\right)^{3/2}} \left( \cL_{1 \leftarrow c_3}^{c_3 c_3} + \xi  \cL_{1 \leftarrow c_3}^{c_7 c_7} \right) \, ,\nn \\
	&  \cL_{1 \leftarrow c_3}^{(2)} =  \frac{\xi ^{3/4}}{\left(\xi ^2+1\right)^{3/2}} \left(  \cL_{1 \leftarrow c_3}^{c_4 c_4} + \xi  \cL_{1 \leftarrow c_3}^{c_8 c_8}\right) \, , \\
	&  \cL_{1 \leftarrow c_3}^{(3)} = \frac{\xi ^{11/4}}{2 \left(\xi ^2+1\right)^{3/2}}\left(\cL_{1 \leftarrow c_3}^{c_3 c_3} + \cL_{1 \leftarrow c_3}^{c_4 c_4} -\frac{1}{\xi} \left(  \cL_{1 \leftarrow c_3}^{c_7 c_7} +  \cL_{1 \leftarrow c_3}^{c_8 c_8}\right) \right) \, , \nn \\ 
	&  \cL_{1 \leftarrow c_3}^{(4)} = \frac{\xi ^{11/4}}{2 \left(\xi ^2+1\right)^{3/2}} \left(\cL_{1 \leftarrow c_3}^{c_3 c_3} - \cL_{1 \leftarrow c_3}^{c_4 c_4} -\frac{1}{\xi} \left(  \cL_{1 \leftarrow c_3}^{c_7 c_7} -  \cL_{1 \leftarrow c_3}^{c_8 c_8}\right)\right)  \, . \nn
\end{align}
Now let us analyze their action on the two Hilbert spaces. First we consider the $\cL^{(3)}$'s and $\cL^{(4)}$'s, which obey the relations
\begin{align}
	\cL_{c_3 \leftarrow 1}^{(3)} \times \hat P_1^{(3)} &= \cL_{c_3 \leftarrow 1}^{(3)}\, ,
	&\cL_{c_3 \leftarrow 1}^{(3)} \times \hat P_1^{(i \neq 3)} &= 0\, , \nn \\
	\cL_{c_3 \leftarrow 1}^{(4)} \times \hat P_1^{(4)} &= \cL_{c_3 \leftarrow 1}^{(4)}\, ,\,
	&\cL_{c_3 \leftarrow 1}^{(4)} \times \hat P_1^{(i \neq 4)} &= 0\, , \nn \\
	\cL_{1 \leftarrow c_3}^{(3)} \times \hat P_{c_3}^{(6)} &= \cL_{1 \leftarrow c_3}^{(3)}\, , 
	&\cL_{1 \leftarrow c_3}^{(3)} \times \hat P_{c_3}^{(i \neq 6)} &= 0\, , \nn \\
	\cL_{1 \leftarrow c_3}^{(4)} \times \hat P_{c_3}^{(7)} &= \cL_{1 \leftarrow c_3}^{(4)}\, , \,
	&\cL_{1 \leftarrow c_3}^{(4)} \times \hat P_{c_3}^{(i \neq 7)} &= 0\, , \\
	\cL_{1 \leftarrow c_3}^{(3)} \times \cL_{c_3 \leftarrow 1}^{(3)} &= \, \hat P_1^{(3)}\, , 
	&\cL_{1 \leftarrow c_3}^{(4)} \times \cL_{c_3 \leftarrow 1}^{(4)} &= \, \hat P_1^{(4)}\, , \nn \\
	\cL_{c_3 \leftarrow 1}^{(3)} \times \cL_{1 \leftarrow c_3}^{(3)} &= \, \hat P_{c_3}^{(6)}\, , 
	&\cL_{c_3 \leftarrow 1}^{(4)} \times \cL_{1 \leftarrow c_3}^{(4)} &= \, \hat P_{c_3}^{(7)}\, . \nn
\end{align}
These identities show that the subspaces selected by $\hat P_1^{(3)}$ and $\hat P_1^{(4)}$ are mapped respectively to those selected by $\hat P_{c_3}^{(6)}$ and $\hat P_{c_3}^{(7)}$, and vice versa. In other words, we have obtained the isomorphisms
\begin{equation}
	\cH_1^{(3)} \sim \cH_{c_3}^{(6)}, \qquad \cH_1^{(4)} \sim \cH_{c_3}^{(7)}\,.
\end{equation}

The $\cL^{(1)}$'s and $\cL^{(2)}$'s are more interesting. They obey the relations
\begin{align}
	\label{lassos13}
	 \hat P_{c_3}^{(1)} \times \cL_{c_3 \leftarrow 1}^{(1)}\times \hat P_1^{(5)}&= \cL_{c_3 \leftarrow 1}^{(1)} \, , & \cL_{c_3 \leftarrow 1}^{(1)} \times \hat P_1^{(i \neq 5)}&= 0 \, , \nn \\
	\hat P_{c_3}^{(1)} \times \cL_{c_3 \leftarrow 1}^{(2)}\times \hat P_1^{(5)}&=\cL_{c_3 \leftarrow 1}^{(2)}  \, , & \cL_{c_3 \leftarrow 1}^{(2)} \times \hat P_1^{(i \neq 5)}&= 0 \, , \\
	\hat P_1^{(5)}  \times \cL_{1 \leftarrow c_3}^{(1)} \times \hat P_{c_3}^{(1)}&= \cL_{1 \leftarrow c_3}^{(1)} \, , & \cL_{1 \leftarrow c_3}^{(1)} \times \hat P_{c_3}^{(i \neq 1)}&= 0 \, , \nn \\
	\hat P_1^{(5)}  \times \cL_{1 \leftarrow c_3}^{(2)} \times \hat P_{c_3}^{(1)}&=\cL_{1 \leftarrow c_3}^{(2)} \, , & \cL_{1 \leftarrow c_3}^{(2)} \times \hat P_{c_3}^{(i \neq 1)}&= 0 \, . \nn
\end{align}
These conditions show that the $\cL^{(1)}$ and $\cL^{(2)}$ lassos map operators in the $\cH_1^{(5)}$ (sub-)Hilbert space to those in $\cH_{c_3}^{(1)}$, and vice versa. At first sight this is puzzling since the irreducible representations of the tube algebra are two-dimensional in $\cH_1^{(5)}$ and one-dimensional in $\cH_{c_3}^{(1)}$. We however also have two maps into the two-dimensional representation. Starting from a single operator $\cO_{c_3}$ in $\cH_{c_3}^{(1)}$, we can compute the inner products of its two images:
\begin{equation}
	\label{map2d}
 \langle \cL_{1 \leftarrow c_3}^{(a)} \cO_{c_3}  |\cL_{1 \leftarrow c_3}^{(b)} \cO_{c_3}\rangle = \langle \cO_{c_3}| \cL_{c_3 \leftarrow 1}^{(a)} \times \cL_{1 \leftarrow c_3}^{(b)} |\cO_{c_3}\rangle = \delta_{ab} \langle \cO_{c_3}| \cO_{c_3}\rangle\, , \quad (a,b = 1,2)\, .
\end{equation}
Since the right-hand side is non-degenerate, the two images form a basis and therefore span the two-dimensional representation.

We can also go the other way. Consider the two states $|\psi_1\rangle$ and $|\psi_2\rangle$ which form an orthonormal basis of an irrep in $\cH_{1}^{(5)}$. There are again two maps into $\cH_{c_3}^{(1)}$, so we have a total of four possible  image states under the lasso map. The matrix of inner products $ \langle \psi_c| \cL_{1 \leftarrow c_3}^{(a)} \times \cL_{c_3 \leftarrow 1}^{(b)} |\psi_d\rangle$ ($a,b = 1, 2$ and $c,d = 1,2$) is however degenerate and only has a single non-zero eigenvalue (equal to 2 in our conventions). We therefore find only a single state in $\cH_{c_3}^{(1)}$.
 
Our conclusion is therefore that the Hilbert space $\cH_{1}^{(5)}$ is ``isomorphic'' to that of $ \cH_{c_3}^{(1)}$ in the sense that there is a two-dimensional irrep in the former for every one-dimensional irrep in the latter, with the same energy and spin. In terms of the partition functions we write:
\begin{equation}
	\label{strangeintertwinerspartitionfn}
	Z_1^{(5)}(u,\bar u) = 2\, Z_{c_3}^{(1)}(u,\bar u)\, .
\end{equation}

\subsubsection{$c_1 \leftrightarrow c_7$}
If we want to map the minimal central idempotents of the untwisted Hilbert space into the ones of the $c_7$ twisted Hilbert space, there are only two maps we can use, namely
\begin{align}
	&   \cL_{c_7 \leftarrow 1}^{(1)} =  \frac{\xi ^{3/2}}{2(\xi ^2+1)} \left(\cL_{c_7 \leftarrow 1}^{c_7 c_7} + \cL_{c_7 \leftarrow 1}^{c_8 c_8}\right)\, , \nn \\
	&  \cL_{c_7 \leftarrow 1}^{(2)} =\frac{\xi ^{3/2}}{2(\xi ^2+1)} \left(\cL_{c_7 \leftarrow 1}^{c_7 c_7} - \cL_{c_7 \leftarrow 1}^{c_8 c_8}\right)\, ,
\end{align}
which have as hermitian conjugates
\begin{align}
	&   \cL_{1 \leftarrow c_7}^{(1)} =  \frac{\xi ^{3/2}}{2(\xi ^2+1)}\left(\cL_{1 \leftarrow c_7}^{c_7 c_7} + \cL_{1 \leftarrow c_7}^{c_8 c_8}\right)\, , \nn \\
	&  \cL_{1 \leftarrow c_7}^{(2)} =\frac{\xi ^{3/2}}{2(\xi ^2+1)} \left(\cL_{1 \leftarrow c_7}^{c_7 c_7} - \cL_{1 \leftarrow c_7}^{c_8 c_8}\right)\, .
\end{align}
The $\cL^{(1)}$’s implement the isomorphism between the Hilbert spaces $\cH_1^{(3)}$ and  $\cH_{c_7}^{(0,1)}$, while the $\cL^{(2)}$’s implement the isomorphism between $\cH_1^{(4)}$ and $\cH_{c_7}^{(0,-1)}$.

\subsubsection{$c_3 \leftrightarrow c_7$}
There are six independent lasso maps between the sectors of the $c_3$ Hilbert space and the ones of the $c_7$ Hilbert space, which are
\begin{align}
	&   \cL_{c_7 \leftarrow c_3}^{(1)} = \frac{\xi ^{7/4}}{2 \left(\xi ^2+1\right)^{3/2}} \left(\cL_{c_7 \leftarrow c_3}^{c_5 c_7} + \cL_{c_7 \leftarrow c_3}^{c_6 c_8} + \cL_{c_7 \leftarrow c_3}^{c_7 c_5} + \cL_{c_7 \leftarrow c_3}^{c_8 c_6} + \frac{1}{\xi^{3/2}}\left(\cL_{c_7 \leftarrow c_3}^{c_7 c_7} + \cL_{c_7 \leftarrow c_3}^{c_8 c_8}\right)\right)\, , \nn  \\
	&  \cL_{c_7 \leftarrow c_3}^{(2)} = \frac{\xi ^{7/4}}{2 \left(\xi ^2+1\right)^{3/2}}  \left(\cL_{c_7 \leftarrow c_3}^{c_5 c_7} - \cL_{c_7 \leftarrow c_3}^{c_6 c_8} + \cL_{c_7 \leftarrow c_3}^{c_7 c_5} - \cL_{c_7 \leftarrow c_3}^{c_8 c_6} + \frac{1}{\xi^{3/2}}\left(\cL_{c_7 \leftarrow c_3}^{c_7 c_7} - \cL_{c_7 \leftarrow c_3}^{c_8 c_8}\right)\right) \, , \nn \\ 
	&  \cL_{c_7 \leftarrow c_3}^{(3)} = \frac{\xi ^{3/4}}{\left(\xi ^2+1\right)^{3/2}}\left( \cL_{c_7 \leftarrow c_3}^{c_5 c_7} + e^{4 \pi i /5} \cL_{c_7 \leftarrow c_3}^{c_7 c_5} + \frac{\sqrt{\xi }}{e^{4 \pi i /5} \xi +1} \cL_{c_7 \leftarrow c_3}^{c_7 c_7}\right) \, , \\
	&  \cL_{c_7 \leftarrow c_3}^{(4)} = \frac{\xi ^{3/4}}{\left(\xi ^2+1\right)^{3/2}} \left(\cL_{c_7 \leftarrow c_3}^{c_6 c_8} + e^{4 \pi i /5} \cL_{c_7 \leftarrow c_3}^{c_8 c_6} + \frac{\sqrt{\xi }}{e^{4 \pi i /5} \xi +1} \cL_{c_7 \leftarrow c_3}^{c_8 c_8}\right) \, , \nn \\
	&  \cL_{c_7 \leftarrow c_3}^{(5)} = \frac{\xi ^{3/4}}{\left(\xi ^2+1\right)^{3/2}} \left(\cL_{c_7 \leftarrow c_3}^{c_5 c_7} + e^{-4 \pi i /5} \cL_{c_7 \leftarrow c_3}^{c_7 c_5} + \frac{\sqrt{\xi }}{e^{-4 \pi i /5} \xi +1} \cL_{c_7 \leftarrow c_3}^{c_7 c_7}\right) \, , \nn \\
	&  \cL_{c_7 \leftarrow c_3}^{(6)} = \frac{\xi ^{3/4}}{\left(\xi ^2+1\right)^{3/2}} \left(\cL_{c_7 \leftarrow c_3}^{c_6 c_8} + e^{-4 \pi i /5} \cL_{c_7 \leftarrow c_3}^{c_8 c_6} + \frac{\sqrt{\xi }}{e^{-4 \pi i /5} \xi +1} \cL_{c_7 \leftarrow c_3}^{c_8 c_8} \right) \, . \nn
\end{align}
Their hermitian conjugates are obtained exchanging simultaneously $c_6$ with $c_4$ and $(c_7 \leftarrow c_3)$ with $(c_3 \leftarrow c_7)$. The maps $\cL^{(1)}$ and $\cL^{(2)}$ implement the isomorphisms $\cH_{c_3}^{(6)} \sim \cH_{c_7}^{(0,1)}$ and $\cH_{c_3}^{(7)} \sim \cH_{c_7}^{(0,-1)}$, while $\cL^{(3)}$ and $\cL^{(4)}$ map the two-dimensional Hilbert space  $\cH_{c_7}^{(8)}$ into two copies of the one-dimensional Hilbert space $\cH_{c_3}^{(5)}$ and vice versa, just as in subsubsection \ref{subsubsec:strangelasso} when we established the equality \eqref{strangeintertwinerspartitionfn}. 

Finally, the maps $\cL^{(5)}$ and $\cL^{(6)}$ realize a similar correspondence between the two-dimensional Hilbert space $\cH_{c_7}^{(7)}$ and the one-dimensional Hilbert space $\cH_{c_3}^{(3)}$.

\subsection{Hilbert spaces with a permutation line}
\label{subsec:lassop}

\subsubsection{$c_2 \leftrightarrow c_4$}
We have two lasso maps to go from the twisted $p$ to the twisted $c_4$ Hilbert space:
\begin{align}
	&  \cL_{c_4 \leftarrow c_2}^{(1)} = \frac{\xi ^{3/4}}{2 \sqrt{\xi ^2+1}} \left( \cL_{c_4 \leftarrow c_2}^{c_7 c_8} + \cL_{c_4 \leftarrow c_2}^{c_8 c_7}   \right) \, , \nn  \\
		&  \cL_{c_4 \leftarrow c_2}^{(2)} =\frac{\xi ^{3/4}}{2 \sqrt{\xi ^2+1}} \left( \cL_{c_4 \leftarrow c_2}^{c_7 c_8} - \cL_{c_4 \leftarrow c_2}^{c_8 c_7}   \right) \, . 
\end{align}
Their hermitian conjugates are obtained by simply exchanging $c_2$ with $c_4$. The $\cL^{(1)}$'s perform the isomorphism $\cH_{p}^{(1)} \sim \cH_{c_4}^{(1)}$, while the $\cL^{(2)}$'s implement $\cH_{p}^{(2)} \sim \cH_{c_4}^{(2)}$.

\subsubsection{$\{c_2, c_4\} \leftrightarrow c_8$}
There are six independent lassos which map the twisted $p$ into the twisted $c_8$ sectors, given by
\begin{align}
 & \cL_{c_8 \leftarrow c_2}^{(1)} = \frac{1}{2 \sqrt{\xi ^2+1}} \left(\cL_{c_8 \leftarrow c_2}^{c_3 c_6} - \cL_{c_8 \leftarrow c_2}^{c_6 c_3} +  \frac{1}{\sqrt{\xi}}\left( \cL_{c_8 \leftarrow c_2}^{c_8 c_7} - \cL_{c_8 \leftarrow c_2}^{c_7 c_8}  \right)\right) \, , \nn \\
  & \cL_{c_8 \leftarrow c_2}^{(2)} = \frac{1}{2 \sqrt{\xi ^2+1}}\left(\cL_{c_8 \leftarrow c_2}^{c_4 c_5} - \cL_{c_8 \leftarrow c_2}^{c_5 c_4} -  \frac{1}{\sqrt{\xi}} \left( \cL_{c_8 \leftarrow c_2}^{c_8 c_7} - \cL_{c_8 \leftarrow c_2}^{c_7 c_8}  \right)\right) \, ,\nn  \\
   & \cL_{c_8 \leftarrow c_2}^{(3)} =\frac{1}{2 \sqrt{\xi ^2+1}} \left(\cL_{c_8 \leftarrow c_2}^{c_3 c_6} + \cL_{c_8 \leftarrow c_2}^{c_6 c_3} -  \frac{1}{\sqrt{\xi}} \left( \cL_{c_8 \leftarrow c_2}^{c_8 c_7} + \cL_{c_8 \leftarrow c_2}^{c_7 c_8}  \right)\right) \, , 
    \\
    & \cL_{c_8 \leftarrow c_2}^{(4)} = \frac{1}{2 \sqrt{\xi ^2+1}} \left(\cL_{c_8 \leftarrow c_2}^{c_4 c_5} + \cL_{c_8 \leftarrow c_2}^{c_5 c_4} - \frac{1}{\sqrt{\xi}} \left( \cL_{c_8 \leftarrow c_2}^{c_8 c_7} + \cL_{c_8 \leftarrow c_2}^{c_7 c_8}  \right) \right)\, , \nn  \\
    & \cL_{c_8 \leftarrow c_2}^{(5)} = \frac{1}{2 \sqrt{\xi } \left(\xi ^2+1\right)} \left(\cL_{c_8 \leftarrow c_2}^{c_3 c_6} - \cL_{c_8 \leftarrow c_2}^{c_6 c_3} + \cL_{c_8 \leftarrow c_2}^{c_4 c_5} - \cL_{c_8 \leftarrow c_2}^{c_5 c_4}  - \sqrt{\xi} \left( \cL_{c_8 \leftarrow c_2}^{c_8 c_7} - \cL_{c_8 \leftarrow c_2}^{c_7 c_8}  \right)\right) \, ,\nn \\
      & \cL_{c_8 \leftarrow c_2}^{(6)} = \frac{1}{2 \sqrt{\xi } \left(\xi ^2+1\right)} \left( \cL_{c_8 \leftarrow c_2}^{c_3 c_6} + \cL_{c_8 \leftarrow c_2}^{c_6 c_3} + \cL_{c_8 \leftarrow c_2}^{c_4 c_5} + \cL_{c_8 \leftarrow c_2}^{c_5 c_4}  + \sqrt{\xi} \left( \cL_{c_8 \leftarrow c_2}^{c_8 c_7} + \cL_{c_8 \leftarrow c_2}^{c_7 c_8}  \right) \right)\,.\nn
\end{align}
Their hermitian conjugates are obtained via the simultaneous exchange $c_4 \leftrightarrow c_6$ and $(c_8 \leftarrow c_2) \leftrightarrow (c_2 \leftarrow c_8)$. 

Analogously, there are eight lasso maps that go from the twisted $c_4$ to the twisted $c_8$ Hilbert space. They can be written with the following independent combinations
\begin{align}
	 \cL_{c_8 \leftarrow c_4}^{(7)} = &\, \frac{1}{2 \xi ^{3/4} \sqrt{\xi ^2+1}} \left( \cL_{c_8 \leftarrow c_4}^{c_3 c_8} - \cL_{c_8 \leftarrow c_4}^{c_6 c_3} + \cL_{c_8 \leftarrow c_4}^{c_7 c_6} - \xi^{-1/2} \cL_{c_8 \leftarrow c_4}^{c_7 c_8} - \xi^{1/2} \cL_{c_8 \leftarrow c_4}^{c_8 c_7} \right)\, , \nn \\
	 \cL_{c_8 \leftarrow c_4}^{(8)} = & \, \frac{1}{2 \xi ^{3/4} \sqrt{\xi ^2+1}} \left(\cL_{c_8 \leftarrow c_4}^{c_4 c_7} - \cL_{c_8 \leftarrow c_4}^{c_5 c_4} + \cL_{c_8 \leftarrow c_4}^{c_8 c_5} - \xi^{1/2} \cL_{c_8 \leftarrow c_4}^{c_7 c_8} - \xi^{-1/2} \cL_{c_8 \leftarrow c_4}^{c_8 c_7}\right) \, , \nn \\
	 \cL_{c_8 \leftarrow c_4}^{(9)} = & \, \frac{1}{2 \xi ^{3/4} \sqrt{\xi ^2+1}} \left( \cL_{c_8 \leftarrow c_4}^{c_3 c_8} + \cL_{c_8 \leftarrow c_4}^{c_6 c_3} + \cL_{c_8 \leftarrow c_4}^{c_7 c_6} - \xi^{-1/2} \cL_{c_8 \leftarrow c_4}^{c_7 c_8} + \xi^{1/2} \cL_{c_8 \leftarrow c_4}^{c_8 c_7} \right)\, , \nn \\
	 \cL_{c_8 \leftarrow c_4}^{(10)} = & \, \frac{1}{2 \xi ^{3/4} \sqrt{\xi ^2+1}} \left( \cL_{c_8 \leftarrow c_4}^{c_4 c_7} + \cL_{c_8 \leftarrow c_4}^{c_5 c_4} + \cL_{c_8 \leftarrow c_4}^{c_8 c_5} + \xi^{1/2} \cL_{c_8 \leftarrow c_4}^{c_7 c_8} - \xi^{-1/2} \cL_{c_8 \leftarrow c_4}^{c_8 c_7}\right)  \, , \nn \\
	 \cL_{c_8 \leftarrow c_4}^{(11)} = & \, \frac{1}{2\xi^{1/4} \left(\xi ^2+1\right)}\Big(\cL_{c_8 \leftarrow c_4}^{c_3 c_8} -  \cL_{c_8 \leftarrow c_4}^{c_4 c_7} + e^{3 \pi i /5} \cL_{c_8 \leftarrow c_4}^{c_5 c_4}  + e^{-2 \pi i /5}  \cL_{c_8 \leftarrow c_4}^{c_6 c_3} \nonumber \\ & + e^{-4 \pi i /5}  \cL_{c_8 \leftarrow c_4}^{c_7 c_6}+ e^{ \pi i /5}  \cL_{c_8 \leftarrow c_4}^{c_8 c_5} +  e^{-2 \pi i /5} \xi^{-1/2} \cL_{c_8 \leftarrow c_4}^{c_7 c_8} +  e^{3 \pi i /5} \xi^{-1/2} \cL_{c_8 \leftarrow c_4}^{c_8 c_7}\Big) \, , \nn \\
	 \cL_{c_8 \leftarrow c_4}^{(12)} = & \, \frac{1}{2\xi^{1/4} \left(\xi ^2+1\right)}\Big(\cL_{c_8 \leftarrow c_4}^{c_3 c_8} +  \cL_{c_8 \leftarrow c_4}^{c_4 c_7} + e^{3 \pi i /5} \cL_{c_8 \leftarrow c_4}^{c_5 c_4}  + e^{3 \pi i /5}  \cL_{c_8 \leftarrow c_4}^{c_6 c_3}  \\ & + e^{-4 \pi i /5}  \cL_{c_8 \leftarrow c_4}^{c_7 c_6}  + e^{ -4 \pi i /5}  \cL_{c_8 \leftarrow c_4}^{c_8 c_5} +  e^{-2 \pi i /5} \xi^{-1/2} \cL_{c_8 \leftarrow c_4}^{c_7 c_8} +  e^{-2 \pi i /5} \xi^{-1/2} \cL_{c_8 \leftarrow c_4}^{c_8 c_7}\Big) \, , \nn \\
    \cL_{c_8 \leftarrow c_4}^{(13)} = & \, \frac{1}{2\xi^{1/4} \left(\xi ^2+1\right)}\Big(\cL_{c_8 \leftarrow c_4}^{c_3 c_8} -  \cL_{c_8 \leftarrow c_4}^{c_4 c_7} + e^{-3 \pi i /5} \cL_{c_8 \leftarrow c_4}^{c_5 c_4}  + e^{2 \pi i /5}  \cL_{c_8 \leftarrow c_4}^{c_6 c_3} \nonumber \\ & + e^{4 \pi i /5}  \cL_{c_8 \leftarrow c_4}^{c_7 c_6}  + e^{- \pi i /5}  \cL_{c_8 \leftarrow c_4}^{c_8 c_5} +  e^{2 \pi i /5} \xi^{-1/2} \cL_{c_8 \leftarrow c_4}^{c_7 c_8} +  e^{-3 \pi i /5} \xi^{-1/2} \cL_{c_8 \leftarrow c_4}^{c_8 c_7} \Big) \, , \nn \\
	\cL_{c_8 \leftarrow c_4}^{(14)} = & \, \frac{1}{2\xi^{1/4} \left(\xi ^2+1\right)} \Big( \cL_{c_8 \leftarrow c_4}^{c_3 c_8} +  \cL_{c_8 \leftarrow c_4}^{c_4 c_7} + e^{-3 \pi i /5} \cL_{c_8 \leftarrow c_4}^{c_5 c_4}  + e^{-3 \pi i /5}  \cL_{c_8 \leftarrow c_4}^{c_6 c_3} \nonumber \\ & + e^{4 \pi i /5}  \cL_{c_8 \leftarrow c_4}^{c_7 c_6}  + e^{ 4 \pi i /5}  \cL_{c_8 \leftarrow c_4}^{c_8 c_5} +  e^{2 \pi i /5} \xi^{-1/2} \cL_{c_8 \leftarrow c_4}^{c_7 c_8} +  e^{2 \pi i /5} \xi^{-1/2} \cL_{c_8 \leftarrow c_4}^{c_8 c_7} \Big) \, . \nn 
\end{align}
Their hermitian conjugates are non-trivial and given by
\begin{align}
	\cL_{c_4 \leftarrow c_8}^{(7)} = &\, \frac{1}{2 \xi ^{3/4} \sqrt{\xi ^2+1}} \left( \cL_{c_4 \leftarrow c_8}^{c_3 c_4} - \cL_{c_4 \leftarrow c_8}^{c_4 c_7} - \cL_{c_4 \leftarrow c_8}^{c_8 c_3} + \xi^{1/2} \cL_{c_4 \leftarrow c_8}^{c_7 c_8} + \xi^{-1/2} \cL_{c_4 \leftarrow c_8}^{c_8 c_7} \right)\, , \nn \\
	\cL_{c_4 \leftarrow c_8}^{(8)} = & \, \frac{1}{2 \xi ^{3/4} \sqrt{\xi ^2+1}} \left(\cL_{c_4 \leftarrow c_8}^{c_6 c_5} - \cL_{c_4 \leftarrow c_8}^{c_5 c_8} - \cL_{c_4 \leftarrow c_8}^{c_7 c_6} + \xi^{-1/2} \cL_{c_4 \leftarrow c_8}^{c_7 c_8} + \xi^{1/2} \cL_{c_4 \leftarrow c_8}^{c_8 c_7}\right) \, , \nn \\
	\cL_{c_8 \leftarrow c_4}^{(9)} = & \, \frac{1}{2 \xi ^{3/4} \sqrt{\xi ^2+1}} \left( \cL_{c_4 \leftarrow c_8}^{c_3 c_4} + \cL_{c_4 \leftarrow c_8}^{c_4 c_7} + \cL_{c_4 \leftarrow c_8}^{c_8 c_3} + \xi^{1/2} \cL_{c_4 \leftarrow c_8}^{c_7 c_8} - \xi^{-1/2} \cL_{c_4 \leftarrow c_8}^{c_8 c_7} \right)\, , \nn \\
	\cL_{c_4 \leftarrow c_8}^{(10)} = & \,\frac{1}{2 \xi ^{3/4} \sqrt{\xi ^2+1}} \left(\cL_{c_4 \leftarrow c_8}^{c_6 c_5} + \cL_{c_4 \leftarrow c_8}^{c_5 c_8} + \cL_{c_4 \leftarrow c_8}^{c_7 c_6} - \xi^{-1/2} \cL_{c_4 \leftarrow c_8}^{c_7 c_8} + \xi^{1/2} \cL_{c_4 \leftarrow c_8}^{c_8 c_7}\right) \, , \nn \\
	\cL_{c_4 \leftarrow c_8}^{(11)} = & \, \frac{1}{2\xi^{1/4} \left(\xi ^2+1\right)}\Big(\cL_{c_4 \leftarrow c_8}^{c_3 c_4} -  \cL_{c_4 \leftarrow c_8}^{c_6 c_5} + e^{3 \pi i /5} \cL_{c_4 \leftarrow c_8}^{c_7 c_6}  + e^{-2 \pi i /5}  \cL_{c_4 \leftarrow c_8}^{c_8 c_3} \nonumber \\ & + e^{2 \pi i /5}  \cL_{c_4 \leftarrow c_8}^{c_4 c_7 }+ e^{-3 \pi i /5}  \cL_{c_4 \leftarrow c_8}^{c_5 c_8} - \xi^{-1/2} \cL_{c_4 \leftarrow c_8}^{c_7 c_8} + \xi^{-1/2} \cL_{c_4 \leftarrow c_8}^{c_8 c_7}\Big) \, , \nn \\
	\cL_{c_4 \leftarrow c_8}^{(12)} = & \, \frac{1}{2\xi^{1/4} \left(\xi ^2+1\right)}\Big(\cL_{c_4 \leftarrow c_8}^{c_3 c_4} +  \cL_{c_4 \leftarrow c_8}^{c_6 c_5} + e^{3 \pi i /5} \cL_{c_4 \leftarrow c_8}^{c_7 c_6}  + e^{3 \pi i /5}  \cL_{c_4 \leftarrow c_8}^{c_8 c_3}  \\ & + e^{-3 \pi i /5}  \cL_{c_4 \leftarrow c_8}^{c_4 c_7 }+ e^{-3 \pi i /5}  \cL_{c_4 \leftarrow c_8}^{c_5 c_8} - \xi^{-1/2} \cL_{c_4 \leftarrow c_8}^{c_7 c_8} - \xi^{-1/2} \cL_{c_4 \leftarrow c_8}^{c_8 c_7}\Big) \, , \nn \\
	\cL_{c_4 \leftarrow c_8}^{(13)} = & \, \frac{1}{2\xi^{1/4} \left(\xi ^2+1\right)}\Big(\cL_{c_4 \leftarrow c_8}^{c_3 c_4} -  \cL_{c_4 \leftarrow c_8}^{c_6 c_5} + e^{-3 \pi i /5} \cL_{c_4 \leftarrow c_8}^{c_7 c_6}  + e^{2 \pi i /5}  \cL_{c_4 \leftarrow c_8}^{c_8 c_3} \nonumber \\ & + e^{-2 \pi i /5}  \cL_{c_4 \leftarrow c_8}^{c_4 c_7 }+ e^{3 \pi i /5}  \cL_{c_4 \leftarrow c_8}^{c_5 c_8} - \xi^{-1/2} \cL_{c_4 \leftarrow c_8}^{c_7 c_8} + \xi^{-1/2} \cL_{c_4 \leftarrow c_8}^{c_8 c_7}\Big) \, , \nn \\
	\cL_{c_4 \leftarrow c_8}^{(14)} = & \frac{1}{2\xi^{1/4} \left(\xi ^2+1\right)}\Big(\cL_{c_4 \leftarrow c_8}^{c_3 c_4} +  \cL_{c_4 \leftarrow c_8}^{c_6 c_5} + e^{-3 \pi i /5} \cL_{c_4 \leftarrow c_8}^{c_7 c_6}  + e^{-3 \pi i /5}  \cL_{c_4 \leftarrow c_8}^{c_8 c_3} \nonumber \\ & + e^{3 \pi i /5}  \cL_{c_4 \leftarrow c_8}^{c_4 c_7 }+ e^{3 \pi i /5}  \cL_{c_4 \leftarrow c_8}^{c_5 c_8} - \xi^{-1/2} \cL_{c_4 \leftarrow c_8}^{c_7 c_8} - \xi^{-1/2} \cL_{c_4 \leftarrow c_8}^{c_8 c_7}\Big) \, . \nn 
\end{align}
Let us focus on the lasso maps between the one-dimensional Hilbert spaces. From $\{c_2 \leftrightarrow c_8\}$ we find that $\cL^{(5)}$ and $\cL^{(6)}$ implement the isomorphisms $\cH_p^{(3)} \sim \cH_{c_8}^{(1,-1)}$ and  $ \cH_p^{(4)}\sim \cH_{c_8}^{(1,1)}$, while from $\{c_4 \leftrightarrow c_8\}$ we have four maps that implement $\cH_{c_4}^{(3)} \sim \cH_{c_8}^{(2,-1)}$, $\cH_{c_4}^{(4)} \sim \cH_{c_8}^{(2,1)}$, $\cH_{c_4}^{(5)} \sim \cH_{c_8}^{(3,-1)}$ and $\cH_{c_4}^{(6)} \sim \cH_{c_8}^{(3,1)}$. With these maps, all the one-dimensional sectors of the $c_8$ Hilbert space can be mapped into one-dimensional sectors in either the twisted $c_2$ or the twisted $c_4$ Hilbert spaces. 

The case of the two-dimensional sectors in the twisted $c_8$ Hilbert space is different. For illustration, let us examine what happens to the operators in the two-dimensional Hilbert space $\cH_{c_8}^{(\text{2d}, -1)}$. The non-trivial part of algebra implies that $\cL^{(1)}$, $\cL^{(2)}$, $\cL^{(7)}$ and $\cL^{(8)}$ map the operators in this Hilbert space to those in the one-dimensional Hilbert space $\cH_{c_2}^{(1)}$ or $\cH_{c_4}^{(1)}$, and vice versa. Since the $\cH_{c_2}^{(1)}$ and $\cH_{c_4}^{(1)}$ Hilbert spaces are themselves related by a lasso transformation, we may restrict our analysis to either one without loss of generality. Repeating the procedures used in subsubsection \ref{subsubsec:strangelasso} with the lasso maps $\cL^{(7)}$ and $\cL^{(8)}$, one finds that $\cH_{c_8}^{(\text{2d}, -1)}$ is mapped into two copies of the one-dimensional Hilbert space $\cH_{c_4}^{(1)}$ and vice versa. At the level of partition functions, this translates into
\begin{equation}
	Z_{c_8}^{(\text{2d}, -1)} (u,\bar u) = 2 \, Z_{c_4}^{(1)}(u, \bar u) =  2 \, Z_{c_2}^{(1)}(u, \bar u) \, .
\end{equation}
An entirely analogous computation using $\cL^{(3)}, \cL^{(4)}, \cL^{(9)}$ and $\cL^{(10)}$ yields
\begin{equation}
	Z_{c_8}^{(\text{2d}, 1)} (u,\bar u) = 2 \, Z_{c_4}^{(2)}(u, \bar u) =  2 \, Z_{c_2}^{(2)}(u, \bar u) \, .
\end{equation}
In summary, all sectors of the twisted $c_8$ Hilbert space can be generated starting from those of $c_2$ and/or $c_4$ and vice versa, meaning that from these $c_8$ sectors we can recover the ones of $c_2$ and/or $c_4$:
\begin{align}
	\cH_{c_8} \sim \cH_{c_2} \cup \cH_{c_4} \, .
\end{align}

\subsection{Drinfeld center}
\label{subsec:drinfeldcenter_complicated}
In section \ref{sec:irreps} we found 52 irreducible representations of $(\text{Fib} \boxtimes \text{Fib}) \rtimes S_2$.\footnote{This counting includes also the twisted $c_5$ and $c_6$ Hilbert spaces.} The lasso maps discussed previously reduce the number of independent ones down to 22, which agrees with the number simple objects in the Drinfeld center. These simple objects $(X, e_X)$ can be indeed classified by
\begin{center}
	\renewcommand{\arraystretch}{2}
	\begin{tabular}{c|c|c}
		 $X$ & $\# e_X$ & sets of minimal central idempotents \\
		\hline
	 $c_1$ &2 & $\{\hat P_1^{(1)}\}$, $\{\hat P_1^{(2)}\}$\\
		\hline
	  $c_1 + c_3 + c_5 + c_7$ & 2 & $\{\hat P_1^{(3)}, \hat P_{c_3}^{(6)},  \hat P_{c_5}^{(6)},  \hat P_{c_7}^{(0,1)}\}$, $\{\hat P_1^{(4)}, \hat P_{c_3}^{(7)},  \hat P_{c_5}^{(7)},  \hat P_{c_7}^{(0,-1)}\}$ \\
	 \hline
 $c_2 + c_4 + c_6 + 2 \, c_8$ & 2 & $\{\hat P_p^{(1)}, \hat P_{c_4}^{(1)},  \hat P_{c_6}^{(1)},  \hat P_{c_8}^{(\text{2d},-1)}\}$, $\{\hat P_p^{(2)}, \hat P_{c_4}^{(2)},  \hat P_{c_6}^{(2)},  \hat P_{c_8}^{(\text{2d},1)}\}$ \\
	\hline
$c_2 + c_8$ & 2 & $\{\hat P_p^{(3)},  \hat P_{c_8}^{(1,-1)}\}$, $\{\hat P_p^{(4)}, \hat P_{c_8}^{(1,1)}\}$ \\
	\hline
 $2\,  c_1 + c_3 + c_5$ & 1 & $\{\hat P_1^{(5)},  \hat P_{c_3}^{(1)},  \hat P_{c_5}^{(1)}\}$ \\
	\hline
	 $c_3 + c_5$ & 2 & $\{ \hat P_{c_3}^{(2)},  \hat P_{c_5}^{(2)}\}$, $\{ \hat P_{c_3}^{(4)},  \hat P_{c_5}^{(4)}\}$\\
	\hline
 $c_3 + c_5 + 2 \, c_7 $ & 2 & $\{ \hat P_{c_3}^{(3)},  \hat P_{c_5}^{(3)},  \hat P_{c_7}^{(7)}\}$, $\{ \hat P_{c_3}^{(5)},  \hat P_{c_5}^{(5)},  \hat P_{c_7}^{(8)}\}$\\
	\hline
 \multirow{2}{*}{$ c_4 + c_6 + c_8$} &  \multirow{2}{*}{4} & $\{ \hat P_{c_4}^{(3)},  \hat P_{c_6}^{(3)},  \hat P_{c_8}^{(2,-1)}\}$, $\{ \hat P_{c_4}^{(4)},  \hat P_{c_6}^{(4)},  \hat P_{c_8}^{(2,1)}\}$,  \\ && $\{ \hat P_{c_4}^{(5)},  \hat P_{c_6}^{(5)},  \hat P_{c_8}^{(3,-1)}\}$, $\{ \hat P_{c_4}^{(6)},  \hat P_{c_6}^{(6)},  \hat P_{c_8}^{(3,1)}\}$\\
	\hline
 $c_7$ & 4 & $\{ \hat P_{c_7}^{(2/5,1)}\}$, $\{ \hat P_{c_7}^{(-2/5,1)}\}$, $\{ \hat P_{c_7}^{(2/5,-1)}\}$, $\{ \hat P_{c_7}^{(-2/5,-1)}\}$ \\
	\hline
 $2 \, c_7$ & 1 & $\{ \hat P_{c_7}^{(9)}\}$ 
	\end{tabular}
\end{center}
We emphasize once again that the computation of the half-braidings is essentially equivalent to solving the tube algebra and identifying the lasso maps that act as intertwiners among the different sectors.

\section{Partition functions and their modular transforms}
By combining the results of the previous section we see that the 22 independent partition functions of $(\text{Fib} \boxtimes \text{Fib}) \rtimes S_2$ can be chosen to be:
\begin{align}
	\label{list}
	\vec Z(u, \bar u) = \Big\{& Z_1^{(1)}\, , \, Z_1^{(2)}\, , \, 	Z_1^{(3)}\, , \,  	Z_1^{(4)}\, , \,  Z_p^{(1)}\, , \, 	Z_p^{(2)}\, , \, Z_p^{(3)}\, , \, 	Z_p^{(4)}\, ,Z_{c_3}^{(1)}\, , \, Z_{c_3}^{(2)}\, , \,  Z_{c_3}^{(3)}\, , \, Z_{c_3}^{(4)}\, , \,  Z_{c_3}^{(5)}\, , \nonumber \\& Z_{c_4}^{(3)}\, , \, Z_{c_4}^{(4)}\, , \, Z_{c_4}^{(5)}\, , \, Z_{c_4}^{(6)}\, , \, Z_{c_7}^{(2/5,1)} \, , \, Z_{c_7}^{(-2/5,1)}  \, , \, Z_{c_7}^{(2/5,-1)}  \, , \, Z_{c_7}^{(-2/5,-1)} \, , \, Z_{c_7}^{(9)} \Big\}\left(u, \bar u\right) \, .
\end{align}
The modular $T$ matrix of this vector can be computed using equation \eqref{Ttransform} and the formulae of the minimal central idempotents found in section  \ref{sec:irreps}. In terms of the spins $\ell$ of the operators in each sector we have:
\begin{align}
 \ell = \Big\{ &0,\,  0, \, 0, \, 0, \, 1/2,\, 0, \, 1/2, \, 0, \, 0, \, 2/5, \, 2/5, \, -2/5, \, -2/5, \nonumber \\ & 1/5, \, -3/10, \, -1/5, \, 3/10, \, 4/5, \, -4/5, \, 4/5  , \, -4/5  , \, 0 \Big\} + \mathbb{Z} \, .
\end{align}
This result agrees with values of the `$T$' networks associated to each minimal central idempotent obtained by solving the algebra.

To obtain the modular $S$ matrix of \eqref{list}, we first express each partition function in \eqref{list} in terms of the $Z_i^{j k}$'s and implement their S transformation \eqref{Stransform}. Then we use the completeness results \eqref{completeness_eq} of subsubsection \ref{sec:completeness} together with the relations between partition functions due to lasso maps to rewrite the resulting partition functions as linear combinations of the ones in \eqref{list}.

To save space we labeled as $D \colonequals 1 + \xi^2 = \mathcal{D}_{\text{Fib}}^2$ the quantum dimension of the Fibonacci category squared. However, given the size of this matrix, we leave it to the reader to assemble it by cutting along the designated marks.

\begin{tikzpicture}
	\node[rotate = -90] at (0,0) {\Large $
\frac{1}{2 D^2}\left(
\begin{array}{ccccccccccc}
	1 & 1 & \xi ^4 & \xi ^4 & D \xi ^2 & D \xi ^2 & D & D & 2 \xi ^2 & 2 \xi  & 2 \xi ^3 \\
	1 & 1 & \xi ^4 & \xi ^4 & -D \xi ^2 & -D \xi ^2 & -D & -D & 2 \xi ^2 & 2 \xi  & 2 \xi ^3 \\
	\xi ^4 & \xi ^4 & 1 & 1 & D & D & D \xi ^2 & D \xi ^2 & 2 \xi ^2 & -2 \xi ^3 & -2 \xi  \\
	\xi ^4 & \xi ^4 & 1 & 1 & -D & -D & -D \xi ^2 & -D \xi ^2 & 2 \xi ^2 & -2 \xi ^3 & -2 \xi  \\
	D \xi ^2 & -D \xi ^2 & D & -D & D \xi ^2 & -D \xi ^2 & D & -D & 0 & 0 & 0 \\
	D \xi ^2 & -D \xi ^2 & D & -D & -D \xi ^2 & D \xi ^2 & -D & D & 0 & 0 & 0 \\
	D & -D & D \xi ^2 & -D \xi ^2 & D & -D & D \xi ^2 & -D \xi ^2 & 0 & 0 & 0 \\
	D & -D & D \xi ^2 & -D \xi ^2 & -D & D & -D \xi ^2 & D \xi ^2 & 0 & 0 & 0 \\
	2 \xi ^2 & 2 \xi ^2 & 2 \xi ^2 & 2 \xi ^2 & 0 & 0 & 0 & 0 & 6 \xi ^2 & 2 \xi ^2 & -2 \xi ^2 \\
	2 \xi  & 2 \xi  & -2 \xi ^3 & -2 \xi ^3 & 0 & 0 & 0 & 0 & 2 \xi ^2 & 2 \xi  & -4 \xi ^2 \\
	2 \xi ^3 & 2 \xi ^3 & -2 \xi  & -2 \xi  & 0 & 0 & 0 & 0 & -2 \xi ^2 & -4 \xi ^2 & 2 \xi  \\
	2 \xi  & 2 \xi  & -2 \xi ^3 & -2 \xi ^3 & 0 & 0 & 0 & 0 & 2 \xi ^2 & 4 \xi ^2 & 2 \xi ^3 \\
	2 \xi ^3 & 2 \xi ^3 & -2 \xi  & -2 \xi  & 0 & 0 & 0 & 0 & -2 \xi ^2 & 2 \xi ^3 & 4 \xi ^2 \\
	D \xi  & -D \xi  & -D \xi  & D \xi  & -D \xi  & D \xi  & D \xi  & -D \xi  & 0 & 0 & 0 \\
	D \xi  & -D \xi  & -D \xi  & D \xi  & D \xi  & -D \xi  & -D \xi  & D \xi  & 0 & 0 & 0 \\
	D \xi  & -D \xi  & -D \xi  & D \xi  & -D \xi  & D \xi  & D \xi  & -D \xi  & 0 & 0 & 0 \\
	D \xi  & -D \xi  & -D \xi  & D \xi  & D \xi  & -D \xi  & -D \xi  & D \xi  & 0 & 0 & 0 \\
	\xi ^2 & \xi ^2 & \xi ^2 & \xi ^2 & -D \xi  & -D \xi  & D \xi  & D \xi  & -2 \xi ^2 & -2 \xi  & 2 \xi  \\
	\xi ^2 & \xi ^2 & \xi ^2 & \xi ^2 & -D \xi  & -D \xi  & D \xi  & D \xi  & -2 \xi ^2 & 2 \xi ^3 & -2 \xi ^3 \\
	\xi ^2 & \xi ^2 & \xi ^2 & \xi ^2 & D \xi  & D \xi  & -D \xi  & -D \xi  & -2 \xi ^2 & -2 \xi  & 2 \xi  \\
	\xi ^2 & \xi ^2 & \xi ^2 & \xi ^2 & D \xi  & D \xi  & -D \xi  & -D \xi  & -2 \xi ^2 & 2 \xi ^3 & -2 \xi ^3 \\
	4 \xi ^2 & 4 \xi ^2 & 4 \xi ^2 & 4 \xi ^2 & 0 & 0 & 0 & 0 & -8 \xi ^2 & 4 \xi ^2 & -4 \xi ^2 \\
\end{array}
\right.
$};

\node at (-7.5,-10) {\Huge\Rightscissors};
\draw[thick, dashed] (-7,-10) -- (-5,-10);
\draw[thick, dashed] (-5,-10) -- (-4,-12);
\draw[thick,dashed] (-4,-12) -- (4.5,-12);
\draw[thick,dashed] (4.5,-12) -- (5.5,-10);
\draw[thick,dashed] (5.5,-10) -- (7.5,-10);
\draw[thick] (-5,-10) -- (5.5,-10);
\node at (0.2,-11) {\Huge \text{glue}};
\end{tikzpicture}

\begin{tikzpicture}
	
	\draw[thick, dashed] (-4,4) -- (4.5,4);
	\draw[thick, dashed] (-5,6) -- (-4,4);
	\draw[thick,dashed] (5.5,6) -- (4.5,4);
	\draw[thick] (-7,6) -- (7.5,6);
	
	\node[rotate = -90] at (0,-5) {\Large $
\left.
\begin{array}{ccccccccccc}
	2 \xi  & 2 \xi ^3 & D \xi  & D \xi  & D \xi  & D \xi  & \xi ^2 & \xi ^2 & \xi ^2 & \xi ^2 & \xi ^2 \\
	2 \xi  & 2 \xi ^3 & -D \xi  & -D \xi  & -D \xi  & -D \xi  & \xi ^2 & \xi ^2 & \xi ^2 & \xi ^2 & \xi ^2 \\
	-2 \xi ^3 & -2 \xi  & -D \xi  & -D \xi  & -D \xi  & -D \xi  & \xi ^2 & \xi ^2 & \xi ^2 & \xi ^2 & \xi ^2 \\
	-2 \xi ^3 & -2 \xi  & D \xi  & D \xi  & D \xi  & D \xi  & \xi ^2 & \xi ^2 & \xi ^2 & \xi ^2 & \xi ^2 \\
	0 & 0 & -D \xi  & D \xi  & -D \xi  & D \xi  & -D \xi  & -D \xi  & D \xi  & D \xi  & 0 \\
	0 & 0 & D \xi  & -D \xi  & D \xi  & -D \xi  & -D \xi  & -D \xi  & D \xi  & D \xi  & 0 \\
	0 & 0 & D \xi  & -D \xi  & D \xi  & -D \xi  & D \xi  & D \xi  & -D \xi  & -D \xi  & 0 \\
	0 & 0 & -D \xi  & D \xi  & -D \xi  & D \xi  & D \xi  & D \xi  & -D \xi  & -D \xi  & 0 \\
	2 \xi ^2 & -2 \xi ^2 & 0 & 0 & 0 & 0 & -2 \xi ^2 & -2 \xi ^2 & -2 \xi ^2 & -2 \xi ^2 & -2 \xi ^2 \\
	4 \xi ^2 & 2 \xi ^3 & 0 & 0 & 0 & 0 & -2 \xi  & 2 \xi ^3 & -2 \xi  & 2 \xi ^3 & \xi ^2 \\
	2 \xi ^3 & 4 \xi ^2 & 0 & 0 & 0 & 0 & 2 \xi  & -2 \xi ^3 & 2 \xi  & -2 \xi ^3 & -\xi ^2 \\
	2 \xi  & -4 \xi ^2 & 0 & 0 & 0 & 0 & 2 \xi ^3 & -2 \xi  & 2 \xi ^3 & -2 \xi  & \xi ^2 \\
	-4 \xi ^2 & 2 \xi  & 0 & 0 & 0 & 0 & -2 \xi ^3 & 2 \xi  & -2 \xi ^3 & 2 \xi  & -\xi ^2 \\
	0 & 0 & -D \xi ^2 & D \xi ^2 & D & -D & -D & D \xi ^2 & D & -D \xi ^2 & 0 \\
	0 & 0 & D \xi ^2 & -D \xi ^2 & -D & D & -D & D \xi ^2 & D & -D \xi ^2 & 0 \\
	0 & 0 & D & -D & -D \xi ^2 & D \xi ^2 & D \xi ^2 & -D & -D \xi ^2 & D & 0 \\
	0 & 0 & -D & D & D \xi ^2 & -D \xi ^2 & D \xi ^2 & -D & -D \xi ^2 & D & 0 \\
	2 \xi ^3 & -2 \xi ^3 & -D & -D & D \xi ^2 & D \xi ^2 & 1 & \xi ^4 & 1 & \xi ^4 & -\xi ^2 \\
	-2 \xi  & 2 \xi  & D \xi ^2 & D \xi ^2 & -D & -D & \xi ^4 & 1 & \xi ^4 & 1 & -\xi ^2 \\
	2 \xi ^3 & -2 \xi ^3 & D & D & -D \xi ^2 & -D \xi ^2 & 1 & \xi ^4 & 1 & \xi ^4 & -\xi ^2 \\
	-2 \xi  & 2 \xi  & -D \xi ^2 & -D \xi ^2 & D & D & \xi ^4 & 1 & \xi ^4 & 1 & -\xi ^2 \\
	4 \xi ^2 & -4 \xi ^2 & 0 & 0 & 0 & 0 & -4 \xi ^2 & -4 \xi ^2 & -4 \xi ^2 & -4 \xi ^2 & 6 \xi ^2 \\
\end{array}
\right)\, .
$};
\end{tikzpicture}

\section{Conclusions}
We have worked out the representation theory of the tube algebra for $(\text{Fib} \boxtimes \text{Fib}) \rtimes S_2$ and used it to calculate the modular S and T matrices. In \cite{toappear} this result will be an essential ingredient of the numerical modular bootstrap search for non-rational Virasoro CFTs with this symmetry. In that work we will also check our S matrix in the explicit example of two disconnected tricritical Ising models.

We expect the category worked out in this paper to be ``sufficiently complicated'', in the sense that it exhibits all the intricacies that distinguish a representation of non-invertible symmetry category from that of an ordinary finite group, including a rich structure of lasso maps between different sub-Hilbert spaces. It is in particular more involved than the Ising example worked out in detail in \cite{Bhardwaj:2023ayw}. In comparing with the literature we should also note the intricate connection between the modular S matrix and the fusion rules of the Verlinde lines in an RCFT, see section 2.2 of \cite{Thorngren:2021yso}. That computation however cannot be used here because our simple lines are not all Verlinde lines. Finally, note that very recently \cite{Albert:2025umy} provided several relatively complicated modular S matrices, once more highlighting the demand for publicly available software to automate this process.

\section{Acknowledgments}
We would like to thank Ant\'onio Antunes, Mathew Bullimore, Cl\'ement Delcamp, Laurens Lootens, Junchen Rong, Francesco Russo and Xi Yin for discussions. The authors would like to thank ICTP-SAIFR (FAPESP grant 2021/14335-0) where part of this work was done. BvR also thanks the Yukawa Institute for Theoretical Physics at Kyoto University, where part of this work was written during ``Progress of Theoretical Bootstrap''. We acknowledge funding from the European Union (ERC “QFTinAdS”, project number 101087025).

\bibliography{biblio}
\bibliographystyle{utphys}

\end{document}